\documentclass[11pt]{article}
\pdfoutput=1 
\usepackage{shorthand}
﻿

\usepackage{mathtools}
\usepackage{booktabs}
\usepackage[english]{babel}
\usepackage{amsmath,amssymb,amsbsy,amstext, amsthm, simplewick, amsfonts,braket}
\usepackage{graphicx}
\usepackage[small]{caption}
\usepackage{siunitx}
\usepackage{upgreek}
\usepackage{framed}
\usepackage{wrapfig}
\usepackage{multirow}
\usepackage{bbm}
\usepackage[numbers,sort&compress]{natbib}
\usepackage[svgnames,dvipsnames,x11names]{xcolor}
\usepackage[utf8x]{inputenc}
\usepackage{selinput}
\usepackage{bm}
\usepackage{float}
\usepackage{dsfont}
\usepackage{caption}
\usepackage{subcaption}
\usepackage{sidecap}
\usepackage{longtable}
\usepackage{anyfontsize}

\usepackage{epstopdf}
\usepackage{cancel}
\usepackage{tcolorbox}
\usepackage{latexsym,amsmath,amssymb,epsfig}
\usepackage{braket}
\usepackage{tensor}
\usepackage{tocloft}

\usepackage{tikz-cd}
\usepackage{ytableau}
\ytableausetup{centertableaux,boxsize=.8em}
\usepackage{genyoungtabtikz}
\Yboxdim{14pt} 
\Ylinethick{.9pt}

﻿
\usepackage{tcolorbox}
\definecolor{greyish2}{rgb}{.96,.96,.96}
﻿
 
﻿
﻿
\def\xyma{\xymatrix@M.7em}
\def\xymas{\xymatrix@M.1em}
﻿
\newcommand{\Comment}[1]{{}}
\definecolor{darkblue}{rgb}{0.15,0.35,0.55}
\definecolor{reddish}{rgb}{0.65, 0.2, 0.2}
\definecolor{darkgreen}{RGB}{50,150,0}
\definecolor{greyish2}{rgb}{.96,.96,.96}
\usepackage[linktocpage=true]{hyperref}
\hypersetup{
colorlinks=true,
citecolor=darkblue,
linkcolor=reddish,
urlcolor=darkblue,
pdfauthor={},
pdftitle={},
pdfsubject={}
}
﻿
﻿

﻿

﻿
﻿
﻿
﻿
 
\DeclareFontFamily{OT1}{rsfs10}{}
\DeclareFontShape{OT1}{rsfs10}{m}{n}{ <-> rsfs10 }{}
\DeclareMathAlphabet{\mathscript}{OT1}{rsfs10}{m}{n}
﻿
﻿
﻿
﻿
\def\gsim{ \lower .75ex \hbox{$\sim$} \llap{\raise .27ex \hbox{$>$}} }
\def\lsim{ \lower .75ex \hbox{$\sim$} \llap{\raise .27ex \hbox{$<$}} }
\def\be{\begin{equation}}
\def\ee{\end{equation}}
\def\bea{\begin{eqnarray}}
\def\eea{\end{eqnarray}}

\newcommand{\baaa}{\begin{eqnarray}}
\newcommand{\eaaa}{\end{eqnarray}}
﻿

﻿

﻿

\newcommand{\rd}{{\rm d}}

﻿
﻿

﻿

﻿

﻿
\DeclareMathOperator{\E}{e}

﻿
﻿
\newcommand{\D}{{\rm d}}

\newcommand{\Mpl}{M_{\rm Pl}} 

\newcommand{\overbar}[1]{\mkern 2.5mu\overline{\mkern-2.5mu#1\mkern-2.5mu}\mkern 2.5mu}
\renewcommand \bar [1] {\overbar{#1}}
﻿

﻿
﻿

\usepackage[letterpaper,margin=1in]{geometry}
﻿
﻿
﻿

﻿
\usepackage{tikz}
\usetikzlibrary{decorations}
\pgfdeclaredecoration{complete sines}{initial}
{
    \state{initial}[
        width=+0pt,
        next state=upsine,
        persistent precomputation={\pgfmathsetmacro\matchinglength{
            \pgfdecoratedinputsegmentlength / int(\pgfdecoratedinputsegmentlength/\pgfdecorationsegmentlength)}
            \setlength{\pgfdecorationsegmentlength}{\matchinglength pt}
        }] {}
    \state{upsine}[width=\pgfdecorationsegmentlength,next state=downsine]{
        \pgfpathsine{\pgfpoint{0.25\pgfdecorationsegmentlength}{0.5\pgfdecorationsegmentamplitude}}
        \pgfpathcosine{\pgfpoint{0.25\pgfdecorationsegmentlength}{-0.5\pgfdecorationsegmentamplitude}}
    }
    \state{downsine}[width=\pgfdecorationsegmentlength,next state=upsine]{
        \pgfpathsine{\pgfpoint{0.25\pgfdecorationsegmentlength}{-0.5\pgfdecorationsegmentamplitude}}
        \pgfpathcosine{\pgfpoint{0.25\pgfdecorationsegmentlength}{0.5\pgfdecorationsegmentamplitude}}
}
    \state{final}{}
}
﻿
﻿
\definecolor{greyish}{rgb}{.90,.90,.90}
\definecolor{greyish2}{rgb}{.96,.96,.96}
\usepackage{xcolor,colortbl}
\usepackage{tcolorbox}
﻿
﻿
\usepackage[all]{xy}
﻿
﻿
﻿
﻿
﻿
﻿
﻿
﻿
\numberwithin{equation}{section}
﻿
﻿
﻿
﻿
\begin{document}
﻿
﻿
%
\renewcommand{\thefootnote}{\fnsymbol{footnote}}
\vspace{0truecm}
\thispagestyle{empty}
﻿
\begin{center}
{\fontsize{20}{24} \bf
Dynamical Tidal Response\\[9pt]
of Schwarzschild Black Holes}
\end{center}
﻿
﻿
\vspace{-.2truecm}
﻿
﻿
\begin{center}
{\fontsize{13.25}{18}\selectfont
Oscar Combaluzier-Szteinsznaider,${}^{\rm a}$
Daniel Glazer,${}^{\rm b}$
Austin Joyce,${}^{\rm b}$\\[4.5pt]
Maria J. Rodriguez,${}^{\rm c,d,e}$
and Luca Santoni${}^{\rm a}$
}
\end{center}
﻿
﻿
\vspace{.1truecm}
﻿
\begin{small}
﻿
\centerline{{\it ${}^{\rm a}$Universit\'e Paris Cit\'e, CNRS, Astroparticule et Cosmologie,}}
\centerline{{\it 10 Rue Alice Domon et L\'eonie Duquet, F-75013 Paris, France}} 
﻿
 \vspace{.3cm}
  
\centerline{{\it ${}^{\rm b}$Kavli Institute for Cosmological Physics, Department of Astronomy and Astrophysics}}
\centerline{{\it The University of Chicago, 
5640 S Ellis Ave, Chicago, IL 60637, USA} } 
 
\vspace{.3cm}
﻿
\centerline{{\it ${}^{\rm c}$Department of Physics, Utah State University,}}
\centerline{{\it 4415 Old Main Hill Road, UT 84322, USA} } 
﻿
 \vspace{.3cm}
 
 \centerline{{\it ${}^{\rm d}$Black Hole Initiative, Harvard University, Cambridge MA 02138, USA}}
 
  \vspace{.3cm}
﻿
 
 \centerline{{\it ${}^{\rm e}$Instituto de Fisica Teorica UAM/CSIC, Universidad Autonoma de Madrid,}}
\centerline{{\it 13-15 Calle Nicolas Cabrera, 28049 Madrid, Spain} } 
﻿
﻿
﻿
\end{small}
  
\vspace{.3cm}
\begin{abstract}
\noindent
Dynamical Love numbers capture the conservative response of an object to a time-dependent external tidal gravitational field. We compute the dynamical Love numbers of Schwarzschild black holes in general relativity within a point-particle effective field theory framework. In addition to the known logarithmic running, we compute the finite scheme-dependent contributions to the Love number couplings.
We do this by matching the renormalized one-point function in the effective theory to the classical field profile computed in general relativity. 
On the general relativity side, we solve the Regge--Wheeler and Zerilli equations perturbatively in a small frequency expansion. In order to match on the effective field theory side we include gravitational interactions using the Born series and employ dimensional regularization to obtain a renormalized field profile.

\end{abstract}
﻿
﻿
﻿
\newpage
﻿
\setcounter{page}{2}
\setcounter{tocdepth}{2}
\tableofcontents
\newpage
\renewcommand*{\thefootnote}{\arabic{footnote}}
\setcounter{footnote}{0}
﻿
﻿
﻿
﻿
﻿
 
﻿
﻿
﻿
\section{Introduction}
\label{sec:intro}

We often figure out what is inside things by shaking them. When presented with objects as diverse as a wrapped present, a piggy bank, or a cereal box, our first instinct is to give them a shake to get an idea of what is inside. It should therefore come as no surprise that this is also an effective strategy to learn about the most mysterious objects in nature: black holes. Unfortunately, black holes are too big and too far away for us to grab, so we must be more flexible in what it means to ``shake” one. Fortunately, the universe performs a version of this experiment for us. 
From far away we can encode the properties of any object in terms of how it responds to external stimuli~\cite{Goldberger:2004jt,Goldberger:2005cd}.
A time-dependent tidal gravitational perturbation effectively makes an object oscillate in response, producing its own gravitational field. The details of this induced gravitational field depend on the make-up of the object, which is encoded in (dynamical) Love numbers~\cite{10.1093/mnras/69.6.476,Poisson_Will_2014}.
When other heavy objects move past black holes, or orbit them, their changing gravitational field induces a dynamical response in the black hole.
This response is physical, and in principle can be measured, for example, in the spectrum of gravitational waves emitted by a binary.
Here we compute these dynamical Love numbers of Schwarzschild black holes.
 
The tidal responses of black holes are well-studied. The leading response of any object to an external tidal gravitational field is to mechanically deform. This response is captured by the object's static Love number. A remarkable---and somewhat counterintuitive---fact is that the static Love numbers of black holes vanish exactly in (four-dimensional) general relativity~\cite{Fang:2005qq,Damour:2009vw,Binnington:2009bb,Kol:2011vg,Hui:2020xxx, Hui:2021vcv,Hui:2022vbh,Rai:2024lho,LeTiec:2020spy,LeTiec:2020bos,Charalambous:2021mea,Gurlebeck:2015xpa}.
 From the perspective of effective field theory (EFT), this result implies that, as far as static perturbations are concerned, black holes in general relativity behave like elementary particles, indistinguishable from point-like objects with no internal structure. 
This emergent simplicity, peculiar to four-dimensional  general relativity~\cite{Kol:2011vg,Hui:2020xxx,Rodriguez:2023xjd,Charalambous:2023jgq,Charalambous:2024tdj,Charalambous:2025ekl,Glazer:2024eyi,Vines:2017hyw,Chung:2018kqs,Arkani-Hamed:2019ymq}, has been the focus of much interest in recent years, not least because it suggests the presence of new symmetries of general relativity~\cite{Hui:2021vcv,Hui:2022vbh,BenAchour:2022uqo,Charalambous:2021kcz,Charalambous:2022rre,Berens:2022ebl,Rai:2024lho,Combaluzier-Szteinsznaider:2024sgb,Lupsasca:2025pnt,Parra-Martinez:2025bcu,Berens:2025okm,Berens:2025jfs}.
Given the vanishing of the static linear Love numbers, it is both theoretically interesting and observationally motivated to ask to what extent this hidden simplicity persists at subleading order. Broadly speaking, two classes of subdominant finite-size effects, beyond the linear static tidal response, can be identified: dynamical (i.e., time-dependent) effects and nonlinearities. Nonlinear corrections to the Love numbers have recently been investigated for both black holes and neutron stars. Interestingly, explicit calculations show that the nonlinear Love numbers of black holes continue to vanish in four-spacetime dimensions~\cite{Gurlebeck:2015xpa,Poisson:2020vap,Poisson:2021yau,Riva:2023rcm,Iteanu:2024dvx,Combaluzier-Szteinsznaider:2024sgb,DeLuca:2023mio,Gounis:2024hcm,Kehagias:2024rtz}, hinting at a fully nonlinear symmetry structure underlying the static sector of general relativity~\cite{Combaluzier-Szteinsznaider:2024sgb,Kehagias:2024rtz,Parra-Martinez:2025bcu}.

In this paper we study subleading-in-frequency tidal effects,
namely the dynamical response. Specifically, we aim to compute the induced time-dependent tidal deformation of Schwarzschild black holes by matching results obtained in general relativity to a point-particle EFT description of black holes~\cite{Goldberger:2004jt,Goldberger:2005cd,Hanson:1974qy,Bailey:1975fe,Porto:2005ac,Steinhoff:2011sya,Delacretaz:2014oxa,Goldberger:2020fot,Porto:2016pyg,Levi:2018nxp,Kol:2011vg,Hui:2020xxx,Charalambous:2021mea}. This provides an unambiguous definition of the dynamical Love numbers~\cite{Goldberger:2004jt,Goldberger:2005cd}.
The general topic of black hole dynamical response is well-studied~\cite{Chakrabarti:2013lua,Hinderer:2016eia,Steinhoff:2016rfi,Poisson:2020vap,Charalambous:2021mea,Pratten:2021pro,Pitre:2023xsr,Chakraborty:2023zed,Perry:2023wmm,Saketh:2023bul,Jakobsen:2023pvx,Mandal:2023hqa,Ivanov:2024sds,Katagiri:2024wbg,HegadeKR:2024agt,Yu:2024uxt,DeLuca:2024ufn,Katagiri:2024fpn,Bhatt:2024rpx,Saketh:2024juq,Chakraborty:2025wvs,Pitre:2025qdf,Kobayashi:2025swn}. In particular, it is well known that frequency-dependent effects are nonzero. For example, a black hole absorbs radiation through its horizon, which, at linear order in frequency, corresponds to a positive imaginary part of the response coefficients~\cite{Chia:2020yla,Charalambous:2021mea,Chia:2024bwc}.
Solutions to the perturbation equations at second order in frequency have also been obtained by various methods. These include the Mano--Suzuki--Takasugi (MST) formalism~\cite{Mano:1996vt,Sasaki:2003xr}---which expresses the solution as a series of special functions truncated at a chosen order in the small-frequency limit---as well as more standard perturbative approaches~\cite{Katagiri:2024wbg,Chakraborty:2025wvs}. It is by now well understood that the dynamical Love numbers, unlike their static counterparts, exhibit a logarithmic dependence on distance, which is interpreted as a classical example of renormalization-group running in the point-particle EFT  description of the  object~\cite{Chakrabarti:2013lua,Charalambous:2021mea,Saketh:2023bul,Ivanov:2024sds,Katagiri:2024wbg,Chakraborty:2025wvs,Caron-Huot:2025tlq}.
However, although the coefficient of the logarithm has  been previously computed using the above methods, as well as other symmetry-based arguments~\cite{Perry:2023wmm}, existing results in the literature remain incomplete for gravitational perturbations (see, however,~\cite{Ivanov:2024sds,Caron-Huot:2025tlq} for a scalar field tidal matching). In particular, a full computation of the (renormalized) dynamical Love numbers cannot be performed at tree level, but  requires evaluating certain (classical) higher-loop diagrams in the point-particle EFT after subtracting ultraviolet divergences.

In this work, we extend previous analyses in several directions. Foremost, we perform (for the first time) the complete matching between general relativity and EFT, so that in addition to the logarithmic running of the 
gravitational dynamical Love numbers, we compute their scheme-dependent finite terms. Unlike the logarithmic coefficient, which is universal, these  terms depend on the chosen renormalization scheme (we will work here in dimensional regularization).
This work also contains a number of technical features of interest. The computation of dynamical Love numbers requires solutions to the equations of black hole perturbation theory at subleading orders in the frequency expansion. We develop a systematic expansion in frequency that allows us to relatively easily extract the terms of interest. In order to account for gravitational effects in the EFT, we follow the elegant approach of~\cite{Correia:2024jgr,Caron-Huot:2025tlq}, though we choose to match the graviton one-point function, as opposed to a scattering amplitude. A benefit of matching this off-shell quantity is that the matching can be performed in a suitably defined near zone. This simplifies both the general-relativistic computation and the EFT matching, obviating the need to resolve the far-zone dynamics. For completeness, we also demonstrate in a toy scalar-field example that, when extended to the far zone, our result reproduces the MST solution and agrees with~\cite{Caron-Huot:2025tlq}.

The most salient aspect of our results is that the dynamical Love numbers of Schwarzschild black holes are indeed nonzero. Aside from this, there are some features worth noting. The responses in the parity even and odd sectors are the same (more properly, there is a renormalization scheme in which they are equal). This can be viewed as a consequence of Chandrasekhar's symmetry that maps these two sectors into each other~\cite{10.2307/78902,1975RSPSA.343..289C,Chandrasekhar:1985kt}. In addition, the responses display intriguing patterns, suggesting there are further insights to be mined from the study of black hole responses.

\noindent
{\bf Outline:} In Section~\ref{sec:scalar}, as a preliminary example, we analyze dynamical scalar response. While our main focus is on gravitational perturbations, the scalar field case allows us to present the underlying logic in a simpler setup and to highlight the technical differences with respect to previous works. In Section~\ref{sec:gravDLNs}, we consider gravitational dynamical tidal effects. First, we solve the Regge--Wheeler and Zerilli equations in a full general-relativistic setup perturbatively in the frequency. We then  perform the matching with the point-particle EFT. In the EFT, we employ dimensional regularization and introduce a renormalization scheme to remove ultraviolet divergences. Through this matching, we obtain the renormalized dynamical Love number couplings up to second order in frequency. Several appendices collect complementary and more technical results that are relevant but lie somewhat outside the main line of the text. In Appendix~\ref{app:scalarkl} we derive general expressions for the running of the  scalar dynamical response for generic multipole number. In Appendix~\ref{app:farzone}, we further analyze the far-zone of the scalar field example, which we use as a crosscheck to show that our approach reproduces the MST solution and previous results. Appendix~\ref{app:l=3-4_solution} contains explicit expressions for the gravitational field solutions and their renormalized counterparts. Finally, Appendix~\ref{app:tidalfield} discusses some technical aspects  of the gravitational tidal field.

﻿
\noindent
{\bf Conventions:} We use the mostly-plus metric convention  $(-,+,\cdots,+)$, and denote the spacetime dimension by $D=d+1$, with $d$ the spatial dimension. Spacetime indices are denoted by Greek letters $\mu,\nu,\cdots$, while spatial indices are denoted by Roman letters $i,j,k,\cdots a,b,c,\cdots$. We will often use multi-index notation, where the multi-index $A_\ell = i_1\cdots i_\ell$. The notation $( {\cdots} )_T$ indicates the trace-subtracted symmetrization of the enclosed indices.
We (anti)symmetrize indices with weight one, so that for example $A_{[ij]} = \frac{1}{2}(A_{ij}-A_{ji})$.
 In many cases we decompose fields using spherical harmonics, where we denote the angular momentum by $\ell$, and the magnetic quantum number by $m$.
﻿
﻿

﻿
﻿
\section{Warm-up: Dynamical scalar response}
\label{sec:scalar}

In order to orient ourselves, we first consider a preliminary example: the dynamical response to an external massless scalar field profile.
From the general relativistic perspective, this problem involves solving the Klein--Gordon equation for a massless scalar field propagating in a Schwarzschild spacetime.
We begin by explaining how to solve this equation perturbatively in the frequency of the field, obtaining a solution that is valid at all spatial distances. Then, following~\cite{Caron-Huot:2025tlq}, we use these solutions to match to worldline EFT.
The final results of this computation are known. The new contribution of the approach taken here is that:
(1) We show that, to perform the EFT matching and obtain the response coefficients, it is not necessary to compute far-field observables; it is  enough to know the solution in a suitably defined intermediate zone. (2) By introducing appropriate far regions,  we show that our perturbative solution reproduces the  MST~\cite{Mano:1996vt,Sasaki:2003xr} result, up to linear order in the small-frequency limit.

While this section is useful for setting up notation, its results are not strictly essential for the remainder of the work. Readers already familiar with these ideas may choose to skip directly to Section~\ref{sec:gravDLNs}.

﻿
\subsection{Small-frequency expansion of relativistic solution}
\label{sec:scalarGR}

We begin by considering the general relativity side. 
We are interested in the dynamics of a massless scalar field, $\Phi$, on a four-dimensional  Schwarzschild spacetime, defined by the line element
\begin{equation}
\D s^2 = -f(r)\D t^2+f(r)^{-1}\D r^2 +r^2(\D \theta^2+ \sin^2\theta \, \D\varphi^2)\,,
\qquad f(r)\equiv 1-\frac{r_s}{r}\,,
\end{equation}
where $r_s\equiv 2G M$ is the Schwarzschild radius.
After decomposing the field in spherical harmonics and in frequency space
\begin{equation}
    \Phi(t,r,\theta,\varphi)=\sum_{\ell=0}^\infty\sum_{m=-\ell}^{\ell}\int \frac{\D\omega}{2\pi}\E^{-i \omega t}\phi_{\omega \ell m}(r)Y_{\ell m}(\theta,\varphi)\,,
\end{equation}
the Klein--Gordon equation $\nabla_\mu\nabla^\mu\Phi=0$ takes the form 
\begin{equation}
\partial_r \big[r(r-r_s) \partial_r\phi(r)\big]   + \left[ \frac{ r^3 \omega ^2}{r-r_s}  - \ell(\ell+1) \right] \phi (r) =0  \,.
\label{eqKGPhi}
\end{equation}
Here we have dropped the $\omega\ell m$ subscript of $\phi(r)$ for convenience, but we will sometimes restore some of these labels when it is useful.

We would like to solve~\eqref{eqKGPhi} at finite frequency $\omega$. In particular, we want to extract the behavior of the solution at asymptotically far distances ($r=\infty$), after imposing appropriate boundary conditions at the black hole horizon ($r=r_s$). From the perspective of the theory of linear differential equations, this type of question is equivalent to deriving the connection formulas for the solutions of~\eqref{eqKGPhi}. As  is well known, at finite frequency, the Klein--Gordon equation~\eqref{eqKGPhi} is of the confluent Heun type: it has regular singularities at $r=0$ and $r=r_s$, and an irregular singularity of rank $1$ at $r=\infty$.\footnote{The Klein--Gordon equation \eqref{eqKGPhi} can be brought to the  standard form of the confluent Heun equation~\cite{MR1392976,slavjanov2000special,Bateman:100233}
\begin{equation}
\frac{\D^2}{\D z^2}w(z)+ \left( \frac{\gamma}{z} + \frac{\delta}{z-1} +\epsilon  \right) \frac{\D}{\D z}w(z) + \frac{\alpha z-q}{z(z-1)}w(z) =0 \,,
\end{equation}
 via the change of variable $\phi(r)=(r-r_s)^{i \omega r_s}\E^{i \omega r} w(z(r))$, with $z\equiv r/r_s$. It has parameters $\gamma=1$, $\delta=1+2i\omega r_s$, $\epsilon=\alpha=2i \omega r_s$ and $q=\ell(\ell+1)$.}
Connection formulas for Heun equations---or, more generally, for Fuchsian equations with more than three regular singularities---are not known generally in closed form. Explicit analytic expressions can be obtained in certain limits, or perturbatively in some parameter.\footnote{See e.g.~\cite{Aminov:2020yma,Bonelli:2021uvf,Bonelli:2022ten,Lisovyy:2022flm,Aminov:2023jve} for recent results in this context.}
This makes solving the problem at finite frequency difficult. However, in some cases it is sufficient to understand the behavior in the limit $\omega \to 0$. Indeed, it is in this limit that the connection to an effective field theory description of compact objects is most transparent.

We are therefore motivated to work in the adiabatic regime. That is, we will assume $\omega r_s\ll1$ and seek solutions to~\eqref{eqKGPhi} perturbatively in $\omega r_s$. 
It is important to note that, regardless of how small the frequency is, the term $r^3 \omega ^2/(r-r_s)$ in~\eqref{eqKGPhi} is not always guaranteed to be small.
For instance, sufficiently close  to the horizon ($r\rightarrow r_s$), this term dominates and eventually becomes the leading component of the potential. Likewise, this term grows at large distances, as $r\rightarrow \infty$.
As such, the small frequency approximation is necessarily more complicated than simply neglecting the $r^3 \omega ^2/(r-r_s)$ term.
A natural way to proceed is to adopt an asymptotic expansion approach, reducing the problem to a simplified set of equations by making approximations which are valid in different regions. 
The complete solution can then be reconstructed by matching the various  solutions at the boundaries of their respective regions of validity.\footnote{This type of approach is sometimes referred to as ``boundary layer theory''. See~\cite{RWhite} for a more rigorous introduction to the theory of boundary layer problems. See also~\cite{Unruh:1976fm,Bezerra:2013iha,Hui:2019aqm,Bucciotti:2023bvw,Hui:2022sri} 
for a non-exhaustive  list of  scalar field applications on various black hole spacetimes.}

We will define three different zones (following~\cite{Unruh:1976fm}):
\vspace{-3pt}
\begin{itemize}

\item {\bf Near Zone (NZ):} Defined by the condition that  $r-r_s\ll r_s$.
\vspace{-1pt}

\item {\bf Intermediate Zone (IZ):} Defined by $r_s \lesssim r \ll \omega^{-1}$.
\vspace{-1pt}

\item {\bf Far Zone (FZ):} Where $r\gg r_s$.

\end{itemize}
In each of these regions, we will be able to neglect certain terms in the original equation~\eqref{eqKGPhi}, allowing us to find an approximate, analytic closed-form solution. A central conceptual point of the analysis is that in order to match to point-particle EFT, it suffices to know the solution in the Intermediate Zone, as this has an overlapping regime of validity with the EFT description, if one matches the one-point function.\footnote{{Note that what we are calling the ``Intermediate Zone" is called the ``Near Zone" in much of the literature. Perhaps what we call the Near Zone would better be called the ``Very Near Zone". We apologize for any confusion.}} 
In~\cite{Caron-Huot:2025tlq} they match to the EFT in the Far Zone, by matching scattering amplitudes to scattering computed in GR using the MST formalism. In order to make contact with this description, we show in Appendix~\ref{app:farzone} that the solution at small frequency reproduces the results of MST when matched to the Far Zone.

We now turn to solving the Klein--Gordon equation in each of these zones and matching across their interfaces.
﻿
﻿
\subsubsection{Near zone}

We begin by considering the Near Zone.
In the vicinity of  the horizon, the potential in~\eqref{eqKGPhi} is dominated by the term $r^3 \omega ^2/(r-r_s)$, and the equation  can be approximated by 
\begin{equation}
\partial_r \big[r(r-r_s) \partial_r\phi_{\rm NZ}\big]   +  \frac{ r^3 \omega ^2}{r-r_s}   \phi_{\rm NZ} =0\,.
    \label{eq:Sch_NZ}
\end{equation}
Up to corrections of order $(r-r_s)/r_s\ll1$, the independent solutions of~\eqref{eq:Sch_NZ} are $\E^{ \pm i \omega r_{\star}}/r$, where $r_\star$ is the tortoise coordinate defined by
\begin{equation}
    \frac{\D r_\star}{\D r} \equiv \left(1-\frac{r_s}{r}\right)^{-1},
    \qquad
    r_\star = r+r_s \log \left(\frac{r}{r_s}-1\right) .
\end{equation}
Imposing ingoing boundary conditions at the horizon,  the physical near-zone solution thus reads
\begin{equation}
    \phi_{\text{NZ}}(r)=B \frac{r_s}{r} \mathrm{e}^{-i \omega\left(r+r_s \log \left(\frac{r}{r_s}-1\right)\right)},
    \label{phiNZ0}
\end{equation}
where $B$ is some arbitrary integration constant.
﻿
﻿
\subsubsection{Intermediate zone}
\label{sec:intermediate-zone-scalar}
﻿
In the  Intermediate Zone, defined such that $r_s \lesssim r \ll \omega^{-1}$, the term involving $\omega^2$ in~\eqref{eqKGPhi}
is genuinely small compared to all the other terms in the equation. 
We can thus treat it perturbatively, and look for a series solution in $\omega$.
To enable this, it is convenient to introduce the  quantities
\begin{equation}
    x\equiv \frac{2r}{r_s}-1\,, \qquad \epsilon\equiv\omega r_s\,,
\label{defxrs}
\end{equation}
where $\epsilon$ is the small parameter we will perturb in.
In terms of these variables~\eqref{eqKGPhi} takes the form
\begin{equation}
    (1-x^2)\partial_x^2\phi_{\rm IZ}-2x\partial_x\phi_{\rm IZ}+\ell(\ell+1)\phi_{\rm IZ}=\epsilon^2\frac{(x+1)^3}{4(x-1)}\phi_{\rm IZ}\,,
    \label{eq:psi_eq}
\end{equation}
where we have simply moved the term proportional to $\epsilon^2$ to the right-hand side, in order to treat it as a source. 
The idea is  to look for a solution of the form
\begin{equation}
    \phi_{\text{IZ}}=\phi^{(0)}+\epsilon \phi^{(1)}+\epsilon^2 \phi^{(2)}+\cdots,
    \label{eq:psi_expansion}
\end{equation}
and solve perturbatively in $\epsilon$.
At each order, the source on the right hand side can be written in terms of lower-order solutions, allowing us to find a solution
using Green's function methods.

Since the right-hand side of~\eqref{eq:psi_eq} starts at order $\epsilon^2$, both $\phi^{(0)}$ and $\phi^{(1)}$ solve the homogeneous equation. 
This equation is Legendre's differential equation, so the solutions can be written in terms of the Legendre polynomials $P_\ell(x)$ and $Q_\ell(x)$.\footnote{We are using the definition of $Q_\ell(x)$ with argument $x>1$. That is,
\begin{equation}
Q_\ell(x) = \frac{1}{2}P_\ell(x)\log\left( \frac{x+1}{x-1}  \right) - \sum_{n=1}^\ell\frac{1}{n} P_{n-1}(x)P_{\ell-n}(x) .
\end{equation}
See also~\eqref{Qell} in Appendix~\ref{app:scalarkl}.}
We can therefore parameterize them as
\begin{align}
    \phi_\ell^{(0)}(x) & =b_{1}^{(0)}P_\ell(x) +b_{2}^{(0)}Q_\ell(x) ,
    \\
    \phi_\ell^{(1)}(x) & =b_1^{(1)}P_\ell(x)+b_2^{(1)}Q_\ell(x),
    \label{eq:psi1}
\end{align}
for some integration constants $b_1^{(k)}$ and $b_2^{(k)}$.\footnote{The constants $b_1^{(k)}$ and $b_2^{(k)}$ are $\ell$-dependent, as we will explicitly see in some examples below. } These can be determined by matching the perturbative solutions $\phi_\ell^{(0)}(x)$ and $\phi_\ell^{(1)}(x)$ to the near-zone solution~\eqref{phiNZ0} across the boundary at $r=r_s$.\footnote{{As noted in~\cite{Unruh:1976fm}, formally the approximations that we have made to define the Near and Intermediate zones have no region of overlap. Nevertheless, this does not preclude us from matching across their interface~\cite{RWhite}.}}
To this end, it is useful to recall the asymptotic behavior of the Legendre polynomials near $x = 1$ (see e.g.,~\cite{NIST:DLMF,Magnus:1966}): 
\begin{equation}
    P_\ell(x) \underset{x\to1}{\sim} 1 \,,
\qquad\quad
    Q_\ell(x) \underset{x\to1}{\sim} -\frac12\log\left(\frac{x-1}{2}\right)-H_\ell\,,
\end{equation}
where $H_\ell$ is the harmonic number, $H_\ell\equiv \sum_{k=1}^\ell \frac{1}{k}=\gamma_E+\psi(\ell+1)$, with  $\psi(z)\equiv \Gamma'(z)/\Gamma(z)$ the digamma function, and $\gamma_E$ the  Euler--Mascheroni  constant.

Let us first compare the $x\rightarrow1$ limit of the intermediate-zone solution $\phi_\ell^{(0)}(x)$ with the near-zone solution \eqref{phiNZ0} expanded at zeroth order in~$\omega$. By matching the two solutions, one readily finds
\begin{equation}
b_1^{(0)}=B , \qquad b_2^{(0)}=0.
\end{equation}
Proceeding similarly for the $\phi_\ell^{(1)}(x) $ solution, and comparing it with $\phi_{\text{NZ}}$  at linear order in $\omega$, we obtain
\begin{equation}
    b_1^{(1)}=i B\left(2H_\ell-1\right) , \qquad\qquad b_2^{(1)}=2iB.
    \label{eq:phi1_cst}
\end{equation}

We now want to find the intermediate-zone solution at second order in the small-$\epsilon$ expansion. Plugging the expansion~\eqref{eq:psi_expansion} into~\eqref{eq:psi_eq}, and truncating at order $\epsilon^2$, we obtain the following inhomogeneous equation for $\phi_\ell^{(2)}$:
\begin{equation}
    (1-x^2)\partial_x^2\phi_\ell^{(2)}(x)-2x\partial_x\phi_\ell^{(2)}(x)+\ell(\ell+1)\phi_\ell^{(2)}(x)=\frac{(1+x)^3}{4(x-1)}\phi_\ell^{(0)}(x)\equiv S_\ell(x),
    \label{eq:Sch_IZO2}
\end{equation}
where the source on the right-hand side is fully determined by the zeroth-order solution $\phi_\ell^{(0)}$.
A general solution to~\eqref{eq:Sch_IZO2} can be written as
\be
\phi_\ell^{(2)}(x) =   \phi_{\ell,\text{p}}^{(2)}(x)+ \phi_{\ell,\text{h}}^{(2)}(x)\,,
\ee
where $\phi_{\ell,\text{h}}^{(2)}(x)$ is the most general homogeneous solution to the equation
\begin{equation}
    \phi_{\ell,\text{h}}^{(2)}(x)=b_1^{(2)}P_\ell(x)+b_2^{(2)}Q_\ell(x),
\label{eq:hsphi2}
\end{equation}
and where $ \phi_{\ell,\text{p}}^{(2)}(x)$ is a particular solution to the equation, which can be 
expressed  as a convolution between the Green's function of the homogeneous equation and the source, as follows:\footnote{We have used that the combination $W\equiv (1-x^2)[P_\ell(x)\partial_x Q_\ell (x)-Q_\ell(x)\partial_x P_\ell (x)]$, proportional to the Wronskian of the homogeneous equation, is independent of $x$ for all $\ell$'s, i.e.~$\partial_x W=0$. In particular, one can check that~$W\equiv 1$.} 
\begin{equation}
    \phi_{\ell,\text{p}}^{(2)}(x)=Q_\ell(x)\int^x_{x_0} \D y \, P_\ell(y)S_\ell(y)-P_\ell(x)\int^x_{x_0} \D y \, Q_\ell(y)S_\ell(y),
\label{eq:psphi2}
\end{equation}
where $x_0$ is an arbitrary constant, which can be fixed to any convenient value when computing the integrals. Different choices of $x_0$ can be reabsorbed into the coefficients $b_1^{(2)}$ and $b_2^{(2)}$ of the homogeneous solution.\footnote{For this reason one can alternatively leave the lower bounds  unspecified and treat the integrals as indefinite. We specify the lower limit to prevent any ambiguity.}

All together, the second-order intermediate-zone solution $\phi_\ell^{(2)}$ reads
\begin{equation}
    \phi_\ell^{(2)}(x)=\left[b_1^{(2)} - \frac{B}{4}\int^x_{x_0}\D y\frac{(1+y)^3}{y-1}P_\ell(y)Q_\ell(y) \right]P_\ell(x) + \left[b_2^{(2)} + \frac{B}{4}\int^x_{x_0}\D y\frac{(1+y)^3}{y-1}P_\ell(y)^2 \right]Q_\ell(x).
    \label{eq:psi2GreenInt}
\end{equation}
Once again, the integration constants $b_1^{(2)}$ and $b_2^{(2)}$ can be fixed by matching the intermediate-zone solution $\phi_\ell^{(2)}(x)$ with the near-zone solution \eqref{phiNZ0} expanded at  second order in the frequency. For instance, in the case  $\ell=0$,  it is straightforward to find (with $x_0=2$)
\begin{equation}
\begin{aligned}
    b_{1,\ell=0}^{(2)} &= 
    \frac{B}{24}  \left[24 \text{Li}_2\left(-\frac{1}{2}\right)-27-90 \log 3+4 \log 2\, (11+\log 8)\right],\\
    b_{2,\ell=0}^{(2)} &= B \left(\frac{35}{6}-\log 4\right),
    \end{aligned}
    \label{eq:bb1200}
\end{equation}
and similarly for higher values of $\ell$.

For sufficiently small $\omega$, the intermediate zone ($r_s \lesssim r \ll \omega^{-1}$) extends  arbitrarily far from $r_s$. For later convenience, it will be useful to write down the large-$r$ expansion of this intermediate-zone solution. 
The solutions are the following, where we keep subleading terms up to $r^{-\ell-1}$ (here we have solved for the $b_1^{(2)}$ and $b_2^{(2)}$ coefficients for $\ell=1, 2$ by matching with the near zone as in~\eqref{eq:bb1200}):
\be
\begin{aligned}
    \phi_{\ell=0}(r)\underset{\frac{r}{r_s}\to\infty}{\sim} B&+Bi\omega r_s\left(\frac{r_s}{r}-1\right) \\
    &+ B\omega ^2r_s^2\bigg[-\frac{r^2}{6r_s^2}-\frac{5r}{6r_s}+\frac{3-\pi^2}{6}-\frac{11}{6}\log\frac{r}{r_s}+\frac{r_s}{r}\left(2+\log\frac{r}{r_s}\right)\bigg]+\cdots\,,
\end{aligned}
\label{eq:phiGRl0}
\ee

\be
\begin{aligned}
    \phi_{\ell=1}(r)\underset{\frac{r}{r_s}\to\infty}{\sim} &B\left(\frac{2r}{r_s}-1\right)+Bi\omega r_s\left(\frac{2r}{r_s}-1+\frac{r_s^2}{6r^2}\right) \\
    &~~+ B\omega^2 r_s^2 \bigg[-\frac{r^3}{5 r_s^3}-\frac{7 r^2}{10 r_s^2}- \frac{r}{r_s}\left(\frac{101+10\pi^2}{30}+\frac{19}{15} \log \frac{r}{r_s}\right)\\
   &~~ \hspace{1.5cm}+ \frac{113+5\pi^2}{30}+ \frac{19}{30} \log\frac{r}{r_s}+\frac{r_s}{2 r}+\frac{r_s^2}{6r^2}\left(\frac{11}{6}+ \log\frac{r}{r_s}\right)\bigg]+\cdots\,,
\label{eq:phiGRl0b}
\end{aligned}
\ee
\be
\begin{aligned}
    \phi_{\ell=2}(r)\underset{\frac{r}{r_s}\to\infty}{\sim} & B\left( \frac{6r^2}{r_s^2}-\frac{6r}{r_s}+1\right)+Bi\omega r_s\left( \frac{12r^2}{r_s^2}-12\frac{r}{r_s} + 2+\frac{r_s^3}{30r^3}\right)\\
    &~~+B \omega^2 r_s^2\bigg[-\frac{3r^4}{7r_s^4}-\frac{8r^3}{7r_s^3}-\frac{r^2}{r_s^2}\left( \frac{4133+210\pi^2}{210}+\frac{79}{35}\log\frac{r}{r_s} \right)\\[2pt]
    &~~~~\hspace{1.5cm}+\frac{r}{r_s}\left( \frac{1003+42\pi^2}{42}+\frac{79}{35}\log\frac{r}{r_s} \right) -\frac{972+35\pi^2}{210}\\
   &~~~~\hspace{1.5cm}-\frac{79}{210}\log\frac{r}{r_s} +\frac{r_s}{6r}+\frac{5r_s^2}{24r^2}+\frac{r_s^3}{30r^3}\left(\frac{287}{60}+\log\frac{r}{r_s}\right)\bigg]+\cdots\,.
\end{aligned}
\label{eq:phiGRl2}
\ee
\vskip1pt
\noindent
Some interesting physical observations can be abstracted from these explicit solutions for particular $\ell$ values.
First of all, as expected, one notices the standard flat-space falloffs $r^\ell$ and $r^{-\ell-1}$ of the Klein--Gordon equation, with $\omega$-dependent coefficients.
Technically, this is because---even though we are parametrically distant from $r_s$---we are not yet in the asymptotically far region (what we called the Far Zone), where the falloffs (at finite frequency) differ (see Appendix~\ref{app:farzone}).  In addition, we see logarithmic scaling in $r$ in the decaying $r^{-\ell-1}$ terms. These are well known and can be interpreted as a classical running of the response in the infrared solution~\cite{Kol:2011vg,Hui:2020xxx,Ivanov:2024sds,Ivanov:2025ozg,Caron-Huot:2025tlq}. The coefficient of the logarithm remains the same at all distances, and is scheme-independent, so there is a sense in which it is universal. This can be computed in full generality for arbitrary $\ell$ from~\eqref{eq:psi2GreenInt}, as we show in Appendix~\ref{app:scalarkl}.  In contrast, finite constant corrections to effective tidal response couplings will in general depend on the regularization prescription in the EFT, and require an explicit matching calculation. 
This will be the subject of the next subsection.

To compute the EFT coefficients, we will match to the  large-$r$ expansion of the intermediate-zone solutions above. The intermediate zone overlaps with the region of validity of the EFT ($r\gg r_s$), so it is not necessary to compute far-zone observables (like scattering amplitudes) for this purpose. Since it will not play a major role in the following, we postpone the discussion of the far zone to Appendix~\ref{app:farzone}, where we will in particular demonstrate that it correctly reproduces previous results obtained via different methods.
﻿

﻿
﻿
\subsection{EFT calculation of scalar dynamical response}
\label{sec:scalarppEFT}
﻿

In order to provide an unambiguous definition of the physical responses of compact objects, we will employ point-particle effective field theory, following~\cite{Goldberger:2004jt,Goldberger:2005cd}. This effective description takes advantage of the fact that any localized object can be approximated as a point particle from sufficiently far away. Its internal composition is encoded in the effective couplings of the particle's worldline to external fields. This approach has twin advantages of being systematic and unambiguous. Our goal in this Section is to derive the effective couplings of a Schwarzschild black hole to a scalar probe. To do so, we employ the formalism of~\cite{Caron-Huot:2025tlq} to treat the effects of coupling the point particle to gravity. One difference in our detailed approach is that we choose to match the off-shell scalar field profile in the intermediate zone between the general relativistic calculation and the EFT description.

The calculation proceeds in three steps. We first derive the response of a point particle to an external scalar probe in the absence of gravity (or near to the point particle). We then couple to gravity using the Born series organization of the interactions~\cite{Correia:2024jgr,Caron-Huot:2025tlq}. This yields a bare solution which must then be renormalized and matched to the general relativistic solutions obtained in Section~\ref{sec:scalarGR}. The end output of the calculation is the scalar response coefficients of the black hole.

In order to regulate some ultraviolet divergences in the EFT calculation, it is useful to conduct the calculation in generic dimension. We therefore need the Schwarzschild metric in $D$ spacetime dimensions:
\begin{equation}
\D s^2 = -f(r)\D t^2+f(r)^{-1}\D r^2 +r^2\D \Omega_{S^{D-2}}^2 ,
\label{eq:bargS}
\end{equation}
where $\rd\Omega_{S^{D-2}}^2$ is the line element on the $(D-2)$-sphere.\footnote{For completeness, the sphere metric in hyperspherical coordinates can be defined recursively using the relation $\rd \Omega_{S^n}^2 = \rd \theta_n^2 + \sin^2\theta_n \rd\Omega_{S^{n-1}}^2$, with $\rd\Omega_{S^1}^2 = \rd\theta_1^2$. In this notation, $\theta_1 \in (0,2\pi)$ while all other angles $\theta_i\in (0,\pi)$.} 
The function $f(r)$ is defined as
\begin{equation}
f(r)\equiv 1-\left(\frac{r_s}{r}\right)^{D-3} = 1-\frac{2 G_\text{N}M n_D \mu^{2\varepsilon}}{r^{D-3}}  ,
\qquad {\rm where}~~n_D\equiv \frac{4\pi^{\frac{3-D}{2}}\Gamma\left(\frac{D-1}{2}\right)}{D-2}.
\label{eq:frD}
\end{equation}
Here we have 
used the relation between the  Schwarzschild radius $r_s$ and  the asymptotically flat black hole mass $M$, and we defined $\varepsilon\equiv \frac{4-D}{2}$, which parameterizes the deviation from four dimensions. In~\eqref{eq:frD}, we introduced a scale $\mu$ with the dimension of energy to ensure that the units of $G$ are independent of $\varepsilon$, and in particular remain the same as in $D=4$~\cite{Caron-Huot:2025tlq}. This is useful as we will eventually take the limit $\varepsilon \rightarrow0$ ($D\rightarrow 4$). In this limit, the scale $\mu$ is analogous to the arbitrary scale introduced in dimensional regularization of standard quantum field theory calculations with Feynman diagrams.

\subsubsection{EFT setup}

Famously, things can fall into black holes and not escape. As such, they are dissipative systems, whose detailed microstate we cannot track. Consequently their effective description is an open EFT~\cite{Schwinger:1960qe,Keldysh:1964ud} (see also~\cite{kamenev2011field,Akyuz:2023lsm,Haehl:2015foa,Crossley:2015evo,Liu:2018kfw,Haehl:2024pqu,Glorioso:2016gsa,Weinberg:2005vy} for recent developments). 
The point-particle EFT action in the Schwinger--Keldysh approach, including dissipative effects, is~\cite{Goldberger:2004jt,Goldberger:2005cd,Goldberger:2020fot,Ivanov:2022hlo,Saketh:2023bul,Glazer:2024eyi,Caron-Huot:2025tlq}
\begin{equation}
    S= S_{\text{bulk}}+S_\text{pp}+S_\text{int} .
\label{eq:Seft}
\end{equation}
The first term, $S_{\text{bulk}}$, is the $D$-dimensional scalar action,
\begin{equation}
S_\text{bulk}=-\int \D^Dx\sqrt{-g}g^{\mu\nu}\partial_\mu\Phi_+\partial_\nu\Phi_-,
\label{eq:Sbulk}
\end{equation}
which describes the scalar field's dynamics in the bulk spacetime. The second term is the worldline action of the object in the point-particle approximation,
\begin{equation}
    S_\text{pp}=-M\int \D\tau = - M\int \D \lambda\sqrt{-g_{\mu\nu}  \frac{\D x^\mu}{\D \lambda} \frac{\D x^\nu}{\D \lambda} }  , 
\label{eq:Spps}
\end{equation}
where $x^\mu(\lambda)$ parametrizes the location of the point particle as a function of the affine parameter $\lambda$. Finally, we introduce couplings between the scalar and the particle worldline 
\begin{equation}
    S_\text{int}= \int \D\tau \sum_{\ell=0}^\infty Q_I^{A_\ell}(\tau)\nabla^{(\ell)}_{A_\ell}\Phi^{I}(x(\tau)) \,.
\label{eq:Sintscalar}
\end{equation}
Here we have defined the traceless combination of derivatives  $\nabla^{(\ell)}_{A_\ell} \equiv \nabla_{( i_1}\cdots \nabla_{i_\ell)_T}$,
where $(\cdots)_T$ denotes the symmetrized trace-free part of the enclosed indices.\footnote{Note that~\eqref{eq:Sintscalar} is written in terms of spatial indices, and so does not look covariant. This can be rectified by defining
the projected covariant derivative $\nabla^\bot_{\mu} \equiv {\mathbb{P}^\nu}_\mu \nabla_\nu$, where ${\mathbb{P}^\nu}_\mu \equiv  \delta_\mu^\nu + u_\mu u^\nu$ projects onto the plane orthogonal to the four velocity $u^\mu\equiv \D x^\mu/\D \tau$ (where $\tau$ is the proper time, and normalized such that $u^\mu u_\mu=-1$).} 
The composite  multipole operator $Q_I^{A_\ell}(\tau)$ is built from internal degrees of freedom $X_\pm$ localized on the worldline of the particle, which physically we can think of as modeling the degrees of freedom into which energy and other charges are dissipating~\cite{Goldberger:2005cd}.\footnote{Introducing the 
auxiliary worldline degrees of freedom $X$ is not strictly necessary. One can instead
 directly write down the effective couplings~\eqref{eq:Kdef} between the $\pm$ fields,
following the Schwinger--Keldysh approach~\cite{Feynman:1963fq,Caldeira:1982iu,Calzetta:1986cq,Kamenev:2009jj,Salcedo:2025ezu}. The two approaches are equivalent and lead to the same description of the open system.} 
Correlation functions of the effective operator $Q$ describe how the point particle reacts to the presence of an external field $\Phi$.

To account for dissipative effects, in \eqref{eq:Sbulk} and \eqref{eq:Sintscalar} we have  introduced two copies of the scalar field $\{\Phi_1,\Phi_2\}$ and  defined $\Phi_+\equiv \frac{1}{2}(\Phi_1+\Phi_2)$ and $\Phi_-\equiv \Phi_1-\Phi_2$. The indices $I,J$ run over these two copies $+,-$ and are contracted with the off-diagonal unit matrix. Each copy lives on a different branch of a two-sided  closed-time contour running from $t=-\infty$ up to some time of interest and then back to $t=-\infty$.

Since we are interested in the response to external $\Phi$ profiles, we can
integrate out the internal degrees of freedom $X$ to obtain an effective action
\begin{equation}
    \E^{\Gamma^\text{in-in}[\Phi_\pm]}=\int \mathcal{D}X_+\mathcal{D}X_- \E^{iS\left[ \Phi_\pm,X_\pm \right]}.
\end{equation}
At leading order, we can do this by replacing $Q$ by its linear response
\begin{equation}
    \langle Q_{I}^{A_\ell}(\tau)\rangle=\int \D\tau' K_{IJ}^{(\Phi)~A_\ell \vert B_{\ell'}}(\tau-\tau')\nabla^{(\ell)}_{B_{\ell'}}\Phi^J(\tau')\,,
\end{equation}
where $K_{IJ}^{(\Phi)\,A_\ell \vert B_{\ell'}}$ is a Green's function of the $Q$ degrees of freedom, corresponding to the two-point function of $Q$ in the Keldysh basis via~\cite{Goldberger:2020fot,Saketh:2022xjb,Saketh:2023bul,Glazer:2024eyi}
\begin{equation}
    \langle Q_{I}^{A_{\ell}}(\tau)Q_{J}^{B_{\ell'}}(\tau')\rangle=-iK_{IJ}^{(\Phi)~A_{\ell} \vert B_{\ell'}}(\tau-\tau')\,.
    \label{eq:Kdef}
\end{equation}
Making this replacement, the effective action takes the form
\begin{equation}
    \Gamma^\text{in-in}_\text{int}=\int \D\tau_1 \D\tau_2\sum_{\ell=0}^\infty K_{{IJ}}^{(\Phi)\,\ell}(\tau_2-\tau_1) \nabla^{(\ell)}_{A_{\ell}}\Phi^I(\tau_2)\nabla^{(\ell)}{}^{A_{\ell}}{\Phi}^{J}(\tau_1)\,,
    \label{scalar_effective_action}
\end{equation}
where we have used the fact that in the case of interest the system is rotationally symmetric.\footnote{This means that the Green's function~\eqref{eq:Kdef} is diagonal in $\ell ,\ell'$ and its index structure can be written in terms of $\delta_{ij}$. The end result is that the indices on the two derivative operators acting on $\Phi$ end up getting contracted. Correspondingly the responses at a given $\ell$ are independent of the magnetic quantum number $m$.} The function $K^{(\Phi)}$ encodes the response of the particle to external sources. In frequency space, real terms capture conservative responses and imaginary terms capture dissipative response.

\subsubsection{Linear response}

To derive the linear response of the classical field $\Phi_+$ to a background $\bar\Phi_+$ induced by the interaction at the location of the worldline~\eqref{scalar_effective_action}, one has to compute its one-point function, expanding $\E^{i\Gamma_\text{int}^\text{in-in}}$ at linear order in the perturbation:
\begin{align}
    \label{scalar_1_pt_function}
    \langle \Phi_+(t,\vec{x})\rangle_\text{in-in} &= \int \mathcal{D}\Phi_+\mathcal{D}\Phi_-\,\Phi_+(t,\vec{x})\E^{i\Gamma^\text{in-in}[\Phi_+,\Phi_-]}\\\nonumber
    &= i\sum_{\ell=0}^\infty (-1)^\ell \int \D\tau_1\D\tau_2 \int \frac{\D{\omega}}{2\pi}\E^{-i\omega(\tau_2-\tau_1)} K^{(\Phi)}_{\ell}({\omega})\nabla^{(\ell)}_{A_{\ell}}\langle \Phi_+(t,\vec{x})\Phi_-(\tau_2)\rangle \nabla^{(\ell)}{}^{A_{\ell}}\Bar{\Phi}_+(\tau_1)\,,
\end{align}
where  $K_{+-}^{(\Phi)}\equiv K^{(\Phi)}$ is related to the retarded (thus causal) Green's function via
\begin{equation}
    K_{+-}^{(\Phi)}(\tau_2-\tau_1)=-G_R^{(Q)}(\tau_2-\tau_1) \equiv i \langle \left[ Q_+(\tau_2),Q_-(\tau_1) \right]\rangle \theta(\tau_2-\tau_1).
\end{equation}
The two-point function of the scalar field appearing in~\eqref{scalar_1_pt_function} is related to the field retarded Green's function by 
\begin{equation}
    \langle \Phi_+(t,\vec{x})\Phi_-(\tau_2,0) \rangle=iG^{(\Phi)}_\text{R}(t-\tau_2,\vec{x}).
    \label{scalar_propagator}
\end{equation}
To further simplify~\eqref{scalar_1_pt_function}, we first note that the fact that
the scalar tidal response coefficients of black holes vanish implies that $K_\ell^{(\Phi)}(\omega)$ starts at linear order in $\omega$ (which can be checked a posteriori). We can therefore neglect order $O(\omega^2)$ terms coming from  $\nabla^{(\ell)}_{i_L}\langle \Phi^+(t,\vec{x})\Phi^-(\tau_2)\rangle \nabla^{(\ell)}{}^{i_L}\Bar{\Phi}^+(\tau_1)$, meaning that
it will be sufficient to treat the propagation of the scalar as instantaneous, so that
\be
iG^{(\Phi)}_\text{R}(t-\tau_2,\vec{x})=-i\delta(t-\tau_2) \int \frac{\D^{D-1}\vec{p}}{(2\pi)^{D-1}}\frac{\E^{i\vec{p}\cdot\vec{x}}}{\vec{p}^2}\,.
\label{scalar_propagator2}
\ee
We additionally need the
explicit form of the tidal field $\Bar{\Phi}_+$. 
Solving the Klein--Gordon equation in flat spacetime, $\square \Bar{\Phi}_+=0$, one can write $\Bar{\Phi}_+$ in cartesian coordinates as a Taylor expansion around $\vec{x}=0$:
\begin{equation}
    \bar{\Phi}_+(\tau_1)=e^{-i\omega \tau_1}\sum_\ell c_{j_1 \cdots j_\ell}x^{j_1}\cdots x^{j_\ell}+O(\omega^2)\,,
\end{equation}
with $c_{j_1\cdots j_\ell}$ traceless and fully symmetric. Using this and~\eqref{scalar_propagator2} in~\eqref{scalar_1_pt_function}, we find
\begin{equation}
    \langle \Phi_+(t,\vec{x})\rangle_\text{in-in} =\E^{-i\omega t}\sum_{\ell=0}^\infty(-i)^\ell \ell!\, K^{(\Phi)}_{\ell}(\omega)c^{i_1\cdots i_\ell} \int \frac{\D^{D-1}\vec{p}}{(2\pi)^{D-1}}\E^{i\vec{p}\cdot\vec{x}}\frac{p_{i_1}\cdots p_{i_\ell}}{\vec{p}^2}\,.
\end{equation}
Fourier transforming back to position space, this is\footnote{We use the Fourier integral
\begin{equation} 
    i^L \int \frac{\rd^d \vec p}{(2 \pi)^d} \E^{i\vec p\cdot \vec x}\, \frac{p^{(i_1} \cdots p^{i_L)_T}}{\vec p^2}=  \frac{\Gamma(\tfrac{d}{2}-1)\Gamma(2-\tfrac{d}{2})}{2^L(4\pi)^{d/2} \Gamma(2-\tfrac{d}{2}-L)}  x^{(i_1} \cdots x^{i_L)_T}\left(\frac{\vec x^2}{4}\right)^{1-\tfrac{d}{2}-L}\,,
    \label{eq:fourierintegral}
\end{equation}
to convert between momentum and position space.
}
\begin{equation}
    \langle \Phi_+(t,\vec{x})\rangle_\text{in-in} = \E^{-i\omega t}\sum_{\ell=0}^\infty (-1)^\ell K^{(\Phi)}_{\ell}(\omega)\frac{2^{\ell-2}\ell!\,\Gamma\left(\frac{D-3}{2}\right)\Gamma\left(\frac{5-D}{2}\right)}{\pi^{\frac{D-1}{2}}\Gamma\left(\frac{5-D}{2}-\ell\right)}c_{i_1\cdots i_\ell}\frac{x^{i_1}\cdots x^{i_\ell}}{\vert\vec{x}\vert^{2\ell+D-3}} \,.
\end{equation}
Simplifying the ratio of gamma functions and writing the tidal field in spherical coordinates as $c_{i_1\cdots i_\ell}x^{i_1}\cdots x^{i_\ell}=c_\text{ext}r^{\ell}Y_\ell^m$,
we obtain\footnote{We suppress indices on $c_\text{ext}$ which depends implicitly on the spherical harmonic quantum numbers $\ell,m$.}
\begin{equation}
    \langle \Phi_+\rangle_\text{in-in} = \E^{-i\omega t}\sum_{\ell=0}^\infty  K^{(\Phi)}_{\ell}(\omega)\frac{2^{\ell-2}\ell!\,\Gamma\left(\ell+\frac{D-3}{2}\right)}{\pi^{\frac{D-1}{2}}}\frac{c_\text{ext} Y_\ell^m}{r^{\ell+D-3}}\,.
\end{equation}
Combining this with the external field and replacing $K_\ell^{(\Phi)}$ by $\mu^{2\varepsilon}K_\ell^{(\Phi)}$ (where $\mu$ is a dimensionful constant  in order to keep the response function dimension independent) yields
\begin{equation}
    \bar{\Phi}_+ + \langle \Phi_+\rangle_\text{in-in}= \E^{-i\omega t}\sum_{\ell=0}^\infty c_\text{ext} Y_\ell^m \left( r^\ell+ \mu^{2\varepsilon}K^{(\Phi)}_{\ell}(\omega)\frac{2^{\ell-2}\ell!\Gamma\left(\ell+\frac{D-3}{2}\right)}{\pi^{\frac{D-1}{2}}}r^{-\ell-D+3}\right).
    \label{eq:scalar_one-point_function}
\end{equation}
This result gives the scalar field response, generated by the interaction term of the action $S_\text{int}$, in the vicinity of the point particle. However, in order to match to the UV result, we have to account for the effects of gravity, which we have so far neglected in the EFT.

\subsubsection{Coupling to gravity via the Born series}

The result~\eqref{eq:scalar_one-point_function} represents the leading response of the point particle in the $G_{\rm N}\to 0$ limit. However, in order to make contact with the UV solution at subleading order in $G_{\rm N}$ we need to include gravitational effects. To do this, we follow the nice approach of~\cite{Correia:2024jgr,Caron-Huot:2025tlq} and utilize the Born series to capture nonlinear $G_{\rm N}$ corrections to the Minkowski spacetime solution.

The philosophy is effectively the following: we consider the full bulk equation of motion following from~\eqref{eq:Sbulk} (including gravity), which can be written as
\begin{equation}
    \vec{\nabla}^2\Phi_+ = \left(  V_G  -\omega^2 \right)\Phi_+,
    \label{eq:scalarEFT}
\end{equation}
where in spherical coordinates we group all terms that depend on $G_\text{N}$ into an effective potential
\begin{equation}
    V_G = \frac{2G_\text{N}M n_D \mu^{2\varepsilon}}{r^{1-2\varepsilon}} \left[ \frac{\D^2}{\D r^2} + \frac{1}{r} \frac{\D}{\D r} 
   - \omega^2 \sum_{n=0}^\infty\left( \frac{2G_\text{N}M n_D \mu^{2\varepsilon}}{r^{1-2\varepsilon}} \right)^n
     \right].
    \label{eq:VGrav}
\end{equation}
We can then imagine solving~\eqref{eq:scalarEFT} perturbatively in $G_{\rm N}$ and $\omega$. Since we are truncating the $G_{\rm N}$ expansion, the solutions will generically be divergent as $r\to 0$. We then view the source of this divergence as the point particle itself. The properties of the long-distance $\Phi$ solution will then be determined by $K(\omega)$ in~\eqref{eq:scalar_one-point_function}.
This will relate the worldline couplings to the parameters of the $\Phi$ solution at large distances. We can then match these solutions (which include gravity) to the GR calculations from Section~\ref{sec:scalarGR} to determine $K(\omega)$, and hence the worldline couplings.

As a practical matter, at each order,~\eqref{eq:scalarEFT} takes the form of an inhomogeneous second-order differential equation. The most general solution is given by a sum of homogeneous and particular solutions. The homogeneous solution has the same form at all orders in $G_\text{N}$ and $\omega$. We will therefore solve it once and allow the integration constants to be generic functions of $G_\text{N}$ and $\omega$. We will then impose boundary conditions at the point particle's location and plug this solution in the source on the right-hand side to obtain the particular solution. In the following, we briefly review the steps of the procedure in the scalar field case, previously discussed in~\cite{Caron-Huot:2025tlq}, which we will later extend to gravitational perturbations in Section~\ref{sec:gravDLNs}.

\paragraph{Homogeneous solution and boundary conditions}~\\
The homogeneous part of~\eqref{eq:scalarEFT} is simply Laplace's equation for a static scalar field in $D$ dimensions
\begin{equation}
\vec{\nabla}^2\Phi_+ =  0.
\label{eq:scalarEFThom}
\end{equation}
If we decompose the field  in  higher-dimensional spherical harmonics as\footnote{We collectively denote by $\vec{\theta}$ the angular coordinates of the hyperspherical harmonics $Y_\ell^m(\vec\theta)$ on the $S^{D-2}$ sphere, and, with a slight abuse of notation,  use $m$ as a multi-index to label all the magnetic quantum numbers. See, e.g., Appendix A of~\cite{Hui:2020xxx} for details. Note that the transformed field $\varphi$ in \eqref{eq:psis} carries implicit $\ell,m$ labels that we suppress for notational simplicity.}
\begin{equation}
    \Phi_+=\sum_{\ell,m}\varphi(r) \,r^{\frac{2-D}{2}} Y_\ell^m(\vec\theta)\,,
 \label{eq:psis}
\end{equation}
then~\eqref{eq:psis} takes the form
\begin{equation}
    \left(\frac{\D^2}{\D r^2}-\frac{(\ell-\varepsilon)(\ell-\varepsilon+1)}{r^2} \right)\varphi^{(h)}(r) = 0.
    \label{eq:EWE}
\end{equation}
The general solution to this equation is
\be
\varphi^{(h)}(r) = B_\text{reg} r^{\ell+1-\varepsilon}+\frac{\mu^{2\varepsilon}B_\text{irr}}{2\ell+1-2\varepsilon} r^{-\ell+\varepsilon}\,,
\label{eq:phiEFT0}
\ee
with $B_\text{reg}$ and $B_\text{irr}$ the two arbitrary integration constants.\footnote{For convenience, we extracted the constant prefactor $\frac{\mu^{2\varepsilon}}{2\ell+1-2\varepsilon}$ from the decaying branch. Implicitly, the $r$-independent coefficients $B_\text{reg}$ and $B_\text{irr}$ should be thought of as expanded in powers of $G$ and $\omega$, as we discussed, and will see explicitly later on.}
The former branch of solution is regular everywhere in space, including the origin, with $B_\text{reg}$ corresponding to the amplitude of the asymptotic external tidal field that we use to probe the point object. On the other hand, the decaying $r^{-\ell+\varepsilon}$ solution is divergent at the location of the particle. We fix its free coefficient $B_\text{irr}$ by demanding that the divergence is sourced precisely by the localized action term $S_\text{int}$ \eqref{eq:Sintscalar}. In practice, we determine the Love numbers $K_\ell^{(\Phi)}(\omega)$ by matching the homogeneous solution for $\Phi$
\begin{equation}
    \Phi_{+}^{(h)}=\E^{-i\omega t}\sum_{\ell=0}^\infty Y_\ell^m\,B_{\rm reg}\left( r^{\ell}+\frac{B_{\rm irr}}{B_{\rm reg}}\frac{\mu^{2\varepsilon}}{2\ell+1-2\varepsilon}r^{-\ell-1+2\varepsilon} \right),
\label{eq:phiphomogsoln}
\end{equation}
to the one-point function \eqref{eq:scalar_one-point_function}. This yields 
\begin{equation}
    \frac{2^{\ell-1}\ell!\Gamma\left(\ell+\frac{D-1}{2}\right)}{\pi^{\frac{D-1}{2}}} K^{(\Phi)}_{\ell}(\omega)=\frac{B_\text{irr}}{B_\text{reg}}\,.
    \label{scalar_EFT_Love_BC}
\end{equation}
As per~\cite{Caron-Huot:2025tlq}, this implies that the ratio of irregular to regular solution is the worldline tidal response. We next include the particular solution to the equation including the gravitational potential.

﻿
\vspace{-10pt}
\paragraph{Particular solution}~\\
The solution~\eqref{eq:phiphomogsoln} with~\eqref{scalar_EFT_Love_BC} is the homogeneous solution to~\eqref{eq:scalarEFT} with the correct boundary conditions at the location of the point particle. We now turn to 
deriving the particular solution in the presence of the $V_G$ potential. To this end, we consider points away from the origin and solve
\begin{equation}
\vec{\nabla}^2\Phi_+ = \left(  V_G  -\omega^2 \right)\Phi_+. 
\label{eq:Sch_phi_grav}
\end{equation}
In terms of the variable $\varphi(r)$ defined in~\eqref{eq:psis}, this equation becomes
\begin{equation}
    \left(\frac{\D^2}{\D r^2}-\frac{(\ell-\varepsilon)(\ell-\varepsilon+1)}{r^2} \right)\varphi(r) = V_\varphi(r)\varphi(r),
    \label{eq:EWE-2}
\end{equation}
with  potential
\begin{equation}
    V_\varphi\equiv\sum_{n=1}^\infty\left( \frac{2G_\text{N}M n_D \mu^{2\varepsilon}}{r^{1-2\varepsilon}} \right)^n\left[ \frac{2\varepsilon-1}{r}\frac{\D}{\D r}+\frac{\ell^2+\ell+1-\varepsilon(3+2\ell)+2\varepsilon^2}{r^2}-(n+1)\omega^2\right]-\omega^2.
    \label{Grav_potential}
\end{equation}
We can solve~\eqref{eq:EWE-2} formally perturbatively in $G_\text{N}$ and $\omega$ via the Born series (see~\cite{Caron-Huot:2025tlq} for details):
\be
\begin{aligned}
\varphi(r)=\varphi^{(h)}(r)&+\int^r\D r' \, G(r,r')V_\varphi(r')\,\varphi^{(h)}(r') \\
    &+\int^r\D r' \, G(r,r')V_\varphi(r')\int^{r'}\D r'' \, G(r',r'')V_\varphi(r'')\,\varphi^{(h)}(r'')+\cdots 
\label{eq:Bornseriesscalar}
\end{aligned}
\ee
with up to $2\ell+4$ insertions of the $V_{\varphi}$ potential, truncated at $n=2\ell+3$.\footnote{Recall that the static Love numbers scale as $r_s^{2\ell+1}$. In a diagrammatic description,  they  correspond to Feynman diagrams with $2\ell$ (classical) loops and $2\ell+1$ worldline mass insertions. Studying quadratic $(GM\omega)^2$-corrections to the tidal response is two orders further in the coupling $G$ in the perturbative expansion.}
In \eqref{eq:Bornseriesscalar}, $\varphi^{(h)}$ is the homogeneous solution~\eqref{eq:phiEFT0},
with $B_\text{reg}$ and $B_\text{irr}$ satisfying \eqref{scalar_EFT_Love_BC}, while  $G(r,r')$ is a Green's function solving
\begin{equation}
\left(\frac{\D^2}{\D r^2}-\frac{(\ell-\varepsilon)(\ell-\varepsilon+1)}{r^2} \right)G(r,r') = \delta(r-r')\,.
\end{equation}
In order to ensure that the particular solution does not alter the boundary condition~\eqref{scalar_EFT_Love_BC} of the homogeneous solution, we will require the Green's function to be proportional to the theta function $\theta(r-r')$, so that it vanishes as $r\rightarrow0$. 
From the continuity of $G(r,r')$ and the jump condition of $\partial_r G(r,r')$ at $r=r'$, it follows that~\cite{Caron-Huot:2025tlq}
\begin{equation}
     G(r,r')=\frac{r^{\varepsilon-\ell}(r')^{1+\ell-\varepsilon}-(r')^{\varepsilon-\ell}r^{1+\ell-\varepsilon}}{2\varepsilon-2\ell-1} \theta(r-r') \,.
\label{eq:Greenfunctionscalar}
\end{equation}
For the same reason, all  integrals in the perturbative series \eqref{eq:Bornseriesscalar} can be treated as indefinite integrals; in other words, we omit the lower integration bound, which would otherwise modify the homogeneous solution $\varphi^{(h)}(r)$.

To illustrate, here we record the solutions for the $\ell=0,2$ multipoles~\cite{Caron-Huot:2025tlq}. Keeping only the terms scaling as  $r^\ell$ and $r^{-\ell-1}$, and taking the  $\varepsilon\rightarrow0$ limit, one has 
\begin{align}
\nonumber
\Phi_{\ell=0}(r) \equiv\varphi_{\ell=0}(r) r^{\frac{2-D}{2}}
= B_{\rm reg}\bigg(1 &-\frac{11 \bar{G}^2 \omega ^2}{6 \varepsilon } +\frac{16 \bar{G}^2 \omega ^2}{9} -\frac{22}{3} \bar{G}^2 \omega ^2 \log (\mu  r)\bigg)\\
\label{eq:phiL0}
   &+B_\text{irr} \left(-\frac{\bar{G} \omega ^2}{\varepsilon } - \frac{\bar{G} \omega ^2}{2}  -4 \bar{G} \omega ^2 \log (\mu  r)\right)\\\nonumber
   &+\frac{B_{\rm reg}}{r} \left(\frac{2 \bar{G}^3 \omega ^2}{\varepsilon }+\frac{80 \bar{G}^3 \omega ^2}{3} +12 \bar{G}^3 \omega ^2 \log (\mu  r)\right) \\\nonumber
   &+\frac{B_\text{irr}}{r} \left(1  +\frac{11 \bar{G}^2 \omega ^2}{6 \varepsilon } +\frac{68}{3} \bar{G}^2 \omega ^2 +  11 \bar{G}^2 \omega ^2 \log (\mu  r)\right) + \cdots,
\end{align}
where we have adopted the convention of~\cite{Caron-Huot:2025tlq}, by defining\footnote{This is just a matter of choice of renormalization scheme. Choosing to work with $G_\text{N}$, instead of ${\bar{G}}$, would result in different finite $O(\varepsilon^0)$ terms in the expanded solutions. 
}
\begin{equation}
\bar{G} \equiv G_\text{N} M n_D.
\label{eq:barGdef}
\end{equation}
As expected, there are UV divergences in the $\varepsilon\rightarrow0$ limit, as well as logarithmic terms in addition to the standard polynomials in $r$.\footnote{The divergent $\frac{1}{\varepsilon}$ terms and the $\log(r)$  result from the integrals  $\int^r\D r' (r')^{-1+a\varepsilon}=\frac{r^{a\varepsilon}}{a\varepsilon}\underset{\varepsilon\to0}{\sim}\frac{1}{a\varepsilon}+\log(r)+ O(\varepsilon)$.}
Similarly, for the quadrupole we have
\begin{equation}
\begin{aligned}
    \Phi_{\ell=2}(r)= \, B_{\rm reg} r^2  \bigg(1 &-\frac{79 \bar{G}^2 \omega ^2}{210 \varepsilon } +\frac{11429 \bar{G}^2 \omega ^2}{11025} -\frac{158}{105} \bar{G}^2 \omega ^2 \log (\mu  r)\bigg)\\
   &+\frac{B_{\rm reg}}{r^3} \left(\frac{8 \bar{G}^7 \omega ^2}{135 \varepsilon }  +\frac{9328 \bar{G}^7 \omega ^2}{2025} +\frac{112}{135} \bar{G}^7 \omega ^2 \log (\mu  r)\right)\\
   &+\frac{B_\text{irr}}{r^3} \left(\frac{1}{5} +\frac{79 \bar{G}^2 \omega ^2}{1050 \varepsilon } +\frac{39416 \bar{G}^2 \omega ^2}{55125}
   +\frac{79}{175} \bar{G}^2 \omega ^2 \log (\mu  r)\right) + \dots\,,
\end{aligned}
\label{eq:phiL2}
\end{equation}
which displays the same characteristic features.

\subsubsection{Renormalization and matching}
The infinities appearing in~\eqref{eq:phiL0} and~\eqref{eq:phiL2} as $\varepsilon\rightarrow0$ are reminiscent of  UV divergences in perturbative calculations in quantum field theory (QFT)---with the difference that here we are dealing with classical worldline  loops, rather than quantum loops~\cite{Goldberger:2004jt,Kol:2011vg,Cheung:2018wkq,Bern:2019crd,Ivanov:2024sds,Ivanov:2025ozg}.
In QFT, one introduces a regularization scheme and a set of renormalization conditions that allow the identification of the physical parameters---those measured in experiments---and their relation to the bare parameters appearing in the Lagrangian. Counterterms are then added to the Lagrangian  to remove the divergences and enforce the renormalization conditions. A byproduct of this perturbative renormalization procedure is the appearance of scale dependence in the coefficients and logarithmic running governed by precise renormalization group equations.
In the present context, the running parameters are $B_\text{reg}$ (the scalar field amplitude) and $B_\text{irr}$ (or, equivalently, the tidal response couplings $K_\ell^{(\Phi)}(\omega)$ via \eqref{scalar_EFT_Love_BC}), and we require  counterterms  for the kinetic and interaction terms (eqs.~\eqref{eq:Sbulk} and \eqref{scalar_effective_action}, respectively). The renormalized coefficients $\Bar{B}_\text{reg}$ and $\Bar{B}_\text{irr}$ can be related to the bare ones via the relations
\begin{equation}
\begin{aligned}
    B_\text{reg}&=\Bar{B}_\text{reg}(1+\omega^2\delta_{11})+\Bar{B}_\text{irr}\omega^2\delta_{12}\,, \\
    B_\text{irr}&=\Bar{B}_\text{irr}(1+\omega^2\delta_{22})+\Bar{B}_\text{reg}\omega^2\delta_{21}\,.
\end{aligned}
\end{equation}
Defining the matrix
\begin{equation}
\label{eq:deltamatrixdef}
\delta = \begin{pmatrix}
        \delta_{11} & \delta_{12} \\
       \delta_{21} & \delta_{22}
    \end{pmatrix} ,
\end{equation}
one finds the following values are needed to cancel the $1/\varepsilon$ divergences~\cite{Caron-Huot:2025tlq}:
\begin{equation}
    \delta^{\ell=0}=
    \begin{pmatrix}
        \frac{11\bar{G}^2}{6\varepsilon} & \frac{\bar{G}}{\varepsilon} \\
       -\frac{2\bar{G}^3}{\varepsilon} &-\frac{11\bar{G}^2}{6\varepsilon} 
    \end{pmatrix}
    \qquad\qquad
        \delta^{\ell=2}=
    \begin{pmatrix}
       \frac{79\bar{G}^2}{210\varepsilon}  & 0 \\
       -\frac{\bar{G}^7}{\varepsilon} &-\frac{79\bar{G}^2}{210\varepsilon}
    \end{pmatrix}.
 \label{eq:deltascalar0}
\end{equation}
Plugging these back into the  scalar field solution~\eqref{eq:Bornseriesscalar} now  yields a finite limit as $\varepsilon\rightarrow0$:\footnote{We stress the substitution should be done not in~\eqref{eq:phiL0} and \eqref{eq:phiL2}, but in $\Phi(r)$ before taking the $\varepsilon \to 0$ limit.}
\begin{equation}
\begin{aligned}
    \Phi^\text{R}_{\ell=0}=&\, \Bar{B}_\text{reg} \left[1+\bar{G}^2\omega^2\left(\frac{16}{9}-\frac{22}{3} \log(\mu r)\right)+\frac{\bar{G}^3\omega^2}{r}\left(\frac{68}{3}+8\log(\mu r)\right)\right]\\
    &+\Bar{B}_\text{irr}\left[\frac{1}{r}-\bar{G}\omega^2 \left(\frac12 + 4\log(\mu r)\right)+\frac{\bar{G}^2 \omega^2}{r}\left(19+\frac{22}{3}\log(\mu r)\right)\right] +\cdots\,,
\end{aligned}
\label{eq:phiRl0}
\end{equation}
and for $\ell=2$:
\begin{equation}
\begin{aligned}
    \Phi^\text{R}_{\ell=2}=&\, \Bar{B}_\text{reg}\left[r^2+\bar{G}^2\omega^2\left(\frac{11429}{11025}-\frac{158}{105}\log(\mu r)\right)+\frac{\bar{G}^7\omega^2}{r^3}\left(\frac{63064}{14175}+\frac{32}{45}\log(\mu r)\right)\right]\\
    &+\Bar{B}_\text{irr}\left[\frac{1}{5r^3}+\frac{\bar{G}^2\omega^2}{r^3}\left(\frac{37757}{55125}+\frac{158}{525}\log(\mu r)\right)\right] + \cdots\,,
\end{aligned}
\label{eq:phiRl2}
\end{equation} 
where $\Phi^\text{R}(r)$ is the renormalized scalar field solution.
As a consistency check, it is worth noting that the conditions \eqref{eq:deltascalar0} are enough to remove all divergences  from the  renormalized solution $\Phi^\text{R}(r)$, not only the ones  in the coefficients of the  $r^\ell$ and $r^{-\ell-1}$  fall-offs, displayed in~\eqref{eq:phiRl0} and~\eqref{eq:phiRl2}.

These renormalized solutions are directly the EFT analogues of the scalar field solutions obtained in Section~\ref{sec:scalarGR}.
Comparing~\eqref{eq:phiRl0} and~\eqref{eq:phiRl2} with the full general relativistic solutions \eqref{eq:phiGRl0}--\eqref{eq:phiGRl2}, we can match  and obtain the EFT parameters: 
\begin{equation}
\begin{aligned}
    \Bar{B}_\text{reg}^{\ell=0} =& B\bigg[1-i\omega r_s+\omega^2r_s^2\left(\frac{1}{18}-\frac{\pi^2}{6}+\frac{11}{6}\log(\mu r_s) \right)\bigg]\\
  \Bar{B}_\text{irr}^{\ell=0} =& B\bigg[i\omega r_s^2-\omega^2r_s^3\left(\frac56+\log(\mu r_s) \right)\bigg]\\[4pt]
    \Bar{B}_\text{reg}^{\ell=1} =& B\bigg[2 + 2i\omega r_s+ \omega ^2 r_s^2 \left(-\frac{1733+150 \pi ^2}{450}+\frac{19}{15} \log (\mu r_s)\right)\bigg]\\
    \Bar{B}_\text{irr}^{\ell=1} =& B\bigg[\frac{i\omega r_s^4}{2} -\omega ^2 r_s^5 \left(\frac{39}{80}+\frac{1}{2}\log(\mu r_s)\right)\bigg]\\[4pt]
    \Bar{B}_\text{reg}^{\ell=2} =& B\bigg[6+12i\omega r_s-\omega^2r_s^2\left(\frac{26014}{1225}+\pi^2-\frac{79}{35}\log(\mu r_s)\right)\bigg]\\
    \Bar{B}_\text{irr}^{\ell=2} =& B\bigg[\frac{i\omega r_s^6}{6}-\omega^2r_s^7\left(\frac{232}{945}+\frac16\log(\mu r_s) \right)\bigg]\,,
    \label{B_UV_fixed}
\end{aligned}
\end{equation}
where we have also included the results for $\ell=1$, though we did not display the intermediate steps.
Note that for $\ell=2$, we redefined $B\to B r_s^2$, so that $B_\text{reg}$ is  a polynomial in $r_s$.
Plugging~\eqref{B_UV_fixed} into~\eqref{scalar_EFT_Love_BC}, we can finally deduce the $\ell=0,1,2$ scalar response coefficients: 
\begin{tcolorbox}[colframe=white,arc=0pt,colback=greyish2]
\begin{equation}
\begin{aligned}
    \frac{1}{4\pi}K^{(\Phi)}_{\ell=0}(\omega)&=i\omega r_s^2+\omega^2r_s^3\left(-\frac{11}{6}-\log(\mu r_s)\right),\\
    \frac{3}{\pi}K^{(\Phi)}_{\ell=1}(\omega)&=i\omega r_s^4+\omega^2 r_s^5\left( \frac{1}{40}-\log(\mu r_s) \right), \\
    \frac{270}{\pi}K^{(\Phi)}_{\ell=2}(\omega)&= i\omega r_s^6+\omega^2r_s^7\left(\frac{166}{315}-\log(\mu r_s)\right)\,,
\end{aligned}
\end{equation}
\end{tcolorbox}
\noindent
which agree with~\cite{Ivanov:2024sds,Caron-Huot:2025tlq}.
Notice that the responses include an imaginary part (corresponding to dissipation) at leading order in the frequency, and a conservative  (real) part at order $\omega^2$, which both has a logarithmic running component, and a particular finite contribution.
As a matter of principle there is no obstruction to carrying out this procedure to subleading orders in either $G_{\rm N}$ or $\omega$, or for higher multipoles. However, we now turn to the analogous computation of gravitational tidal responses (Love numbers).

\section{Dynamical Love numbers}
\label{sec:gravDLNs}
﻿

The responses of objects to tidal gravitational perturbations are encoded in the Love numbers of the object. Here we compute the dynamical Love numbers of a Schwarzschild black hole.

The dynamics of perturbations of Schwarzschild black holes in general relativity is  described by the Regge--Wheeler~\cite{Regge:1957td} and Zerilli~\cite{Zerilli:1970wzz} equations. Similar to the Schwarzschild Klein--Gordon equation for a massless scalar, both the Regge--Wheeler and Zerilli equations feature an irregular singular point at $r=\infty$, and  can be recast into the standard form of the confluent Heun equation.\footnote{In the way it is commonly written down \cite{Zerilli:1970wzz} (see eq.~\eqref{eq:ddimZeq} with $D=4$), the Zerilli equation  exhibits an additional regular singular point at a negative value of $r$. This singularity can, however, be eliminated through a suitable field redefinition. Indeed, it is well known that the Zerilli equation can be mapped to the Regge--Wheeler equation via the Chandrasekhar transformation~\cite{10.2307/78902,1975RSPSA.343..289C,Chandrasekhar:1985kt}.} 
For the boundary value problem of interest here---where the field asymptotically approaches a tidal-field profile at large distances---small-frequency corrections to the static solution have been computed using a range of semi-analytical techniques, approximation schemes, and symmetry arguments (see, e.g.~\cite{Chakrabarti:2013lua,Poisson:2020vap,Charalambous:2021mea,Creci:2021rkz,Bonelli:2021uvf,Saketh:2023bul,Perry:2023wmm,Chakraborty:2023zed,Pitre:2023xsr,Katagiri:2024wbg,DeLuca:2024ufn,Bhatt:2024yyz,Bhatt:2024rpx,HegadeKR:2024agt,Chakraborty:2025wvs}).

We first construct perturbative-in-frequency solutions to the Regge--Wheeler and Zerilli equations, paralleling the scalar analysis of Section~\ref{sec:scalar} (see also~\cite{Katagiri:2024wbg,Chakraborty:2025wvs} for related treatments). We then match these solutions to the worldline EFT in dimensional regularization, which allows us to explicitly compute the dynamical Love number coefficients, including, for the first time, the finite scheme-dependent terms.

﻿
﻿
\subsection{General relativistic solution via small-frequency expansion}
\label{sec:GRsols}
﻿
﻿

We begin by considering the general relativistic computation of tidal responses.
We first set up notation and 
 briefly recall the derivation of the Regge--Wheeler and Zerilli equations. For later convenience, we present these expressions in general $D$ dimensions. In the general relativistic part of our calculation we will exclusively work in $D=4$, but keeping the equations in generic $D$ will be important for  implementing dimensional regularization within the EFT framework.

We are interested in  the dynamics of perturbations around a background Schwarzschild geometry~\eqref{eq:bargS}. Denoting it with $\bar{g}_{\mu\nu}$, we  perturb the metric as ${g}_{\mu\nu}=\bar{g}_{\mu\nu}+2h_{\mu\nu}/\Mpl$. 
To make maximal use of the ${\rm SO}(D-1)$ background symmetry, it is convenient to decompose  $h_{\mu\nu}$  into scalar, vector, and tensor spherical harmonics. Concretely, we can write $h_{\mu\nu}$ as follows~\cite{Kodama:2000fa,Kodama:2003jz,Ishibashi:2003ap,Hui:2020xxx,Rosen:2020crj}:
\be
\begin{aligned}
h_{\mu \nu} = ~&\sum_{\ell, m} \begin{pmatrix}
f(r)H_0(t,r) & H_1(t,r) & {\cal H}_0(t,r)\nabla_i 
\\
* & f(r)^{-1}H_2(t,r)  & {\cal H}_1(t,r)\nabla_i 
\\
* & * & r^2 \big[ {\cal K}(t,r)\gamma_{ij}  +  G(t,r)\nabla_{( i}\nabla_{j)_T} \big]
\end{pmatrix} Y_\ell^m 
\\[4pt]
&+ \sum_{\ell, m} \begin{pmatrix}
0 & 0 & h_0(t,r)Y_i^{(T)}{}_\ell^m
\\
* & 0  & h_1(t,r)Y_i^{(T)}{}_\ell^m
\\
* & * & r^2 h_2(t,r)\nabla_{( i}Y_{j )_T}^{(T)}{}_\ell^m
\end{pmatrix} 
+ \sum_{\ell, m} \begin{pmatrix}
0 & 0 & 0
\\
* & 0 & 0
\\
* & * &  r^2 h_T(t,r) 
\end{pmatrix} Y^{(TT)}_{ij}{}_\ell^m \,,
\end{aligned}
\label{eq:hdecomp}
\ee
where the entries denoted by $*$ are the same as the entries across the diagonal because $h_{\mu\nu}$ is symmetric. In~\eqref{eq:hdecomp},
 $Y_\ell^m$ are scalar spherical harmonics,  $Y_i^{(T)}{}_\ell^m$ are  (transverse) vector harmonics, and $Y_{ij}^{(TT)}{}_\ell^m$ are  (transverse and traceless) tensor harmonics. They are orthogonal to each other and satisfy standard eigenvalue equations of the Laplace operator on the $S^{D-2}$ sphere (see~\cite{Hui:2020xxx} for details).
The term proportional to the scalar harmonics $Y_\ell^m$ corresponds to the even (or, polar) sector in $D=4$, while the term proportional to $Y_i^{(T)}{}_\ell^m$ describes the odd (or, axial) sector of perturbations in $D=4$.
Since we are ultimately interested  in the $D=4$ limit,  we will henceforth ignore the last term in~\eqref{eq:hdecomp}, proportional to $Y_{ij}^{(TT)}$, which corresponds to the tensor sector that is present only in higher-dimensional spacetimes. Owing to the background symmetry, the component $h_T(t,r)$ decouples from the other metric fluctuations and is non-dynamical in $D=4$.

After fixing the gauge  $h_2={\cal H}_0 = {\cal K} = G = 0$,\footnote{Note that this choice slightly differs  from what is commonly referred to as the Regge--Wheeler gauge \cite{Regge:1957td}.}
and solving for the constraint variables,
one finds the following equations for the physical degrees of freedom $\Psi_{\text{RW}}$ and $\Psi_{\text{Z}}$ \cite{Hui:2020xxx}:
\begin{align}
\frac{\rd^2\Psi_{\rm RW}}{\rd r_\star^2} + \Big(\omega^2 - V_{\rm RW}(r)\Big)\Psi_{\rm RW} &= 0
\label{eq:ddimRWeq}
\\
\frac{\rd^2\Psi_{\rm Z}}{\rd r_\star^2} + \Big(\omega^2 - V_{\rm Z}(r)\Big)\Psi_{\rm Z} &= 0,
\label{eq:ddimZeq}
\end{align}
where $\D r_\star/\D r\equiv f^{-1}$, and where the potentials $V_{\rm RW}$ and $V_{\rm Z}$ are given by
\begin{align}
V_{\rm RW}(r) &= f\frac{(\ell+1)(\ell+D-4)}{r^2}+f^2\frac{(D-4)(D-6)}{4r^2}-ff'\frac{(D+2)}{2r}\,,
\label{RWpotentialDD}
\\[4pt]
\nonumber
V_\text{Z}(r) &= \bigg[ 4(D-4)(D-2)^4 f^3-8(D-2)^2(D-2)(D-6)\ell(\ell+D-3)f^2\\\nonumber
&~~~~~+4(D-2)(D-2)(D-12)\ell^2(\ell+D-3)^2f\\[2pt]
&~~~~~+2(D-2)^3(D+2)r^3 f'{}^3-4(D-2)^2 (D-6)\ell(\ell+D-3)r^2f'{}^2\\[2pt]\nonumber
&~~~~~-8(D-2)^2\ell^2(\ell+D-3)^2rf'+12(D-2)^5 rf^2f'+(D-2)^3(D(D+10)-32)r^2 ff'{}^2\\[2pt]\nonumber
&~~~~~-4(D-2)^2(D-2)(3D-8)\ell(\ell+D-3)rff'\\[2pt]\nonumber
&~~~~~+16\ell^2(\ell+D-3)^2(D-2)\ell(\ell+D-3)
\bigg]
\frac{f [2\ell(\ell+D-3)+(D-2)(rf'-2f)]^{-2}}{4(D-2)r^2 } \,.
\end{align}
In $D=4$, the equations \eqref{eq:ddimRWeq} and \eqref{eq:ddimZeq} reduce to the Regge--Wheeler and Zerilli equations, respectively \cite{Regge:1957td,Zerilli:1970wzz}.
The exact relations between the Regge--Wheeler and Zerilli fields $\Psi_{\text{RW}}$ and $\Psi_{\text{Z}}$, and the metric perturbations in \eqref{eq:hdecomp} can be found in \cite{Hui:2020xxx}.

We next move on to solving~\eqref{eq:ddimRWeq} and~\eqref{eq:ddimZeq} perturbatively in $\omega$. As in the scalar case, we identify three distinct regions: a near zone ($r-r_s\ll r_s$), an intermediate zone ($r_s \lesssim r \ll \omega^{-1}$), and a far zone ($r\gg r_s$). In each region, we employ an approximation scheme that yields an analytic solution, valid within that regime.
From the scalar example, we learned---and explicitly showed---that the far zone solution is not strictly required for matching to the EFT and determining the Love number couplings. Consequently, in the present general relativistic setup described by~\eqref{eq:ddimRWeq} and~\eqref{eq:ddimZeq}, we restrict our attention to computing the near and intermediate-zone solutions only.
For simplicity in the remainder of this subsection we set $D=4$ in~\eqref{eq:ddimRWeq} and~\eqref{eq:ddimZeq}. We will return to the case of arbitrary $D$ in the EFT discussion of Section~\ref{sec:GR-EFT}.
﻿
﻿
﻿\subsubsection{Odd sector}
  
We first focus on the odd-parity sector of gravitational perturbations, which is described by 
the Regge--Wheeler equation \eqref{eq:ddimRWeq} in $D=4$:
\begin{equation}
    \partial_r\big(f(r)\partial_r\Psi_\text{RW}(r)\big)+\left( \frac{\omega^2}{f(r)}-\frac{\ell(\ell+1)}{r^2}+\frac{3r_s}{r^3} \right)\Psi_\text{RW}(r)=0.
    \label{eq:RW_d=4}
\end{equation}
We define two overlapping zones (near and intermediate) where we can solve this equation and use boundary conditions at the black hole horizon to fix the solution.

\vspace{-10pt}
\paragraph{Near zone:}~\\
In the near-zone limit $r\rightarrow r_s$ (where $f\rightarrow0$), the potential is dominated by the $\omega^2$ term. By employing the tortoise coordinate, the resulting equation can be written in the form of the usual wave equation
\begin{equation}
    \left(\frac{\D^2}{\D r_\star^2}+\omega^2\right)\Psi_\text{RW}=0\,.
\label{eq:RW_NZ}
\end{equation}
The independent solutions are $\mathrm{e}^{ \pm i \omega r_{\star}}$. As in the scalar field case, imposing standard  infalling boundary conditions at the horizon selects  the following near-zone  solution:
\begin{equation}
    \Psi_{\text{RW}}^{{\rm NZ}}(r)=B  \E^{-i \omega\left(r+r_s \log \left(\frac{r}{r_s}-1\right)\right)},
 \label{eq:psiRWNZ}
\end{equation}
with $B$ an arbitrary integration constant. The solution \eqref{eq:psiRWNZ} is needed to set the correct  boundary conditions for the intermediate zone solution by matching across their region of overlap, which we now discuss.

\vspace{-10pt}
\paragraph{Intermediate zone:}~\\
For values of $r$ satisfying $r_s \lesssim r \ll \omega^2$, the $\omega^2$ term is always small if $\omega$ is, and it can be treated perturbatively. Defining $ \epsilon \equiv \omega r_s$, we can  expand the Regge--Wheeler field as
\begin{equation}
    \Psi_\text{RW}=\Psi^{(0)}+\epsilon \Psi^{(1)}(r)+\epsilon^2 \Psi^{(2)}(r)+\cdots,
\end{equation}
and then solve order-by-order in $\epsilon$.
It is convenient to define a new coordinate $z\equiv r_s/r$ and to redefine the field $\Psi$ 
\begin{align}
    u_\ell(z(r)) &\equiv \sqrt{\frac{(\ell+2)(\ell-1)}{2}}\frac{r^\ell}{r_s^{\ell+1}}\Psi_\text{RW}(r)\, ,
\label{eq:psiRWuL}
\end{align}
in order to recast~\eqref{eq:RW_d=4} into
\begin{equation}
    z(1-z)u_\ell''+\big[2\ell+2-(2\ell+3)z\big]u_\ell'-(\ell+3)(\ell-1)u_\ell=-\frac{\epsilon^2}{z^3(1-z)}u_\ell \,.
    \label{eq:RW_u}
\end{equation}
The differential operator on the left hand side is hypergeometric, so that the homogeneous equation is in the standard form of a hypergeometric equation.

At zeroth and first order in $\epsilon$, we can set the right-hand side of~\eqref{eq:RW_u} to zero and solve the un-sourced hypergeometric equation for $u_\ell$. Denoting with $u_{\ell}^{(n)}$ the  $O(\epsilon^n)$ intermediate-zone solution for $u_\ell$, we have
\begin{equation}
    u_{\ell}^{(n)}(z)=b_1^{(n)}u_\text{reg}(z)+b_2^{(n)}u_\text{irr}(z), \qquad \text{for } n=0,1,
\label{eq:uellIZ}
\end{equation}
with $b_1^{(n)}$ and $b_2^{(n)}$ arbitrary integration constants, to be determined by matching to the near-zone solution~\eqref{eq:psiRWNZ}. The solutions $u_\text{reg}$ and $u_\text{irr}$ are the two independent hypergeometric functions solving the homogeneous version of~\eqref{eq:RW_u}:\footnote{
Note that the regular solution \eqref{eq:ureg0} can be equivalently written as
\begin{equation}
    u_\text{reg}(z)=-\frac{24(2\ell)!}{\left((\ell+2)!\right)^2}z^{-2\ell-1}\,{}_2F_1\left[\begin{array}{c}
    -\ell-2\,,\,-\ell+2\\[-3pt]
    -2\ell
    \end{array}\Big\rvert \,z\,\right]\,,
\label{eq:ureg1}
\end{equation}
by using hypergeometric identities.
} 
\begin{align}
    u_\text{irr}(z) &= \,{}_2F_1\left[\begin{array}{c}
    \ell-1\,,\,\ell+3\\[-3pt]
    2\ell+2
    \end{array}\Big\rvert \,z\,\right] , \label{eq:uirr0} \\
u_\text{reg}(z) &= (-z)^{-\ell-3}\,{}_2F_1\left[\begin{array}{c}
    2-\ell\,,\,\ell+3\\[-3pt]
    5
    \end{array}\Big\rvert \,\frac{1}{z}\,\right]
\nonumber
\\[1pt]
& = 24(-z)^{-\ell-3}\sum_{k=0}^{\ell-2}\frac{(-1)^k(\ell-2)!}{(\ell-2-k)!k!} \frac{\Gamma(\ell  +k+3)}{\Gamma(\ell+3)\Gamma(k+5)} z^{-k} ,
\label{eq:ureg0}
\end{align}
where $u_\text{reg}$ is regular at $z=\infty$ ($r=0$), while $u_\text{irr}$ is singular there. In particular, $u_\text{reg}$ is a finite polynomial in $1/z$, while $u_\text{irr}$  contains a $\log(1-z)$.

Comparing  the $z\rightarrow1$ ($r\to r_s$) limit of the intermediate zone solution~\eqref{eq:uellIZ} with the small-$\epsilon$ expansion of the near-zone solution~\eqref{eq:psiRWNZ}, which in these variables reads
\begin{equation}
    u_\text{NZ}(z)=\frac{A_\ell}{z^\ell} \mathrm{e}^{-i\epsilon\left(\frac1z+\log\left(\frac{1-z}{z}\right) \right)},
    \qquad
A_\ell\equiv\frac{ B}{r_s}\sqrt{\frac{(\ell+2)(\ell-1)}{2}} ,
\label{eq:NZu0}
\end{equation}
we find the following matching conditions for the  constant coefficients appearing in~\eqref{eq:uellIZ}:\footnote{In doing this, we used  the limit  of the hypergeometric function
\begin{equation}
\lim_{z\rightarrow 1^-}
\,{}_2F_1\left[\begin{array}{c}
    a\,,\,b\\[-3pt]
    a+b
    \end{array}\Big\rvert \,z\,\right]
= - \frac{\Gamma(a+b)}{\Gamma(a)\Gamma(b)} \left[ \log(1-z) - 2 \gamma_E +\psi(a)+\psi(b) \right] + O(1-z)\,,
\end{equation} 
with $a=\ell-1$ and $b=\ell+3$, and the Chu--Vandermonde identity
\begin{equation}
\,{}_2F_1\left[\begin{array}{c}
    -m\,,\,b\\[-3pt]
    c
    \end{array}\Big\rvert \,1\,\right]
= \frac{(-1)^m(b-c)!}{(b-c-m)!}  \frac{\Gamma(c)}{\Gamma(c+m)} 
\qquad\qquad \text{for } m=0,1,2,\cdots ,
\end{equation}
with $m=\ell-2$, $b=\ell+3$ and $c=5$.} 
\begin{align}
    b_1^{(0)} &= -\frac{A_\ell}{24}\frac{(\ell+2)!}{(\ell-2)!},& \qquad b_2^{(0)} &= 0, \\
    b_1^{(1)} &= -\frac{iA_\ell}{24}\frac{(\ell+2)!}{(\ell-2)!}\left(H_{\ell+2}+H_{\ell-2}-1\right),& \qquad b_2^{(1)} &= iA_\ell\frac{(\ell+2)!(\ell-2)!}{(2\ell+1)!}\,.
\end{align}
At second order in $\epsilon$, the  intermediate-zone solution $u^{(2)}$ satisfies
\begin{equation}
    z(1-z)\partial_z^2u_\ell^{(2)}+\left[2\ell+2-(2\ell+3)z\right]\partial_zu_\ell^{(2)}-(\ell+3)(\ell-1)u_\ell^{(2)}=-\frac{u_\ell^{(0)}(z)}{z^3(1-z)}\equiv S^{(u)}_{\ell}(z).
    \label{eq:RW_Sch_IZO2}
\end{equation}
This is an inhomogeneous equation with source $S^{(u)}_{\ell}(z)$, which can be solved using a Green's function.
The general solution  reads:\footnote{We used the Green's function
\begin{equation}
G(x,y)= \big[u_\text{irr}(x)u_\text{reg}(y) - u_\text{reg}(x)u_\text{irr}(y) \big]\frac{y^{2\ell+1}}{W_\ell} \theta(x-y)\,.
\end{equation}
It is also useful to recall the standard  formula for the Wronskian: $W\left[f_1(y),f_2(y)\right]=(1-c)y^{-c}(1-y)^{c-a-b-1}$, where 
\be
f_1(y) = \,{}_2F_1\left[\begin{array}{c}
    a\,,\,b\\[-3pt]
    c
    \end{array}\Big\rvert \,y\,\right] ~~~~{\rm and}~~~~ f_2(y) = y^{1-c}\,{}_2F_1\left[\begin{array}{c}
   a-c+1\,,\,b-c+1\\[-3pt]
    2-c
    \end{array}\Big\rvert \,y\,\right]\,.
\ee
Here $f_1(z)$ and $f_2(z)$ correspond to the homogeneous solutions $u_\text{irr}(z)$ and $u_\text{reg}(z)$ respectively (with $u_\text{reg}(z)$ given in~\eqref{eq:ureg1}), for $a=\ell-1$, $b=\ell+3$ and $c=2\ell+2$.}
\begin{equation}
\begin{aligned}
    u_\ell^{(2)}(z) &= \left[b_1^{(2)} + b_1^{(0)}\int^z\D y \frac{y^{2\ell-2}}{(1-y)W_\ell}u_\text{reg}(y)u_\text{irr}(y) \right]u_\text{reg}(z) \\
    &~~~\,~+ \left[b_2^{(2)} - b_1^{(0)}\int^z \D y \frac{y^{2\ell-2}}{(1-y)W_\ell}u_\text{reg}(y)^2 \right]u_\text{irr}(z)\,,
    \label{eq:u2GreenInt}
\end{aligned}
\end{equation}
where  $ W_\ell = -24(2\ell+1)!/((\ell+2)!)^2$.
In~\eqref{eq:u2GreenInt} we are omitting the lower integration bound---see the discussion around equation~\eqref{eq:psi2GreenInt}.

Evaluating  the integrals in~\eqref{eq:u2GreenInt} and matching the result to the near zone solution at second order in $\epsilon$, one can fix the coefficients $b_1^{(2)}$ and $b_2^{(2)}$ (see~\eqref{eq:bb1200} for an example in the scalar field case).
Plugging the expressions back into \eqref{eq:psiRWuL}, one finds the following $\ell=2$ intermediate-zone solution for the Regge--Wheeler field:
\begin{equation}
\begin{aligned}
    \Psi^{\rm IZ}_{\text{RW},\ell=2}(r) \underset{\frac{r}{r_s}\to\infty}{=} &~B\frac{r^3}{r_s^3}+iB\omega r_s\left(\frac{13r^3}{12 r_s^3}+\frac{r_s^2}{5 r^2}\right) \\
    &+ B\omega^2 r_s^2 \bigg[-\frac{r^5}{14r_s^5}-\frac{13r^4}{42r_s^4}+\frac{r^3}{r_s^3} \left(- \frac{95+8\pi ^2}{48}+\frac{107}{210}\log \frac{r_s}{r}\right)+\frac{319r^2}{420r_s^2}\\
    &\hspace{1.7cm}+\frac{153r}{280r_s}
    +\frac{223}{420}+\frac{363r_s}{560 r}+\frac{r_s^2}{25 r^2}\left(6-5 \log \frac{r_s}{r}\right)\bigg]+O\left(\frac{r_s^3}{r^3}\right)\,.
\end{aligned}
\label{eq:psiRWl2}
\end{equation}
The analogous solutions for $\ell=3,4$ are provided in Appendix~\ref{app:l=3-4_solution}. We will match these general relativistic solutions to point-particle EFT in Section~\ref{sec:GR-EFT}. Before proceeding, we present the analogous analysis for the Zerilli equation in the even sector.
﻿
﻿
\subsubsection{Even sector}
\label{sec:GRsolseven}

The even-parity sector is governed by the Zerilli equation, which in $D=4$ is~\cite{Zerilli:1970wzz}
\begin{equation}
    \partial_r\left(f(r)\partial_r\Psi_\text{Z}(r)\right)+\left( \frac{\omega^2}{f(r)}-\frac{r_s}{r^3}-\frac{2\lambda}{3r^2}-\frac{8 \lambda^2(2\lambda+3)}{3(2\lambda r+3r_s)^2} \right)\Psi_\text{Z}(r)=0\,,
    \label{eq:Zerilli_D4}
\end{equation}
where $ \lambda \equiv (\ell-1)(\ell+2)/2$.
As before, we split space up into
a near zone ($r-r_s\ll r_s$) and an intermediate zone ($r_s \lesssim r \ll \omega^{-1}$). (The far zone is again not needed for the matching calculation we are undertaking.) We consider each of these in turn.

\paragraph{Near zone:}~\\
In the near zone regime, ($r-r_s\ll r_s$), the $\omega$-independent piece of the potential is subdominant, and can be neglected. The Zerilli equation then takes the same form as the Regge--Wheeler equation in this region:
\begin{equation}
    \left(\frac{\D^2}{\D r_\star^2}+\omega^2\right)\Psi_\text{Z}=0.
\label{eq:Z_NZ}
\end{equation}
Imposing  ingoing  boundary conditions at the horizon determines the following solution:
\begin{equation}
    \Psi_\text{Z}^{\rm NZ}(r)=C \E^{-i\omega\left(r+ r_s\log\left(\frac{r}{r_s}-1\right)\right)},
    \label{eq:Zerilli_NZ}
\end{equation}
with $C$ a free integration constant.

\vspace{-10pt}
\paragraph{Intermediate zone:}~\\
In the intermediate zone, one can derive a solution perturbatively in $\omega$, as in the scalar and the gravitational odd cases. However, a more direct and efficient approach is to take advantage of Chandrasekhar's symmetry~\cite{10.2307/78902,1975RSPSA.343..289C,Chandrasekhar:1985kt}. The Chandrasekhar duality is a symmetry---present only in $D=4$---that  relates the Regge--Wheeler and Zerilli equations. Mathematically, it belongs to the class of Darboux transformations of second-order ordinary differential  equations, and can be used as a way of generating solutions~\cite{1999physics...8003D,Glampedakis:2017rar,Rosen:2020crj}. In particular, if we know a solution $\Psi_\text{RW}(r)$ of the Regge--Wheeler equation~\eqref{eq:RW_d=4},  the Chandrasekhar duality guarantees  that~$\Psi_\text{Z}(r)$, defined by
\begin{equation}
\Psi_\text{Z}(r)=\Big( f(r)\partial_r-{\cal W}(r) \Big) \Psi_\text{RW}(r) ,
\label{eq:Chandrasekhar_transformation}
\end{equation}
where $ {\cal W}$ is given by
\begin{equation}
    {\cal W}(r)\equiv \frac{3r_s(r_s-r)}{r^2(3r_s+2\lambda r)}-\frac{2\lambda(\lambda+1)}{3r_s}\,,
    \label{eq:superpotential}
\end{equation}
solves the $D=4$ Zerilli equation~\eqref{eq:Zerilli_D4}. We can therefore easily generate solutions to the Zerilli equation by mapping our known solutions to the Regge--Wheeler equation.

To carry this out in practice, 
let us thus redefine $\Psi_\text{RW}(r)$ in \eqref{eq:Chandrasekhar_transformation} using the field redefinition \eqref{eq:psiRWuL}, with $u_\ell(z)$ given by
\begin{equation}
    u_\ell(x)=u_\ell^{(0)}(z)+\epsilon u_\ell^{(1)}(z)+\epsilon^2 u_\ell^{(2)}(z) ,
\end{equation}
where $u_\ell^{(0)}$ and  $u_\ell^{(1)}$ are (cf.~\eqref{eq:uellIZ})
\begin{equation}
    u_{\ell}^{(n)}(z)=c_1^{(n)}u_\text{reg}(z)+c_2^{(n)}u_\text{irr}(z), \qquad \text{for } n=0,1,
\label{eq:uellIZ-Zerilli}
\end{equation}
while  $u_\ell^{(2)}$ is (cf.~\eqref{eq:u2GreenInt})
\begin{equation}
\begin{aligned}
    u_\ell^{(2)}(z) =& \left[c_1^{(2)} + \int^z\D y \frac{y^{2\ell-2}}{(1-y)W_\ell}u_\ell^{(0)}(y)u_\text{irr}(y) \right]u_\text{reg}(z) \\
    &+ \left[c_2^{(2)} - \int^z \D y \frac{y^{2\ell-2}}{(1-y)W_\ell}u_\ell^{(0)}(y)u_\text{reg}(y)\right]u_\text{irr}(z).
    \label{eq:u2GreenIntGal}
\end{aligned}
\end{equation}
Here $u_\text{irr}$ and $u_\text{reg}$ can be read off from~\eqref{eq:uirr0}--\eqref{eq:ureg0}. Then, it is straightforward to check that the field $\Psi_\text{Z}$ constructed in \eqref{eq:Chandrasekhar_transformation} satisfies \eqref{eq:Zerilli_D4} up to second order in $\omega$. We stress that, up to this point, we are not yet making any statements about the physical implications of the symmetry. We are merely using the duality  as a solution-generating technique. In particular, the integration constants $c_j^{(n)}$ appearing in~\eqref{eq:uellIZ-Zerilli} and~\eqref{eq:u2GreenIntGal} are unrelated to the~$b_j^{(n)}$ in \eqref{eq:uellIZ} and~\eqref{eq:uellIZ}. We will discuss the significance of the duality, thought of as a symmetry, for the Love numbers later on.

By matching to  the near zone solution~\eqref{eq:Zerilli_NZ}, we can then easily determine the $c_j^{(n)}$ constants in terms of the amplitude $C$. 
For $\ell=2$, this yields 
the following  result for the intermediate-zone solution to the Zerilli equation:
\begin{equation}
\begin{aligned}
   \Psi_{\text{Z},\ell=2}^{\rm IZ}(r) \underset{\frac{r}{r_s}\to\infty}{=}  &C\left(\frac{r^3}{r_s^3}+\frac{3 r^2}{4 r_s^2}-\frac{9 r}{16 r_s}-\frac{21}{64}+\frac{63 r_s}{256 r}-\frac{189 r_s^2}{1\,024 r^2}\right)\\
   &+iC\omega r_s\left(\frac{4 r^3}{3 r_s^3}+\frac{r^2}{r_s^2}-\frac{3 r}{4 r_s}-\frac{7}{16}+\frac{21 r_s}{64 r}-\frac{59 r_s^2}{1\,280 r^2}\right)\\
   &+C\omega^2 r_s^2 \bigg[-\frac{r^5}{14 r_s^5}-\frac{67 r^4}{168 r_s^4}-\frac{r^3}{r_s^3} \left(\frac{\pi ^2}{6}+\frac{1\,711}{672}-\frac{107}{210} \log \frac{r_s}{r}\right)\\
   &\hspace{1.75cm}-\frac{r^2}{r_s^2}\left(\frac{\pi ^2}{8}+\frac{11\,119}{13\,440}-\frac{107}{280} \log \frac{r_s}{r}\right) +\frac{r}{r_s}\left(\frac{3 \pi ^2}{32}+\frac{8\,739}{3\,584}-\frac{321}{1\,120}\log \frac{r_s}{r}\right)\\
   &\hspace{1.75cm}+\frac{7\pi^2}{128}+\frac{34\,591}{30\,720}-\frac{107}{640} \log \frac{r_s}{r}-\frac{r_s}{r}\left(\frac{21 \pi ^2}{512}+\frac{6\,431}{40\,960}-\frac{321}{2\,560} \log \frac{r_s}{r}\right)\\
   &\hspace{1.75cm}+\frac{r_s^2}{r^2}\left(\frac{63 \pi ^2}{2\,048}+\frac{431\,313}{819\,200}-\frac{3\,011}{10\,240} \log \frac{r_s}{r}\right)\bigg]+O\left(\frac{r_s^3}{r^3}\right).
   \label{eq:Z_UV_sol}
\end{aligned}
\end{equation}
One can carry out the same exercise for higher multipoles. See
Appendix~\ref{app:l=3-4_solution} for the explicit $\ell=3,4$ solutions.
﻿
﻿
\subsection{EFT calculation of dynamical Love numbers}
\label{sec:GR-EFT}
﻿

We now match the  general relativistic solutions we have just obtained to the  worldline EFT of a point particle coupled to gravity, order by order in frequency. As argued in the toy example of the scalar field, the matching can be carried out at large $r$, but still within the intermediate zone of the full solution, where the EFT is well defined.

The point-particle EFT in $D$ spacetime dimensions, to quadratic order in the bulk graviton fluctuation $g_{\mu\nu} = \eta_{\mu\nu} + 2h_{\mu\nu}/M_{\rm Pl}^{(D-2)/2}$, can be written as~\cite{Goldberger:2004jt,Goldberger:2005cd,Hui:2020xxx,Hadad:2024lsf}
\begin{equation}
\begin{aligned}
S  = S_{\rm pp} 
& + \int {\rm d}^Dx\sqrt{-g}\left[ -\frac{1}{2}\nabla_\lambda h_{\mu\nu} \nabla^\lambda h^{\mu\nu}+\nabla_\lambda h_{\mu\nu} \nabla^\nu h^{\mu\lambda}-\nabla_\mu h\nabla_\nu h^{\mu\nu} +\frac{1}{2} \nabla_\mu h\nabla^\mu h
\right]
\\
&  + \int\rd \tau \sum_{\ell=2}^\infty \left[
Q_{B}^{A_\ell\, j}(\tau)  B^{(\ell)}_{A_\ell\, j}
+ Q_{E}^{A_\ell }(\tau)  E^{(\ell)}_{A_\ell} \right] ,
\end{aligned}
\label{eq:ppEFTgrav}
\end{equation}
where $S_{\rm pp}$ is the worldline action~\eqref{eq:Spps} of the point particle, and where $Q_{B,E}$ are composite operators describing the induced response of the object to external gravitational $B$ and $E$  fields, built from some internal degrees of freedom, $X$. In~\eqref{eq:ppEFTgrav}, we introduced the multi-index notation $A_\ell\equiv i_1\cdots i_\ell$ and the operators~\cite{Goldberger:2020fot,Saketh:2022xjb,Saketh:2023bul,Glazer:2024eyi}
\begin{align}
B^{(\ell)}_{i_1\cdots i_\ell \,j} & \equiv \partial_{( i_1}\cdots \partial_{i_{\ell-2}} B_{i_{\ell-1}i_\ell )_T\, j}
\\
E^{(\ell)}_{i_1\cdots i_\ell} & \equiv \partial_{( i_1}\cdots \partial_{i_{\ell-2}} E_{i_{\ell-1}i_\ell)_T}\,,
\end{align}
which are symmetrized derivatives of the gravito-magnetic and gravito-electric fields, which are themselves defined in terms of the Weyl tensor as\footnote{In $D>4$, $B_{i_1 i_2 j}$ and $E_{i_1 i_2} $ do not exhaust all independent components of the Weyl tensor. While in $D=4$ $C_{ijkl}$ can be re-expressed in terms of  the electric and magnetic components, this is not the case in higher spacetime dimensions.  This implies that in \eqref{eq:ppEFTgrav} one must add an additional action term describing the response of a purely tensorial degree of freedom~\cite{Hui:2020xxx}. Since this extra sector is decoupled  from the $B$ and $E$ operators at quadratic order, and will play no role in the following discussion, we have omitted it from~\eqref{eq:ppEFTgrav}.
}
\begin{align}
B_{i_1 i_2 j}  & \equiv C_{0i_1i_2 j} ,
\label{eq:Bijk}
\\
E_{i_1 i_2}  & \equiv C_{0i_10i_2} .
\label{eq:Eij}
\end{align}
In~\eqref{eq:ppEFTgrav} we implicitly put ourselves in the rest frame of the point particle. If desired, the expressions can be covariantized  using the projector ${\mathbb{P}^\nu}_\mu \equiv  \delta_\mu^\nu + u_\mu u^\nu$, as discussed in Section~\ref{sec:scalarppEFT}.

We will follow the same strategy as in the scalar field case. First, we will obtain the induced gravitational field sourced by the point-particle in the presence of a background field, neglecting the nonlinearities of gravity. 
This will fix the homogeneous solution, up to an overall amplitude. Then, we will include the coupling to gravity using the Born series~\cite{Correia:2024jgr,Caron-Huot:2025tlq} to determine the particular solution induced by the  Schwarzschild corrections to the bulk potential.
Finally we will match this EFT solution to the general relativity calculations performed in Section~\ref{sec:GRsols}. As before, we treat each parity sector separately, beginning with the parity odd case.
﻿
﻿
\subsubsection{Odd sector}
﻿
First consider the odd  sector of the point-particle EFT action~\eqref{eq:ppEFTgrav}. We double the number of fields on a closed-time contour and 
denote the graviton fields on the two-sided path by $h_1^{\mu\nu}$ (inserted on the forward part of the contour) and $h_2^{\mu\nu}$ (inserted on the backward part of the contour). As before we define the Keldysh basis
\begin{equation}
h_+\equiv \frac{1}{2}\left( h_1+h_2 \right),
\qquad
h_-\equiv  h_1-h_2.
\label{eq:keldyshbasispm}
\end{equation}
We obtain an effective action for $h_{\pm}$ by integrating out the $X$ degrees of freedom on which the $Q$ depend
\begin{equation}
\label{eq:PIXX}
    \E^{i\Gamma^\text{in-in}_\text{int}[h_\pm]} = \int\mathcal{D}X_+\mathcal{D}X_- \, \E^{iS\left[h_\pm,X_\pm\right]} .
\end{equation}
Perturbatively we can replace $Q_B$, at leading order, by its linear response
\begin{equation}
    \langle Q_{B,\,I}^{A_\ell \,j}(\tau)\rangle=\int \D \tau'\,K_{IJ}^{(B)}{}^{A_\ell\, j\vert A_{\ell'}\, j'}(\tau-\tau'){B}^J_{A_{\ell'}\, j'}(\tau')\,,
\end{equation}
where $I,J=\{+,-\}$, and $K_{IJ}^{(B)}{}^{A_\ell\, j\vert A_{\ell'}\, j'}$ is a response kernel, related to the two-point function of $Q_B$ in the Keldysh basis~\cite{Goldberger:2020fot,Saketh:2022xjb,Saketh:2023bul,Glazer:2024eyi} by
\begin{equation}
    \langle Q_{B,I}^{A_{\ell}\, j}(\tau)Q_{B,J}^{A_{\ell'}\, j'}(\tau')\rangle=-iK_{IJ}^{(B)}{}^{A_\ell\, j\vert A_{\ell'}\, j'}(\tau-\tau')\,.
\end{equation}
As we discuss below, we remain agnostic about its detailed form, which we will parametrize in the most general way, compatible with the symmetries of the problem.\footnote{We are assuming that the kernel admits an expansion in small frequency, which relies on the characteristic time scale of the black hole's internal dynamics being much faster than the timescales on which we are probing the system. In this case, this is natural because we are considering $\omega \ll r_s^{-1}$.}
For non-spinning objects, the tensorial structure of the kernel in the spatial indices reduces to a product of Kronecker deltas, leading to the following  in-in effective action:
\begin{equation}
    \Gamma^\text{in-in}_\text{int}[h_\pm,X_\pm]=\int \D\tau_1 \D\tau_2\sum_{\ell=2}^\infty K_{{IJ},\ell}^{(B)}(\tau_2-\tau_1) {B}^I_{A_\ell\, j}(\tau_2){B}^{{J}}{}^{ A_\ell\, j}(\tau_1) ,
\label{eq:gammainineft}
\end{equation}
which we can use to compute the one-point function of the gravito-magnetic field.

\vspace{-10pt}
\paragraph{Magnetic one-point function}~\\
The effective action~\eqref{eq:gammainineft} depends on the gravitational field $h$ through the magnetic component of the Weyl tensor. 
Thus, rather than computing the one-point function
of the graviton field $h$, it is more convenient to evaluate the one-point function of $B$ itself. We therefore wish to compute:
\begin{equation}
    \langle B_{+\,abc}(t,\vec{x})\rangle_\text{in-in}=\int \mathcal{D}h_+\mathcal{D}h_- B_{+\,abc}(t,\vec{x})\E^{i\Gamma^\text{in-in}_\text{int}[h_\pm,X_\pm]} ,
    \label{eq:1-pt_fct_def}
\end{equation}
in the presence of a background  $\bar B$ for the magnetic field. We are computing the expectation value of $B^+$ because this is the field combination that has a classical interpretation. For the external classical source we fix $h^+_1=h^+_2\equiv h^+$ (equivalently, $B^+_1=B^+_2\equiv B^+$ for the linearized Weyl tensor).
Using the explicit form of the in-in effective action~\eqref{eq:gammainineft}, we obtain the following expression for the one-point function
\begin{equation}
    \langle B_{+\,abc}(t,\vec{x})\rangle_\text{in-in}= i\sum_{\ell=2}^\infty\int \D\tau_1 \D\tau_2 \, K_\ell^{(B)}(\tau_2-\tau_1)\langle B_{+\,abc}(t,\vec{x}) {B}_{-\,A_\ell\, j}\left(\tau_2\right) \rangle\bar{{B}}_+^{A_\ell\, j}\left(\tau_1\right).
    \label{eq:1-pt_fct}
\end{equation}
where $K_\ell^{(B)} \equiv K_{+-,\ell}^{(B)}$ is the response function 
\begin{equation}
    K_{+-}^{(B)}(\tau_2-\tau_1)=-G_R^{(Q_B)}(\tau_2-\tau_1) \equiv i \langle \left[ Q_{B,+}(\tau_2),Q_{B,-}(\tau_1) \right]\rangle \theta(\tau_2-\tau_1).
\end{equation}
The UV information is stored in the response function $K_{\ell}^{(B)}$, which depends on the microscopic details of the object (the boundary conditions at the surface, the object's internal dynamics, etc.). Assuming that the timescale of the object's dynamics is parametrically faster than the timescale on which we probe the system (so that we can think of the internal dynamics as instantaneous), we can expand the Fourier transform of the response function
\begin{equation}
\begin{aligned}
    K_{\ell}^{(B)}(\tau_2-\tau_1) &= \int \frac{\D {\omega}}{2\pi}\E^{-i{\omega}(\tau_2-\tau_1)} K_{\ell}^{(B)}(\omega) \\
    &= \frac{1}{\ell!}\frac{\ell}{\ell+1}\int \frac{\D{\omega}}{2\pi}\E^{-i{\omega}(\tau_2-\tau_1)}\left[ \lambda_{0,\ell}^B+i{\omega}r_s \lambda_{1,\ell}^B+ ({\omega} r_s)^2\lambda_{2,\ell}^B+\cdots\right]\,,
\end{aligned}
\label{eq:EFTLNsodd}
\end{equation}
where the parameters $\lambda_i$ are the response coefficients {and the overall normalization is chosen for later convenience}.
 Terms with odd powers of $\omega$ (which are time-reversal odd) capture the dissipative response of the object, while those with even powers of $\omega$ parametrize the conservative tidal deformability. In particular, $\lambda_{0,\ell}^B$ is related to the magnetic static Love numbers, while $\lambda_{2,\ell}^B$ corresponds to the quadratic-in-frequency dynamical Love numbers that we wish to compute.

To determine the one-point function \eqref{eq:1-pt_fct}, we need to compute  the two-point function of $B$. We first express the magnetic component of the Weyl tensor in terms of the graviton field. Since the linearized Weyl tensor on Minkowski space is a gauge-invariant quantity, it is useful to fix a gauge. A convenient choice is the de Donder gauge, defined by $\partial^\mu(h_{\mu\nu}-\frac{1}{2}\eta_{\mu\nu}h)=0$, with $h\equiv \eta^{\mu\nu}h_{\mu\nu}$. In this gauge, the linearized equations of motion are $\square h_{\mu\nu}=0$, and the on-shell magnetic component of the Weyl tensor reads\footnote{Recall that $h_{\mu\nu}$ is the canonically normalized graviton field.}
\begin{equation}
B_{abc}=C_{0abc}
= \frac{2}{\Mpl}\left( \partial_a\partial_{[b}h_{c]0}-\partial_0\partial_{[b}h_{c]a} \right).
\label{eq:Babc}
\end{equation}
Similarly, we can write the symmetrized derivative of the Weyl tensor as
\begin{equation}
   B_{A_\ell\, j}= \frac{1}{\Mpl} \left[ \partial_{i_L}h_{0j}-\partial_j\partial_{( i_1}\cdots \partial_{i_{\ell-1}}h_{i_\ell)_T0}+\partial_0 F_{A_\ell\, j} \right] \,,
\label{eq:BiLj}
\end{equation}
where we have defined the combination
\begin{equation}
    F_{A_\ell j} \equiv -\partial_{( i_1}\cdots\partial_{i_{\ell-1}}h_{i_{\ell} )_T\, j}+\partial_j\partial_{( i_1}\cdots\partial_{i_{\ell-2}}h_{i_{\ell-1}i_{\ell} )_T}\, ,
\label{eq:AiLj}
\end{equation}
which is symmetric and traceless in the first $i_1\cdots i_\ell$ indices.

Let us  start by computing the product ${B}^-_{A_\ell\, j}\bar{{B}}^{+\, A_\ell j}$. It is convenient to unpack it using \eqref{eq:BiLj} and  \eqref{eq:AiLj}, and evaluate term by term. First, notice the following identity:
\be
\begin{aligned}
\Big(\partial_{A_\ell}h_{j0}-\partial_j\partial_{( i_1}\cdots \partial_{i_{\ell-1}}h_{i_\ell)0}\Big)
&\Big(\partial^{A_\ell}\bar h^{j}{}_{0}-\partial^j\partial^{( i_1}\cdots \partial^{i_{\ell-1}}\bar h^{i_\ell)}{}_{0}\Big)
\\
&= \frac{\ell+1}{\ell} \partial_{A_\ell} h_{0j} \left( 
 \partial^{A_\ell}\bar h^{j}{}_0
-  \partial^j\partial^{ i_1}\cdots \partial^{i_{\ell-1}}\bar h^{i_\ell}{}_{0}
\right),
\label{eq:identity1}
\end{aligned}
\ee
which holds when the $\ell$ indices denoted by the multi-index $A_\ell$ are symmetrized on the left-hand side (we suppress the $\pm$ superscripts  on $h$ and $\bar h$ in these intermediate steps for simplicity, and restore them at the end). Removing the traces from \eqref{eq:identity1}---as prescribed by \eqref{eq:BiLj}---amounts to correcting the right-hand side of \eqref{eq:BiLj} by additional terms with the schematic form
$\partial_k\partial^k h$, $\partial_k\partial^k \bar h$, or $\partial^k h_{k0}\partial^j \bar h_{j0}$. From the equations of motion in de Donder gauge, it follows that all these terms are $O(\omega^2)$. We can neglect these terms since
the static Love numbers of black holes vanish,  $\lambda_{0,\ell}^B=0$ so $K_{\ell}^{(B)}(\omega)$ starts linearly in $\omega$. Therefore, such terms would correspond to at least $O(\omega^3)$ contributions in the one-point function \eqref{eq:1-pt_fct}, which we neglect. Similarly, the combination  $\partial_0 F_{A_\ell\,j}\partial_0 \bar{F}^{A_\ell\,j}$, arising from the product ${B}^-_{A_\ell\, j}\bar{{B}}^{+\,A_\ell\, j}$, is of $O(\omega^2)$ and gives subdominant corrections to the dynamical Love numbers for black holes. Therefore, we are left with 
\begin{equation}
\begin{aligned}
{B}_{-\,A_\ell j}\bar{{B}}_+^{A_\ell\, j} =  \frac{1}{\Mpl^2}\bigg[ &
      2\frac{\ell+1}{\ell}\partial_{A_\ell}h_{0j}\partial^{A_{\ell-1}}\partial^{[i_\ell}\bar{h}^{j]}{}_{0}
  +\partial_{A_\ell}h_{0j}\partial_0\bar{F}^{A_\ell\,j}-(\partial_j\partial_{( A_{\ell-1}}h_{i_{\ell})_T0})\partial_0\bar{F}^{A_\ell \,j}  \\
     &+\partial_{A_\ell}\bar{h}_{0j}\partial_0 F^{A_\ell\,j}-(\partial_j\partial_{( A_{\ell-1}}\bar{h}_{j_{\ell})_T0})\partial_0F^{A_\ell \,j}
     + O(\omega^2) \bigg].
     \label{Magnetic_square}
\end{aligned}
\end{equation}
Let us focus for a moment on the terms involving $F$ in~\eqref{Magnetic_square}.
For the same reasons as above, removing traces from $\partial_{ A_{\ell-1}}{h}_{j_{\ell}0}$ and $\partial_{ A_{\ell-1}}\bar{h}_{j_{\ell}0}$ corresponds to subtracting  terms of order $\omega$, as a consequence of the equations of motion and gauge conditions.
Moreover, since the $A_\ell$ indices in $F^{A_\ell\, j}$ and $\bar F^{A_{\ell} j}$ are already symmetrized, we can drop the brackets $(\cdots)_T$ from $(\partial_j\partial_{( A_{\ell-1}}{h}_{j_{\ell})_T0})\partial^0\bar F^{A_\ell j}$ and $(\partial_j\partial_{( A_{\ell-1}}\bar{h}_{j_{\ell})_T0})\partial^0F^{i_L j}$, up to $O(\omega^3)$ corrections, which we neglect. As a result,~\eqref{Magnetic_square} is
\begin{equation}
\begin{aligned}
{B}_{-\,A_\ell\, j}\bar{{B}}_+^{A_\ell\, j} =
      \frac{1}{\Mpl^2}\bigg[ & 2\frac{\ell+1}{\ell}\partial_{A_\ell}h_{0j}\partial^{A_{\ell-1}}\partial^{[i_\ell}\bar{h}^{j]}{}_{0}
  \\
     & -2\frac{\ell+1}{\ell} \left[ \partial_{A_\ell}h_{0j} \left(\partial_0\partial^{A_{\ell-2}}\partial^{[i_{\ell-1}}  
      \bar{h}^{j]i_\ell}\right)+(h \leftrightarrow \bar{h}) \right]
     + O(\omega^2).
     \label{Magnetic_square-2}
\end{aligned}
\end{equation}
Combining this expression with~\eqref{eq:Babc}, we  thus have
\begin{equation}
\begin{aligned}
   \langle B_{+\,abc}{B}_{-\,A_\ell\, j} \rangle\bar{{B}}_+^{A_\ell \,j} =& \,\, \frac{4(\ell+1)}{\ell \Mpl^3} \langle (\partial_a\partial_{[b}h_{+\,c]0})( \partial_{A_\ell}h_{-\,j0} )\rangle \left[\partial^{A_{\ell-1}}\partial^{[i_\ell}\bar{h}_+^{j]}{}_{0}-\partial_0\partial^{A_{\ell-2}}\partial^{[i_{\ell-1}}\bar{h}_+^{j]i_\ell} \right] \\
   &- \frac{4(\ell+1)}{\ell \Mpl^3} \langle (\partial_a\partial_{[b}h_{+\,c]0})( \partial_0\partial_{A_{\ell-2}}\partial_{[i_{\ell-1}}h_{-\,j]i_\ell} )\rangle \partial^{A_\ell}\bar{h}_+^{j}{}_{0} \\
   &- \frac{4(\ell+1)}{\ell\Mpl^3}  \langle (\partial_0\partial_{[b}h_{+\,c]a}) ( \partial_{A_\ell}h_{-\,0j}) \rangle \partial^{A_{\ell -1}}\partial^{[i_\ell}\bar{h}_+^{j]}{}_{0} 
   + O(\omega^2)
    \label{eq:Simp_corr}
\end{aligned}
\end{equation}
To compute this up to linear order in $\omega$, we can use the instantaneous propagator for the graviton in  de Donder gauge~\cite{Hui:2020xxx}: 
\begin{equation}
    \langle h_{+\,\mu\nu}(t,\vec{x})h_{-\,\rho\sigma}(\tau_1, \vec{0})\rangle=i\delta(t-\tau_1)\mathcal{P}_{\mu\nu\rho\sigma}^\text{dD}\int \frac{\D^{D-1}\vec{p}}{(2\pi)^{D-1}}\frac{\E^{i\vec{p}\cdot \vec{x}}}{\vec{p}^2},
    \label{graviton_propagator}
\end{equation}
where the propagator numerator is
\begin{equation}
    \mathcal{P}_{\mu\nu\rho\sigma}^\text{dD}\equiv -\frac12\left( \eta_{\mu\sigma}\eta_{\nu\rho}+\eta_{\mu\rho}\eta_{\nu\sigma}-\frac{2}{D-2}\eta_{\mu\nu}\eta_{\rho\sigma} \right).
    \label{graviton_propagator_tensor}
\end{equation}
Upon substituting \eqref{graviton_propagator} into \eqref{eq:Simp_corr}, the tensor structure of $\mathcal{P}_{\mu\nu\rho\sigma}^\text{dD}$ forces  the two-point functions on the second and third lines of \eqref{eq:Simp_corr} to vanish. 
We are thus left with 
\begin{equation}
    \langle B_{+\,abc}{B}_{-\,A_\ell\, j} \rangle\bar{{B}}_+^{A_\ell\, j} = \frac{4(\ell+1)}{\ell \Mpl^3}\langle (\partial_a\partial_{[b}h_{+\,c]0})( \partial_{A_\ell}h_{-\,j0} )\rangle \left[\partial^{A_{\ell-1}}\partial^{[i_\ell}\bar{h}_+^{j]}{}_{0}-\partial_0\partial^{ A_{\ell-2}}\partial^{[i_{\ell-1}}\bar{h}_+^{j]i_\ell} \right] + O(\omega^2).
\label{eq:BBBbar-2}
\end{equation}

It is convenient at this point to use the explicit form of the external tidal field $\bar{h}_+$. Recall that we are interested in  solving the Minkowski spacetime equation of motion in de Donder gauge, $\square \bar h_{\mu\nu}= (\vec{\nabla}^2+\omega^2) \bar h_{\mu\nu} =0$, up to linear order in $\omega$. At this order, the tidal field solution can be expressed as a Taylor expansion around $\vec{x}=0$, with the time dependence  factored out as $ \E^{-i\omega t}$. In cartesian coordinates,   the component $\bar{h}_{0j}$ can be written as
\begin{equation}
    \bar h_{0 j}(t,\vec{x}) 
    = \Mpl\E^{-i\omega t}\sum_{\ell} c_{j\vert j_1\cdots j_{\ell}}x^{j_1}\cdots x^{j_{\ell}} + O(\omega^2) ,
    \label{dD_tidal_solution}
\end{equation}
with $c$  symmetric and traceless  in its $j_1\cdots j_{\ell}$ indices (and vanishing when totally symmetrized).

We can write the other components of $h_{\mu\nu}$ similarly.
Notice that the last term in~\eqref{eq:BBBbar-2} is already of order  $O(\omega)$. Therefore, it is sufficient to evaluate it using the static solution for $\bar{h}_+^{ij}$. From the form of this solution, it is easy to show that this term vanishes identically. We provide the details in Appendix~\ref{app:tidalfield} (see in particular~\eqref{eq:oddcond1}).
As a result,~\eqref{eq:Simp_corr} reduces to its first term, and the one-point function for the magnetic component of the Weyl tensor  reads 
\begin{multline}
    \langle B_{+\,abc}(t,\vec x)\rangle_\text{in-in} = \frac{4i}{\Mpl^3}\sum_{\ell=2}^\infty\frac{(-1)^\ell(\ell+1)}{\ell}\int \D\tau_1 \D\tau_2\int\frac{\D{\omega}}{2\pi}\E^{-i{\omega}(\tau_2-\tau_1)}K^{(B)}_\ell({\omega})\\
    \times  \partial_{A_\ell}\partial_a\partial_{[b}\langle h_{+\,c]0}(t, \vec x)h_{-\,j0}(\tau_2)\rangle \partial^{A_{\ell-1}}\partial^{[i_\ell}\bar{h}_+^{j]}{}_{0}(\tau_1)\,,
    \label{1_pt_fct-grav_propagators}
\end{multline}
where the $(-1)^\ell$ prefactor arises from moving the derivatives $\partial_{i_L}$ out of the correlator. Plugging in the expressions for the propagator~\eqref{graviton_propagator} and the external field~\eqref{dD_tidal_solution}, we obtain 
\begin{equation}
     \langle B_{+\,abc}(t, \vec x)\rangle_\text{in-in} = -\frac{\E^{-i\omega t}}{\Mpl^2}\sum_{\ell=2}^\infty\frac{4(-i)^\ell(\ell+1)!}{\ell}K^{(B)}_\ell(\omega)c^{[j\vert i_1]\cdots i_\ell}\partial_a \mathcal{P}_{0j0[c}^\text{dD}\partial_{b]}\int \frac{\D^{D-1}\vec{p}}{(2\pi)^{D-1}}\E^{i\vec{p}\cdot\vec{x}}\frac{p_{i_1}\cdots p_{i_\ell}}{\vec{p}^2}\,.
\label{1_pt_fct-grav_propagators-2}
\end{equation}
We can Fourier transform this back into position space
\begin{equation}
     \langle B_{+abc}(t, \vec x)\rangle_\text{in-in}\! 
     = \frac{\E^{-i\omega t}}{\Mpl^2}\sum_{\ell=2}^\infty\frac{(-1)^{\ell+1}(\ell+1)!}{\ell}K^{(B)}_\ell(\omega)\frac{2^{\ell-1}\Gamma\left(\frac{D-3}{2}\right)\Gamma\left(\frac{5-D}{2}\right)}{\pi^{\frac{D-1}{2}}\Gamma\left(\frac{5-D}{2}-\ell\right)} c_{[j\vert i_1]\cdots i_\ell} \partial_a \delta^j_{[c}\partial_{b]} \frac{x^{i_1}\cdots x^{i_\ell}}{\vert\vec{x}\vert^{2\ell+D-3}}
\label{1_pt_fct-grav_propagators-3}
\end{equation}
and simplify the tensor structure using\footnote{Note that in deriving~\eqref{eq:oddidt} we  assume that  $\partial_b$ is a derivative with respect to coordinates on the $(D-2)$-sphere---which we make explicit in~\eqref{eq:Brijplus}. This allows us to  factor out $1/\vert\vec{x}\vert^{2\ell+D-3}$ in~\eqref{eq:oddidt}.}
 \begin{align}
\label{eq:oddidt}
    c_{[j\vert i_1]\cdots i_\ell} & \delta^j_{[c}\partial_{b]}  \left(x^{i_1}\cdots x^{i_\ell}\right) = \frac14\left(\partial_b c_{c\vert A_\ell}-\partial_c c_{b\vert A_\ell}-\partial_b c_{i_\ell\vert c A_{\ell-1}}+\partial_c c_{i_\ell\vert b A_{\ell-1}}\right)x^{i_1}\cdots x^{i_\ell} \nonumber \\
    &=\frac14 \left[\partial_{b}(x^{i_1}\cdots x^{i_\ell}) c_{c\vert A_\ell}+\partial_c\left(x^i c_{i\vert A_\ell} \frac1\ell\partial_b(x^{i_1}\cdots x^{i_\ell})\right)-(c \leftrightarrow b)\right] \nonumber \\
    &=\frac14 \left[\partial_{b}(x^{i_1}\cdots x^{i_\ell}) c_{c\vert A_\ell}+\frac1\ell\partial_c\partial_b\left(x^i c_{i\vert A_\ell}  x^{i_1}\cdots x^{i_\ell}\right)-\frac1\ell\partial_c\left(c_{b\vert A_\ell} x^{i_1}\cdots x^{i_\ell}\right)-(c \leftrightarrow b)\right] \nonumber \\
    &=\frac14 \left[\partial_{b}(x^{i_1}\cdots x^{i_\ell}) c_{c\vert A_\ell}+\frac1\ell\partial_b\left(x^{i_1}\cdots x^{i_\ell}\right)c_{c\vert A_\ell}-(c \leftrightarrow b)\right] \nonumber \\
    &= \frac{\ell+1}{2\ell} \partial_{[b}\left( x^{i_1}\cdots x^{i_\ell}\right) c_{c]\vert A_\ell}.
\end{align}
 Considering the $B_{rAB}$ component of~\eqref{1_pt_fct-grav_propagators-2}, where $AB$ denote angular coordinates, and writing the tidal field in spherical coordinates as 
\begin{equation}
    c_{c\vert i_1\cdots i_\ell}x^{i_1}\cdots x^{i_\ell}=c_\text{ext}r^{\ell+1}Y_c^{(T)}{}_\ell^m,
\end{equation}
with $Y_c^{(T)}{}_\ell^m$ a vector spherical harmonic~\cite{Hui:2020xxx}, we finally obtain
\begin{equation}
\label{eq:Brijplus}
    \langle B_{+\,r AB}(x)\rangle_\text{in-in} = -\frac{\E^{-i\omega t}}{\Mpl^2}\sum_{\ell=2}^\infty\frac{\ell+1}{2\ell}\frac{(\ell+1)!}{\ell}K^{(B)}_\ell(\omega)\frac{2^{\ell-1}\Gamma\left(\ell+\frac{D-3}{2}\right)}{\pi^{\frac{D-1}{2}}}\nabla_{[A}Y_{B]}^{(T)}{}_{\ell}^{m}r^2\partial_r\left( r^{-2}\frac{c_\text{ext}}{r^{\ell+D-4}}\right).
\end{equation}
We can also write the background tidal field in spherical coordinates
\begin{equation}
\begin{aligned}
    \bar{B}_{rAB} &= \frac{2}{\Mpl}\left[ \nabla_r\nabla_{[A}\bar h_{B]0}-\partial_t\nabla_{[A}\bar h_{B]r} \right] \\
    &= 2 \E^{-i\omega t}\sum_{\ell=2}^\infty c_\text{ext}  r^2\partial_r\big(r^{\ell-1}\big)\big(1+O(\omega r)\big)\nabla_{[A}Y_{B]}^{(T)}{}_{\ell}^{m}.
\end{aligned}
\end{equation}
The $O(\omega r)$ term in $\bar{B}_{rAB}$ represents a far-zone, frequency-dependent  correction to the static tidal field. It exhibits a different scaling in $r$ compared to the standard $r^\ell$ and $r^{-\ell-1}$ falloffs. Therefore, as long as we are concerned with matching the EFT in the intermediate zone, we can consistently neglect this contribution. Putting everything together and replacing $K_\ell^{(B)}$ by $\mu^{2\varepsilon}K_\ell^{(B)}$ (as we did in the scalar section), we obtain
\be
\begin{aligned}
    \bar{B}_{rAB}+\langle B_{+,rAB}\rangle_\text{in-in} =  ~&2\E^{-i\omega t} \sum_{\ell=2}^\infty c_{\rm ext}(\ell-1)\nabla_{[A}Y_{B]}^{(T)}{}_{\ell}^{m} \\ 
  &  \times\left( r^\ell+\mu^{2\varepsilon} \frac{ K^{(B)}_\ell(\omega)}{\Mpl^2}\frac{\ell+1}{2\ell}\frac{(\ell+1)!(D+\ell-2)}{\ell(\ell-1)}\frac{2^{\ell-2}\Gamma\left(\ell+\frac{D-3}{2}\right)}{\pi^{\frac{D-1}{2}}}r^{-\ell+3-D} \right).
\end{aligned}
    \label{Brij_Love}
\ee
The result~\eqref{Brij_Love} gives the one-point function of the magnetic component of the Weyl tensor in the presence of a tidal field, up to quadratic order in $\omega$, which we will use to match to the full black hole  solution. 
Before doing this, we must introduce the coupling to gravity and regulate the UV divergences.

\paragraph{Gravitational corrections}~\\
The solution~\eqref{Brij_Love} captures the leading in $r$ behavior of the $B$ field, but in order to match to the relativistic solution, we must include the coupling to gravity which generates subleading terms. To do this, we work in dimensional regularization and 
a Born series expansion following~\cite{Correia:2024jgr,Caron-Huot:2025tlq} to capture nonlinear $G_{\rm N}$-corrections to the Minkowski-spacetime solution.
This procedure is more easily carried out in Regge--Wheeler gauge. However, the one-point function above was derived using the graviton solution in de Donder gauge. So, we must 
first relate the function $K_\ell^{(B)}(\omega)$ to the integration constants of the homogeneous solution of the Regge--Wheeler equation. This will ensure that the homogeneous solution for $\Psi_\text{RW}$ used below  satisfies the correct boundary condition at the point particle's location, with the worldline delta function sourcing its decaying falloff at infinity.

\noindent
{\it Homogeneous solution and boundary conditions:}
Away from the origin, and neglecting all $G_{\rm N}$-corrections, the $D$-dimensional Regge--Wheeler equation~\eqref{eq:ddimRWeq} in the bulk reads
\begin{equation}
    \left(\frac{\D^2}{\D r^2}-\frac{(\ell-\varepsilon)(\ell-\varepsilon+1)}{r^2} \right)\Psi_\text{RW}(r) = 0\,.
    \label{eq:RW_EWE-flat}
\end{equation}
This is analogous to the  free scalar equation~\eqref{eq:EWE} on Minkowski space, and admits the following standard growing and decaying solutions:
\begin{equation}
    \Psi_\text{RW}^{(h)}(r)=\mu^{-\varepsilon}B_\text{reg}r^{\ell+1-\varepsilon}+\frac{\mu^{\varepsilon} B_\text{irr}}{2\ell+1-2\varepsilon}r^{-\ell+\varepsilon}.
    \label{eq:EFT_RW_homog}
\end{equation}
In order for $\Psi_\text{RW}$ to have the same mass dimension in $D>4$, we rescaled the growing and decaying branches  by $\mu^{-\varepsilon}$ and $\mu^{\varepsilon}$, respectively (cf.~the scalar field case, in particular~\eqref{eq:psis} and~\eqref{eq:phiEFT0}).

Now we recall that the metric perturbations  $h_0$ and $h_1$, defined in~\eqref{eq:hdecomp}, are related to the Regge--Wheeler variable via (see~\cite{Hui:2020xxx} for details)
\begin{equation}
\begin{aligned}
    h_0 &=-q_\ell \E^{-i\omega t}r \mu^{\varepsilon} \left[(D-2)r^{\frac{2-D}{2}}\Psi_\text{RW}+r\partial_r\left(r^{\frac{2-D}{2}}\Psi_\text{RW} \right) \right], \\
    h_1 &= i\omega q_\ell \E^{-i\omega t}\mu^{\varepsilon}r^{3-\frac{D}{2}}\Psi_\text{RW},
\end{aligned}
\label{eq:h0h1eftodd}
\end{equation}
where $q_\ell\equiv1/\sqrt{2(\ell-1)(D-2+\ell)}$. The components~\eqref{eq:h0h1eftodd} can be used to evaluate the Weyl tensor in  Regge--Wheeler gauge. In particular, we find $B_{rAB}$ to be
\begin{equation}
\begin{aligned}
    B_{rAB}^\text{RW} & = \frac{1}{\Mpl} \sum_{\ell=2}^\infty\left[ r^2\partial_r\left( r^{-2}h_0(t,r)\right)-\partial_0h_1(t,r)\right]\nabla_{[A}Y_{B]}^{(T)}{}_{\ell}^{m}
\\
&	= -\frac{1}{\Mpl}\sum_{\ell=2}^\infty\frac{q_\ell \E^{-i\omega t}}{4}\mu^{\varepsilon}r^{1-\frac{D}{2}}\left[ \left(2D-D^2+4r^2\omega^2\right)\Psi_\text{RW}(r)+4r^2\Psi_\text{RW}''(r) \right] \nabla_{[A}Y_{B]}^{(T)}{}_{\ell}^{m} .
\end{aligned}
    \label{eq:Brij_RW}
\end{equation}
Plugging in the homogeneous solution~\eqref{eq:EFT_RW_homog} for $\Psi_\text{RW}$, we get
\begin{equation}
     B_{rAB}^\text{RW}=-\frac{1}{\Mpl}\sum_{\ell=2}^\infty q_\ell \E^{-i\omega t} \left[ (\ell-1) (\ell+2-2 \varepsilon)+r^2 \omega ^2 \right]\left( B_\text{reg}r^\ell +\frac{\mu^{2\varepsilon} B_\text{irr}}{2\ell+1-2\varepsilon}r^{-\ell-1+2\varepsilon}\right) \nabla_{[A}Y_{B]}^{(T)}{}_{\ell}^{m}.
\end{equation}
Since the linearized Weyl tensor is gauge invariant in flat space, we can directly compare $B_{rAB}^\text{RW}$ with~\eqref{Brij_Love}, obtained in de Donder gauge.  By matching the $r^\ell$ and $r^{-\ell-1+2\varepsilon}$ coefficients in the two expressions, we can identify
\begin{equation}
\frac{\ell+1}{2\ell}\frac{(\ell+1)!(D+\ell-2)}{\ell(\ell-1)}\frac{2^{\ell-1}\Gamma\left(\ell+\frac{D-1}{2}\right)}{\pi^{\frac{D-1}{2}}} \frac{K^{(B)}_\ell(\omega)}{\Mpl^2} =  \frac{B_\text{irr}}{B_\text{reg}} .
\label{eq:KBtoBB}
\end{equation}
This formula relates the response $K^{(B)}(\omega)$ to the coefficients of the homogeneous solution to the Regge--Wheeler equation.

\noindent
{\it Particular solution:}
Now that we have obtained the homogeneous solution~\eqref{eq:EFT_RW_homog} for the Regge--Wheeler field, with the constants related to the worldline couplings~\eqref{eq:EFTLNsodd} via~\eqref{eq:KBtoBB}, we can proceed to compute the $G_{\rm N}$-corrections to that solution and determine the renormalized coefficients.
Expanding in powers of $G_{\rm N}$, the Regge--Wheeler equation~\eqref{eq:ddimRWeq} can be cast in the form
\begin{equation}
    \left(\frac{\D^2}{\D r^2}-\frac{(\ell-\varepsilon)(\ell-\varepsilon+1)}{r^2} \right)\Psi_\text{RW}(r) = V_{\Psi_\text{RW}}(r)\Psi_\text{RW}(r),
    \label{eq:RW_EWE}
\end{equation}
where we placed all $G_{\rm N}$ terms on the right-hand side of the equation:
\begin{equation}
\begin{aligned}
    V_{\Psi_\text{RW}} &=\sum_{n=1}^\infty\left( \frac{2G_\text{N}M n_D \mu^{2\varepsilon}}{r^{1-2\varepsilon}} \right)^n \left[ \frac{2\varepsilon-1}{r}\frac{\D }{\D r}+\frac{\ell^2+\ell-3-\varepsilon(2\ell-5)-2\varepsilon^2}{r^2}-(n+1)\omega^2\right]-\omega^2,
    \label{RW_Grav_potential}
\end{aligned}
\end{equation}
with $n_D$ defined in \eqref{eq:frD}.
These terms will be treated perturbatively in the Born series, in analogy with the scalar field example~\cite{Caron-Huot:2025tlq}.

We are interested in the perturbative particular solution to~\eqref{eq:RW_EWE}, sourced by $V_{\Psi_\text{RW}}$, up to the order $O(\omega^2G_\text{N}^2G_\text{N}^{2\ell+1})$. This solution can be obtained from the Born series with up to $2\ell+4$ insertions of the $V_{\Psi_\text{RW}}$ potential:
\be
\begin{aligned}
    \Psi_\text{RW}(r)=\Psi_\text{RW}^{(h)}(r)&+\int^r\D r'G(r,r')V_{\Psi_\text{RW}}(r')\Psi_\text{RW}^{(h)}(r') \\
    &+\int^r\D r' G(r,r')V_{\Psi_\text{RW}}(r')\int^{r'}\D r'' G(r',r'')V_{\Psi_\text{RW}}(r'')\Psi_\text{RW}^{(h)}(r'')+\dots
\end{aligned}
\ee
where $\Psi_\text{RW}^{(h)}$ is the  homogeneous solution~\eqref{eq:EFT_RW_homog}  and $G(r,r')$ is the same Green's function as in~\eqref{eq:Greenfunctionscalar}.
Taking the $\varepsilon\to0$ limit, we obtain (for $\ell=2$):  
\begin{equation}
\begin{aligned}
    \Psi_{\text{RW}}^{\ell=2}(r)=& \, \, r^3 B_\text{reg} \left(1+\frac{468 \bar{G}^2 \omega^2}{1225}-\frac{107 \bar{G}^2 \omega ^2}{210 \varepsilon}-\frac{107 \bar{G}^2 \omega^2}{70}\log(\mu r)\right)\\
   &+\frac{B_\text{reg}}{r^2} \left(\frac{17704 \bar{G}^7 \omega ^2}{315} +\frac{32 \bar{G}^7 \omega ^2}{15 \varepsilon}+\frac{416 \bar{G}^7 \omega^2}{15}\log(\mu r) \right)\\
   &+\frac{B_\text{irr}}{r^2} \left(\frac15+\frac{26998\bar{G}^2 \omega^2}{55125} +\frac{107 \bar{G}^2 \omega^2}{1050 \varepsilon }+\frac{107 \bar{G}^2 \omega^2}{210}\log(\mu r)\right),
\end{aligned}
\label{eq:bareRWsoln}
\end{equation}
where we have introduced $\bar G$, as in~\eqref{eq:barGdef}.

\vspace{-10pt}
\paragraph{Renormalization and matching}~\\
We now renormalize and match~\eqref{eq:bareRWsoln} to the full solution of the Regge--Wheeler equation.
To subtract the infinities, we introduce the renormalized coefficients $\Bar{B}_\text{reg}$ and $\Bar{B}_\text{irr}$ as follows:
\begin{equation}
\begin{aligned}
    B_\text{reg}&=\Bar{B}_\text{reg}(1+\omega^2\delta_{11})+\Bar{B}_\text{irr}\omega^2\delta_{12}, \\
    B_\text{irr}&=\Bar{B}_\text{irr}(1+\omega^2\delta_{22})+\Bar{B}_\text{reg}\omega^2\delta_{21},
\end{aligned}
\label{eq:rencoeffsRW}
\end{equation}
where we have defined the coefficients
\begin{equation}
    \delta^{\ell=2}=
    \begin{pmatrix}
       \frac{107 \bar{G}^2}{210\varepsilon}  & 0 \\
       -\frac{32 \bar{G}^7}{3\varepsilon} &-\frac{107 \bar{G}^2}{210\varepsilon}
    \end{pmatrix}.
\end{equation}
Plugging~\eqref{eq:rencoeffsRW} back into the expression for $\Psi_\text{RW}$ before taking the $\varepsilon \to 0$ limit yields the following renormalized solution:
\begin{equation}
\begin{aligned}
    \Psi_\text{RW}^{\text{R},\ell=2}=&~\Bar{B}_\text{reg}\left[r^3+r^3\bar{G}^2\omega^2\left(\frac{468}{1225}-\frac{214}{105}\log(\mu r)\right)+\frac{\bar{G}^7\omega^2}{r^2}\left(\frac{79472}{1575}+\frac{128}{5}\log(\mu r)\right)\right]\\
    &+\Bar{B}_\text{irr}\left[\frac{1}{5r^2}+\frac{\bar{G}^2\omega^2}{r^2}\left(\frac{24751}{55125}+\frac{214}{525}\log(\mu r)\right)\right].
\end{aligned}
\end{equation}
Comparing $\Psi_\text{RW}^{\text{R},\ell=2}$ with~\eqref{eq:psiRWl2}, we match the renormalized  constants  $\Bar{B}_\text{reg}$ and $\Bar{B}_\text{irr}$ as functions  of the frequency $\omega$, the Schwarzschild radius $r_s$, the renormalization scale $\mu$, and the amplitude $B$ of the UV solution~\eqref{eq:psiRWNZ}:
\begin{equation}
\begin{aligned}
    \Bar{B}_\text{reg}^{\ell=2} =&\, B\left[ 1+i\omega r_s\frac{13}{12}-\omega^2r_s^2\left(\frac{121991}{58800}+\frac{\pi^2}{6}-\frac{107}{210}\log(\mu r_s) \right) \right],\\
    \Bar{B}_\text{irr}^{\ell=2} =& \, B\left[i\omega r_s^6-\omega^2r_s^7 \left( \frac{1943}{2520}+\log (\mu r_s)\right)\right]. 
\end{aligned}
\end{equation}
(Note that we rescaled $B\to B r_s^2$ so that $B_\text{reg}$ is a polynomial in $r_s$.)
Using~\eqref{eq:KBtoBB}, we  finally obtain 
\begin{tcolorbox}[colframe=white,arc=0pt,colback=greyish2]
\begin{equation}
    \frac{3}{2}\frac{45}{ 2\pi\Mpl^2}K^{(B)}_{\ell=2}(\omega)=i\omega r_s^6+\omega^2r_s^7\left(\frac{787}{2520}-\log (\mu r_s)\right).
    \label{eq:magl0}
\end{equation}
\end{tcolorbox}
\noindent
Repeating the same procedure for $\ell=3$ and $\ell=4$ (see Appendix~\ref{app:l=3-4_solution} for details), we get:
\begin{tcolorbox}[colframe=white,arc=0pt,colback=greyish2]
\begin{equation}
\begin{aligned}
    \frac{4}{3}\frac{18900}{2\pi\Mpl^2}K^{(B)}_{\ell=3}(\omega) &= i \omega r_s^8+\omega^2 r_s^9\left( \frac{1\,727}{2\,520}-\log  (\mu r_s)\right), \\
   \frac{5}{4} \frac{11113200}{2\pi\Mpl^2}K^{(B)}_{\ell=4}(\omega) &=i \omega r_s^{10}+\omega^2 r_s^{11}\left( \frac{237\,529}{277\,200} -\log (\mu r_s)\right).
\end{aligned}
\label{eq:magl12}
\end{equation}
\end{tcolorbox}
\noindent
We see that, as expected, the magnetic dynamical Love numbers are nonzero. It is also worth noting that the relative coefficient between the dissipative response coefficient and the running of the dynamical Love numbers is $-1$.

\subsubsection{Even sector}

We now turn to the parity-even sector of perturbations. Following the same procedure as for the odd case, we first solve for the expectation value of composite operator $Q_{E}(\tau)$, which appears in point-particle EFT~\eqref{eq:ppEFTgrav}, in a gravito-electric background using linear response theory:
\begin{equation}
    \langle Q_{E,I}^{A_\ell}(\tau)\rangle=\int \D \tau'K_{IJ}^{(E)A_\ell \vert A_{\ell'} }(\tau-\tau'){E}^J_{A_{\ell'} }(\tau'),
\end{equation}
where $I,J=\{+,-\}$ again label (polar) gravitational fields~\eqref{eq:keldyshbasispm} in the Keldysh basis, and $K_{IJ}^{(E)\,A_\ell\vert A_{\ell'}}$ is related to the two-point function of $Q_E$ by
\begin{equation}
    \langle Q_{E,I}^{A_\ell}(\tau)Q_{E,J}^{A_{\ell'}}(\tau')\rangle=-iK_{IJ}^{(E)\,A_\ell\vert A_{\ell'}}(\tau-\tau')\,,
\end{equation}
as in the previous section~\cite{Goldberger:2020fot,Saketh:2022xjb,Saketh:2023bul,Glazer:2024eyi}. This yields the following in-in effective action:\footnote{Here we are again using the spherical symmetry of the black hole.}
\begin{equation}
    \Gamma^\text{in-in}_\text{int}[h_\pm]=\int \D\tau_1 \D\tau_2\sum_{\ell=2}^\infty K_{{IJ},\ell}^{(E)}(\tau_2-\tau_1) E^{I}_{A_\ell}(\tau_2)E^{{J}\,A_\ell}(\tau_1) .
    \label{eq:ininactionE}
\end{equation}

\vspace{-10pt}
\paragraph{Electric one-point function}~\\
Using $\Gamma^\text{in-in}_\text{int}$, we can  evaluate the  one-point function of the electric component of the  Weyl tensor in the presence of the external tidal field $\bar{E}_+$: 
\begin{equation}
    \langle E_{+\,ab}(t,\vec{x})\rangle_\text{in-in}=i\sum_{\ell=2}^\infty\int \D\tau_1 \D\tau_2\, K_{\ell}^{(E)}(\tau_2-\tau_1)\langle E_{+\,ab}(t,\vec x) E_{-\,A_\ell}\left(\tau_2\right) \rangle\bar{E}_+^{A_\ell }\left(\tau_1\right)\,,
    \label{eq:1-pt_fct_electric}
\end{equation}
with $K_{\ell}^{(E)}\equiv K_{+-,\ell}^{(E)}$ and
\begin{equation}
    K_{+-}^{(E)}(\tau_2-\tau_1)=-G_R^{(Q_E)}(\tau_2-\tau_1) \equiv i \langle \left[ Q_{E\,+}(\tau_2),Q_{E\,-}(\tau_1) \right]\rangle \theta(\tau_2-\tau_1).
\end{equation}
As in the magnetic sector, we can expand the response kernel in Fourier space similarly to~\eqref{eq:EFTLNsodd},  
\begin{equation}
\begin{aligned}
    K_{{+-},\ell}^{(E)}(\tau_2-\tau_1) &= \int \frac{\D {\omega}}{2\pi}\E^{-i{\omega}(\tau_2-\tau_1)} K_{\ell}^{(E)}(\omega) \\
    &= \frac{1}{\ell!}\int \frac{\D{\omega}}{2\pi}\E^{-i\omega(\tau_2-\tau_1)}\left[ \lambda_{0,\ell}^E+i{\omega}r_s\lambda_{1,\ell}^E+ ({\omega} r_s)^2\lambda_{2,\ell}^E+\cdots\right],
\end{aligned}
\label{eq:EFTLNseven}
\end{equation}
where $\lambda_{0,\ell}^E$ are the static Love numbers, $\lambda_{1,\ell}^E$ capture the linear-in-frequency dissipative response of the object, and $\lambda_{2,\ell}^E$ are the dynamical Love numbers at order $O(\omega^2)$. Since we are interested in computing  $\lambda_{2,\ell}^E$ for Schwarzschild black holes in four-dimensional general relativity, we will set $\lambda_{0,\ell}^E=0$ in the following. This will allow us to consistently neglect  $O(\omega^2)$ terms in the expressions of $\bar{E}_{+}$ and $\langle E_{+\,ab}(t,\vec x) E_{- \,A_\ell}\left(\tau_2\right) \rangle$ in \eqref{eq:1-pt_fct_electric}.

From the definition of the Weyl tensor,  we have on shell:
\begin{equation}
\begin{aligned}
    E_{ab} &= C_{0a0b}  = \frac{2}{\Mpl} \left[ \partial_a\partial_{[0}h_{b]0}-\partial_0\partial_{[0}h_{b]a} \right] = \frac{1}{\Mpl}\left[ -\partial_a\partial_b h_{00}+2\partial_0\partial_{(a}h_{b)0}+O(\omega^2) \right], \\
     E_{A_\ell} &= \frac{1}{\Mpl} \left[-\partial_{A_\ell}h_{00}+2\partial_0\partial_{( A_{\ell-1}}h_{i_\ell)_T0}+O(\omega^2)\right].
    \end{aligned}
\end{equation}
As in the odd sector, we fix de Donder gauge and  expand $\langle E^+_{ab}(t,\vec x) E^-_{A_\ell}\left(\tau_2\right) \rangle\bar{E}^{+\,A_\ell }\left(\tau_1\right)$ in~\eqref{eq:1-pt_fct_electric} up to order $O(\omega)$. From the gauge constraint $\partial^\mu(h_{\mu\nu}-\frac{1}{2}\eta_{\mu\nu}h)=0$ and equations of motion  $\square h_{\mu\nu}=0$, it follows that  subtracting  traces in the definition of $E^-_{A_\ell}$ and $\bar{E}^{+\,A_\ell }$  corresponds to terms of order $O(\omega^2)$ in the product~\eqref{eq:1-pt_fct_electric}. Therefore, in what follows, we can simply replace $( {\cdots} )_T$ by $({\cdots})$.
Then, from the contraction of the Weyl  $E$-operators we find
\begin{equation}
     E_{A_\ell} \bar{E}^{A_\ell} = \frac{1}{\Mpl^2} \left[\partial_{A_\ell}h_{00}\partial^{A_\ell}\bar h^{00}-2\partial_{A_\ell}h_{00}\partial^0\partial^{A_{\ell-1}}\bar{h}^{i_\ell0}-2\partial_0\partial_{A_{\ell-1}}h_{i_\ell0}\partial^{A_\ell}\bar{h}^{00} +O(\omega^2) \right] .
     \label{Electric_square}
\end{equation}
Putting this together, we find the contraction
\be
   \langle E_{+\,ab}(t,\vec x) E_{-\,A_\ell}\left(\tau_2\right) \rangle\bar{E}_+^{A_\ell }\left(\tau_1\right) 
   = -\frac{1}{\Mpl^3}\langle  \partial_a\partial_bh_{+\,00} \partial_{A_\ell}h_{-\,00}\rangle\left(\partial^{A_\ell}\bar{h}_+^{00}-2\partial^0\partial^{A_{\ell-1}}\bar{h}_+^{i_\ell0} \right)+O(\omega^2),
\label{eq:Simp_corr_electric}
\ee
where we dropped terms that are either of order $O(\omega^2)$ or proportional to $\mathcal{P}^{\rm dD}_{0a00}=0$, with $\mathcal{P}_{\mu\nu\rho\sigma}^\text{dD}$ defined as in~\eqref{graviton_propagator_tensor}. We stress that $h_{\mu\nu}$ and $\bar{h}_{\mu\nu}$ are the response and tidal fields, respectively, in de Donder gauge. It is useful to note that the combination in parenthesis is also\footnote{To see this, it suffices to write the transformation that relates the fields in the two gauges: $h_{\mu\nu}^\text{dD} = h_{\mu\nu}^\text{RW} + \partial_\mu \xi_\nu + \partial_\nu \xi_\mu$, with $\xi$ the gauge parameter (for simplicity, we omit here all Keldysh indices as well as the bar symbol denoting the tidal field).  Taking the linear combination~\eqref{eq:dDtoRWeven}, one obtains $\partial_{i_\ell}{h}_{00}^\text{dD}-2\partial_0{h}_{0i_\ell}^\text{dD} = \partial_{i_\ell}{h}_{00}^\text{RW}-2\partial_0{h}_{0i_\ell}^\text{RW}-2\partial_0^2\xi_{i_\ell} $. Using that ${h}_{0i_\ell}^\text{RW}=O(\omega)$ in the even sector of perturbations (see, e.g., \cite{Riva:2023rcm}),~\eqref{eq:dDtoRWeven}  then follows.}
\begin{equation}
\partial_{A_\ell}\bar{h}_{00}-2\partial_0\partial_{A_{\ell-1}}\bar{h}_{i_\ell0} = \partial_{A_\ell}\bar{h}_{00}^{\text{RW}} + O(\omega^2),
\label{eq:dDtoRWeven}
\end{equation}
where $\bar{h}_{00}^{\text{RW}}$ is the $tt$-component of the tidal field metric perturbation in Regge--Wheeler gauge. We can express this as~\cite{Poisson_Will_2014}
\begin{equation}
\bar{h}_{00}^{\text{RW}} (t,\vec{x})= \Mpl \E^{-i\omega t}\sum_\ell c_{\rm ext} Y_\ell^m r^\ell= \Mpl  \E^{-i\omega t}\sum_\ell c_{j_1\cdots j_\ell}x^{j_1}\cdots x^{j_\ell} + O(\omega^2) ,
\end{equation}
where $c^{i_1\cdots i_\ell}$ is a traceless symmetric tensor.
Plugging this into~\eqref{eq:1-pt_fct_electric} for the one-point function and using the expression~\eqref{graviton_propagator} for the graviton propagator, we obtain  
\be
    \langle E_{+\,ab}(t,\vec x)\rangle_\text{in-in} 
    = -\frac{\E^{-i\omega t}}{\Mpl^2}\frac{D-3}{D-2}\sum_{\ell=2}^\infty \ell! (-i)^\ell K_\ell^{(E)}(\omega) c^{i_1\cdots i_\ell}  \partial_a\partial_b\int \frac{\D^{D-1}\vec{p}}{(2\pi)^{D-1}}\E^{i\vec{p}\cdot\vec{x}}\frac{p_{i_1}\cdots p_{i_\ell}}{\vec{p}^2}.
\ee
Going back to position space, this is
\be
    \langle E_{+\,ab}(t,\vec x)\rangle_\text{in-in} 
    = -\frac{\E^{-i\omega t}}{\Mpl^2}\sum_{\ell=2}^\infty K_\ell^{(E)}(\omega)\frac{2^{\ell-2}\ell! \, \Gamma\left(\ell+\frac{D-3}{2}\right)}{\pi^{\frac{D-1}{2}}}\frac{D-3}{D-2} c_{i_1\cdots i_\ell}\partial_a\partial_b  \frac{x^{i_1}\cdots x^{i_\ell}}{\vert \vec{x}\vert^{2\ell+D-3}} .
\label{eq:evenEplus}
\ee
We can similarly write the background expression of the  Weyl tensor. Focusing on the $rr$-component,
\begin{equation}
\begin{aligned}
    \bar{E}_{rr} & =\frac{1}{\Mpl}\left(-\partial_r^2\bar{h}_{00}+2\partial_0 \partial_r   \bar{h}_{0r}\right)+O(\omega^2)
    \\
& = -\frac{1}{\Mpl}\partial_r^2 \bar{h}_{00}^{\text{RW}}  +O(\omega^2)
    = -\E^{-i\omega \tau}\sum_{\ell=2}^\infty c_{\rm ext} \,  \ell(\ell-1) r^{\ell-2} Y_\ell^m  +O(\omega^2).
\end{aligned}
    \label{tidal_Weyl_even}
\end{equation}
 Using $c_{j_1\cdots j_\ell}x^{j_1}\cdots x^{j_\ell} = c_{\rm ext} r^\ell Y_\ell^m $ with $r\equiv \vert \vec x\vert$, and   $\nabla_r\nabla_r=\partial_r\partial_r$ in~\eqref{eq:evenEplus}, and adding it to~\eqref{tidal_Weyl_even}, we get 
 \be
\begin{aligned}
    \bar{E}_{rr}+\langle E_{+\,rr}\rangle_\text{in-in}&=-\E^{-i\omega t}\sum_{\ell=2}^\infty c_{\rm ext} \ell(\ell-1)Y_{\ell}^{m}\\
 &\!\!\!  \times\left( r^{\ell-2} + \mu^{2\varepsilon}\frac{K^{(E)}_\ell(\omega)}{\Mpl^2}\frac{D-3}{D-2}\frac{(\ell+D-3)(\ell+D-2)}{\ell(\ell-1)}\frac{2^{\ell-2}\ell! \, \Gamma\left(\ell+\frac{D-3}{2}\right)}{\pi^{\frac{D-1}{2}}}r^{-\ell-D+1} \right),
    \label{EFT_BC_solution}
\end{aligned}
\ee
where we replaced $K^{(E)}_\ell$ by $\mu^{2\varepsilon}K^{(E)}_\ell$. As before, we need to include gravity, renormalize, and match.

\vspace{-10pt}
\paragraph{Including gravity}~\\
We employ the same methodology to include the nonlinearities of gravity. As before it is convenient to solve the gravitational equations away from the particle in Zerilli variables, and match the (gauge-invariant) Weyl tensor to relate the parameters of the homogeneous solution to the worldline response coefficients.

\noindent
{\it Homogeneous solution and boundary conditions:}
At zeroth order in the frequency and to leading order in flat-space limit, the Zerilli equation~\eqref{eq:ddimZeq} in the bulk simply reduces to
\begin{equation}
    \left(\frac{\D^2}{\D r^2}-\frac{(\ell-\varepsilon)(\ell-\varepsilon+1)}{r^2} \right)\Psi_\text{Z}(r) = 0.
    \label{eq:Z_EWE-flat}
\end{equation}
As in the previous cases, this admits the  standard growing and decaying solutions
\begin{equation}
    \Psi_\text{Z}^{(h)}(r)=\mu^{-\varepsilon}B_\text{reg}r^{\ell+1-\varepsilon}+\frac{\mu^{\varepsilon} B_\text{irr}}{2\ell+1-2\varepsilon}r^{-\ell+\varepsilon},
    \label{eq:EFT_Z_homog}
\end{equation}
where the $\mu$ factors ensure that the field maintains the correct dimension away from $D=4$.
Let us express the  Weyl tensor in terms of the even metric perturbations \eqref{eq:hdecomp}. In particular, to leading order in the flat-space limit, 
\begin{equation}
    E_{rr}=-\frac{1}{\Mpl}\left(\partial_r^2 H_0 +2i\omega\partial_r  H_1\right).
 \label{eq:ErrevenZ}
\end{equation}
Using the constraint equations (see eq.~(3.54) of~\cite{Hui:2020xxx}),  
\begin{align}
H_0 &=-\frac{\mu^\varepsilon r^{\frac{2-D}{2}}}{2[(D-2)(D-3)\ell(\ell-1)(\ell+D-3)(\ell+D-2)]^{1/2}}\bigg[2(D-3)(D-2)r\partial_r\Psi_\text{Z}  \nonumber \\
       &\qquad\qquad\qquad\quad + (D-3) \left(\left(D^2+2 D (\ell-3)+2\ell (\ell-3) +8\right)-2(D-2)r^2\omega^2\right)\Psi_\text{Z} \bigg] ,
        \\
    H_1 &= \frac{i\omega \mu^\varepsilon r^{\frac{4-D}{2}}}{4[(D-2)(D-3)\ell(\ell-1)(\ell+D-3)(\ell+D-2)]^{1/2}}\bigg[2(D-2)r\partial_r\Psi_\text{Z} \nonumber  \\
       &\qquad\qquad\qquad\qquad\qquad\qquad\qquad\qquad\qquad\qquad
        + \left(D^2-2\ell(\ell-3)-2D(\ell+1)\right)\Psi_\text{Z}\bigg] , 
\end{align}
we can express~\eqref{eq:ErrevenZ} in terms of $\Psi_{\text{Z}}$. Comparing the resulting Weyl tensor of the homogeneous solution~\eqref{eq:EFT_Z_homog} with~\eqref{EFT_BC_solution}, we can extract $K^{(E)}_\ell$ as a function of the ratio $B_\text{irr}/B_\text{reg}$:
\begin{equation}
\frac{D-3}{D-2}\frac{(\ell+D-3)(\ell+D-2)}{\ell(\ell-1)}\frac{2^{\ell-1}\ell! \, \Gamma\left(\ell+\frac{D-1}{2}\right)}{\pi^{\frac{D-1}{2}}} \frac{K^{(E)}_\ell(\omega)}{\Mpl^2} =   \frac{B_\text{irr}}{B_\text{reg}}.
\label{eq:KEBs}
\end{equation}
This relates the worldline response to the parameters of the Zerilli solution.

 \noindent
{\it Particular solution:}
Given the homogeneous solution \eqref{eq:EFT_Z_homog}, in which the coefficients $B_\text{irr}$ and $B_\text{reg}$ are related to the effective coupling $K^{(E)}_\ell(\omega)$ via~\eqref{eq:KEBs}, we can now solve for the particular solutions of the perturbative Zerilli equation
\begin{equation}
    \left(\frac{\D^2}{\D r^2}-\frac{(\ell-\varepsilon)(\ell-\varepsilon+1)}{r^2} \right)\Psi_{\text{Z}}(r) = V_{\Psi_\text{Z}}(r)\Psi_\text{Z}(r),
    \label{eq:Z_EWE}
\end{equation}
where the right-hand side contains all corrections in $G_{\rm N}$ and in the frequency:
\begin{equation}
    V_{\Psi_\text{Z}}=\sum_{n=1}^\infty\left( \frac{2G_\text{N} M n_D \mu^{2\varepsilon}}{r^{1-2\varepsilon}} \right)^n \left[ \frac{2\varepsilon-1}{r}\frac{\D}{\D r}+\frac{V^\text{Z}_{\ell,n}}{r^2}-(n+1)\omega^2\right]-\omega^2,
    \label{Z_Grav_potential}
\end{equation}
where we introduced the (fairly complicated) potential
\begin{align}
    V^\text{Z}_{\ell,n} & \equiv \frac{1}{(\ell-1)^2 (\ell-2 \varepsilon +2)^2}\bigg[\ell (1-2 \varepsilon )^2 \left(\varepsilon ^2-1\right) (\ell-2 \varepsilon +1) \zeta (n-2) (\delta _{1,n}-1) \nonumber \\
    &-(\varepsilon -1)^2 \left(4 \varepsilon ^4-16 \varepsilon ^3+5 \varepsilon ^2+11 \varepsilon -6\right) s(n-2) (\delta _{1,n}-1) \nonumber \\
    &+(\varepsilon -3) (\varepsilon -1)^2 (2 \varepsilon -1)^3 \zeta (n-3) (\delta _{1,n}-1) (\delta _{2,n}-1) \nonumber \\
    &+\ell^3 (\ell-2 \varepsilon +1)^3 \zeta(n)-\ell^2 \left(2 \varepsilon ^2-3 \varepsilon +1\right) (\ell-2 \varepsilon +1)^2 \zeta(n-1) \nonumber \\
    &-2 (\varepsilon -1)^2 \left(\ell^2 \left(6 \varepsilon ^2-5 \varepsilon +4\right)+\ell \left(-12 \varepsilon ^3+16 \varepsilon ^2-13 \varepsilon +4\right)+12 \varepsilon ^3-26 \varepsilon ^2+20 \varepsilon -6\right) s(n-1) \nonumber \\
    &+(\varepsilon -1) \Big[\ell^4 (\varepsilon +4)-2 \ell^3 \left(2 \varepsilon ^2+7 \varepsilon -4\right)+\ell^2 \varepsilon  \left(4 \varepsilon ^2+16 \varepsilon -15\right) \nonumber \\
    &+\ell \left(-8 \varepsilon ^3+4 \varepsilon ^2+8 \varepsilon -4\right)+4 (\varepsilon -1)^2 \varepsilon \Big] s(n) \bigg] ,
\end{align}
which involves the functions
\begin{equation}
    s(n) \equiv (n+1)\left(\frac{(1-\varepsilon) (2 \varepsilon -3)}{(\ell-1) (\ell-2 \varepsilon +2)}\right)^n , 
    \qquad  \zeta(n) = \sum_{k=0}^n s(k).
\end{equation}
In order to display explicit expressions
we  now set $\ell=2$. By solving for the particular solution using the Born series and taking the $\varepsilon\to0$ limit, we obtain the following expression for $\Psi_\text{Z}^{\ell=2}$:
\begin{equation}
\begin{aligned}
    \Psi_\text{Z}^{\ell=2}(r) = \, \,  & r^3B_\text{reg}\left[1+\omega^2 \bar{G}^2\left(-\frac{107}{210\varepsilon} -\frac{2731}{9800} -\frac{107}{210}\log(\mu r)\right)\right] \\
    &+ \frac{B_\text{reg}}{r^2}\left[ -\frac{189 \bar{G}^2}{32}+\omega^2\bar{G}^7\left(\frac{4937}{960\varepsilon}+  \frac{2802329}{403200} +\frac{64181}{960}\log(\mu r)\right)\right] \\
    &+ \frac{B_\text{irr}}{r^2}\left[1+\omega^2\bar{G}^2\left(\frac{107}{1050\varepsilon}+ \frac{129359}{441000} +\frac{107}{210}\log(\mu r)\right) \right].
\end{aligned}
\end{equation}
This bare solution has divergences that must be renormalized so that we can match to the relativity calculation.

\vspace{-10pt}
\paragraph{Renormalization and matching}~\\
To subtract off the $1/\varepsilon$ divergences, we introduce the renormalized coefficients $\bar{B}_\text{reg},\bar{B}_\text{irr}$ as defined in~\eqref{eq:rencoeffsRW},  but where now 
\begin{equation}
    \delta^{\ell=2}=
    \begin{pmatrix}
       \frac{107 \bar{G}^2}{210\varepsilon}  & 0 \\
       -\frac{32 \bar{G}^7}{3\varepsilon} &-\frac{107 \bar{G}^2}{210\varepsilon}
    \end{pmatrix}.
\end{equation}
Plugging this back into the expression of  $\Psi_\text{Z}^{\ell=2}$ before taking the $\varepsilon\to0$ limit, we obtain the renormalized Zerilli solution
\begin{align}
    \Psi_\text{Z}^{\text{R},\ell=2} =\bar{B}_\text{reg} & \Bigg[r^3-\bar{G}^2\omega^2 r^3\left(\frac{2731}{9800}+\frac{214}{105} \log(\mu r)\right)
    +\frac{\bar{G}^7 \omega ^2 }{r^2}\left(\frac{1545209}{57600}+\frac{3011}{80} \log(\mu r)\right)
    -\frac{189 \bar{G}^5}{32 r^2}  \Bigg]  \nonumber
    \\
   & + \bar{B}_\text{irr}\left[\frac{1}{5r^2}+\frac{\bar{G}^2 \omega ^2 }{r^2}\left(\frac{111383}{441000}+\frac{214}{525} \log(\mu r)\right)\right].
\end{align}
By 
matching this to the full solution~\eqref{eq:Z_UV_sol}, we find
\begin{equation}
\begin{aligned}
    \bar{B}_\text{reg} &= C+\frac43 iC\omega r_s+C\omega^2r_s^2\left(-\frac{9101}{3675}-\frac{\pi^2}{6}+\frac{107}{210}\log(\mu r_s)\right) \\
    \bar{B}_\text{irr} &= iC\omega r_s^6 - C\omega^2 r_s^7\left(\frac{883}{1260}+\log(\mu r_s)\right),
\end{aligned}
\end{equation}
where we rescaled $C\to C r_s^3$, similar to the odd sector.
Combining this with~\eqref{eq:KEBs}, 
this implies the following value for the $\ell=2$ electric type Love number
\begin{tcolorbox}[colframe=white,arc=0pt,colback=greyish2]
\begin{equation}
    \frac{45}{2\pi \Mpl^2}K^{(E)}_{\ell=2}(\omega) = i\omega r_s^6-\omega^2r_s^7\left(-\frac{797}{1260}+\log(\mu r_s)\right).
    \label{eq:elecl0}
\end{equation}
\end{tcolorbox}
\noindent
Note that the linear term in the frequency and the coefficient of the logarithm agree with~\cite{Chakrabarti:2013lua} (see also~\cite{Goldberger:2005cd}) after accounting for the different conventions.\footnote{In particular, in the notation of~\cite{Chakrabarti:2013lua},  $\tilde F(\omega)= 4 K^{(E)}_{\ell=2}(\omega)$, where the factor of $4$ follows from canonical normalization of the graviton field.}
Repeating the same procedure for $\ell=3$ and $\ell=4$ (see Appendix~\ref{app:l=3-4_solution} for details)  we get: 
\begin{tcolorbox}[colframe=white,arc=0pt,colback=greyish2]
\begin{equation}
\begin{aligned}
    \frac{18900}{2\pi\Mpl^2}K^{(E)}_{\ell=3}(\omega) &= i \omega r_s^8-\omega^2 r_s^9\left( -\frac{709}{840}+\log  (\mu r_s)\right), \\
    \frac{11113200}{2\pi{\Mpl^2}}K^{(E)}_{\ell=4}(\omega) &= i \omega r_s^{10}-\omega^2 r_s^{11}\left( -\frac{5\,501}{5\,775} +\log (\mu r_s)\right).
\end{aligned}
\label{eq:elecl12}
\end{equation}
\end{tcolorbox}
\noindent
We see that the dynamical Love numbers are nonzero, and their logarithmic running is given by the dissipative response coefficients.

\subsection{Symmetries}

Aside from the fact that they are nonzero, the dynamical Love numbers have some properties worth remarking on. Most notably, if we expand the response kernels $K$ as in~\eqref{eq:EFTLNsodd} and~\eqref{eq:EFTLNseven} and then use~\eqref{eq:magl0} and~\eqref{eq:elecl0} along with~\eqref{eq:magl12} and~\eqref{eq:elecl12} to determine the effective worldline couplings,
we see that $\lambda_{1,\ell}^E = \lambda_{1,\ell}^B$ and the coefficients of the logs in $\lambda_{2,\ell}^E$ are exactly the same as those in $\lambda_{2,\ell}^B$.
That is, the dynamical Love numbers in the even and odd sectors are equal. This can be understood as a consequence of Chandrasekhar's duality.




As we mentioned in Section~\ref{sec:GRsolseven}, Chandrasekhar discovered a mapping between the Regge--Wheeler and Zerilli equations~\cite{10.2307/78902,1975RSPSA.343..289C,Chandrasekhar:1985kt}. There we used this fact as a solution-generating technique to obtain solutions in the Zerilli sector from those in the Regge--Wheeler sector. Here we note that this relation can also be read as a symmetry~\cite{Hui:2020xxx,Solomon:2023ltn}. Specifically, both the Regge--Wheeler potential and Zerilli potential in $D=4$ can be written as
\be
V_{\rm RW} = {\cal W}^2 + f(r)\frac{{\rm d} {\cal W}}{{\rm d} r}+\beta, \qquad\qquad V_{\rm Z} = {\cal W}^2 - f(r)\frac{{\rm d} {\cal W}}{{\rm d} r}+\beta,
\ee
with ${\cal W}$ as in~\eqref{eq:superpotential} and $\beta \equiv -4\lambda^2(\lambda+1)^2/(9r_s^2)$. This implies that the two potentials are supersymmetric partners~\cite{Cooper:1994eh}. One can further check that the following transformation is a symmetry of the action written solely in terms of the physical Regge--Wheeler and Zerilli fields~\cite{Hui:2020xxx}:
\be
\label{eq:psisym}
\delta\Psi_\mathrm{Z} = \left(\frac{\partial}{\partial r_\star}-{\cal W}(r)\right)\Psi_\mathrm{RW}, \qquad\qquad 
\delta\Psi_\mathrm{RW} = \left(\frac{\partial}{\partial r_\star}+{\cal W}(r)\right)\Psi_\mathrm{Z}.
\ee
This is an off-shell symmetry with a corresponding Noether current. At large distances it interchanges the electric and magnetic parts of the Weyl tensor, and so is a Schwarzschild generalization of electric-magnetic duality.\footnote{Note that this transformation keeps the background fixed, it does not map Schwarzschild into Taub--NUT. It is therefore directly a symmetry of linearized perturbations around Schwarzschild.}
In the present context, this symmetry maps the worldline response operators~\eqref{eq:ppEFTgrav} in the electric and magnetic sectors into each other, relating their coefficients.  
More precisely, it acts on $E_{ab}$ and $B_{abc}$ as $\delta E_{ab}=-\frac{1}{2}{\epsilon_{0a}}^{cd}B_{bcd}$ and $\delta B_{abc} = {\epsilon^{0d}}_{bc} E_{ad}$~\cite{Hui:2020xxx,Solomon:2023ltn}. 
Applying this transformation to~\eqref{Brij_Love} and \eqref{EFT_BC_solution} yields the relation $K^{(B)}_\ell(\omega)=\frac{\ell}{\ell+1}K^{(E)}_\ell(\omega)$. Using the definitions \eqref{eq:EFTLNsodd} and \eqref{eq:EFTLNseven}, this  in turn implies an equality  among the $\lambda_\ell$ coefficients.
The astute reader will note however that, while the dissipative coefficients $\lambda_{1,\ell}^E$ and $\lambda_{1,\ell}^B$, and the coefficients of the logs in $\lambda_{2,\ell}^E$ and $\lambda_{2,\ell}^B$ are indeed found to coincide, the finite contributions in e.g.,~\eqref{eq:magl0} and~\eqref{eq:elecl0} {\it differ}. This would seem to be in conflict with the symmetry. However, we computed these quantities via dimensional regularization, and the symmetry~\eqref{eq:psisym} exists only in $D=4$. Since we have chosen a regularization scheme that breaks the symmetry, it is not surprising that the scheme-dependent parts of the final answers do not respect it. The existence of the symmetry implies the existence of a scheme that preserves the symmetry. Indeed, an important consistency check is that the scheme-independent parts of the response---the imaginary dissipative response and the coefficient of the logarithms---match exactly between the two sectors. In addition, we can certainly pick the scales $\mu$ in the two sectors differently so that the finite pieces agree, which would be a scheme that preserves the duality.

﻿
﻿\newpage
\section{Conclusions}
\label{sec:Concl}

In this paper, we have computed the dynamical tidal Love numbers of Schwarzschild black holes in both the even (gravito-electric) and odd (gravito-magnetic) sectors.
In order to do this unambiguously, we have employed an approach that synthesizes computations in full general relativity with those done in an effective description of a black hole as a point particle. This latter description allows us to give the tidal responses a gauge-invariant definition in terms of couplings between the worldline and external gravitational fields in the Schwinger--Keldysh effective action.

In order to enable the systematics of the calculation, we have solved the equations of black hole perturbation theory in a small frequency expansion. This expansion is amenable to matching with the EFT description in an intermediate zone, bypassing many of the complexities of the problem associated with the far zone. We dealt with the nonlinearities of gravitational interactions in the EFT by utilizing the Born series approach of~\cite{Correia:2024jgr,Caron-Huot:2025tlq}, and we also find that the field profiles in the EFT must be regularized and renormalized before they can be matched to the general relativity solution.

After matching to define the dynamical Love numbers of black holes, we find they have several interesting properties. 
They are nonzero, and have a characteristic running, where the coefficient of the logarithm in all cases is $-\omega r_s$ times the corresponding dissipative response coefficient. One way to read this relation is that the beta function for the dynamical Love numbers is determined by the dissipative response.\footnote{{This is rather surprising because the long-distance running of the conservative response is sourced by the Schwarzschild potential and hence is {\it universal}, while the dissipative response is a finite-size effect particular to black holes. This relation appears to become more complicated at subleading order~\cite{Caron-Huot:2025tlq}.}} Another reading is that the discontinuity of the conservative part of $K$ is $-\beta\omega/2$ times the dissipative part of $K$, where $\beta = 4\pi r_s$ is the inverse black hole temperature.\footnote{This looks suspiciously like a fluctuation--dissipation relation. However, it is a relation involving just the causal Green's function.} (Interestingly this relation continues to hold for the most transcendental parts of the response kernel at subleading orders in frequency, at least in the examples of~\cite{Caron-Huot:2025tlq}.) 
It would be nice to understand these features of the logarithmic terms better. 
Another interesting feature is that, 
as expected from the symmetries of black hole perturbations, there exists a scheme in which the electric and magnetic dynamical Love numbers (along with the dissipative responses) are equal.

There are a number of future directions suggested by this study that it would be interesting to pursue. Perhaps the most obvious would be to extend the computation done here to general $\ell$ and (relatedly) to higher dimension. In principle there are only technical complications to doing so, and it would be very interesting to see if the patterns seen in the four-dimensional case survive there. Relatedly, one could compute dynamical responses to external electromagnetic fields, and consider the generalization to charged black hole responses. Perhaps the most phenomenologically interesting generalization would be to compute the full dynamical Love numbers 
for a spinning Kerr black hole. The main advances required are technical. In particular, resumming and regulating gravitational effects in the EFT is simplest in dimensional regularization, but gravitational couplings to spinning black holes in higher dimensions are complicated.
At leading order, we can extract a Kerr black hole's dynamical Love number by exploiting the fact that a Kerr black hole is a spinning Schwarzschild black hole. The coordinate transformation between the (rotating) black hole's rest frame and the laboratory frame captures the leading-in-spin effect (similar to how a Kerr black hole's zero frequency dissipation is related to the finite frequency Schwarzschild dissipation~\cite{Goldberger:2020fot,Glazer:2024eyi}).
The computation of the full response for Kerr is a problem that we hope to return to.

One of the most compelling reasons to study the material properties of black holes is that these quantities are in principle measurable.
However, there is a practical consideration of understanding how the dynamical Love numbers actually enter into gravitational waveforms, which it would be nice to study in more detail (see~\cite{Chakraborty:2025wvs} for a recent analysis).

An important frontier is to push the computations that we have done to subleading order. This can be done in several directions. Most straightforwardly, one could imagine computing further subleading-in-frequency corrections to black hole tidal responses. In the scalar context, this was done in~\cite{Caron-Huot:2025tlq}. In order to try to further understand the structure of black hole responses, it would be useful to amass further theoretical data.
In addition to this, it has recently become clear that the nonlinear static tidal responses of black holes share many of the features of their linear counterparts~\cite{Poisson:2020vap,Riva:2023rcm,Iteanu:2024dvx,Combaluzier-Szteinsznaider:2024sgb,Kehagias:2024rtz,Parra-Martinez:2025bcu}. As such, it would be very interesting to see whether nonlinear dynamical tides have similar features to the linear dynamical tides considered here.

The goal of precision computation of gravitational observables is insight into the structure of gravity.
Studying the properties  of black holes teaches us how gravity organizes itself, and what features are universal. Most intriguingly, the remarkable simplicity of black holes has suggested the presence of symmetries, at least of black hole perturbation theory, that were not previously known. In practice, these symmetries appear when one truncates to some near region or to zero frequency, and so one might therefore naturally think that these symmetries are at best approximate. However, we might take solace in the fact that the dynamical Love numbers seem to share at least a piece of the magic of their static counterparts. In particular the fact that the running of dynamical tides is related in a simple way at leading order to the dissipative response is intriguing, not least because the dissipative responses themselves can be understood as conserved charges of symmetries of a near zone~\cite{Hui:2022vbh}. One might therefore hope that there is some regularity to the pattern of black hole properties that reflects some further hidden symmetries of gravity.

The black holes in our universe are a unique window into physics in extreme environments, but black holes serve as equally valuable theoretical laboratories. There is some irony that black holes are perhaps the most mysterious objects, but nevertheless provide us with continual new lessons about gravity. We look forward to seeing what else we learn.

﻿
\vspace{-18pt}
\paragraph{Acknowledgements:} Thanks to John Joseph Carrasco, Lam Hui, Riccardo Penco, Chia-Hsien Shen, Mikhail Solon, and Zihan Zhou for helpful conversations, and to Giulia Isabella, Massimiliano M.~Riva, and Jan Steinhoff for comments on the draft. 
 AJ and DG are supported in part by DOE (HEP) Award DE-SC0025323. 
 The work of MJR is supported through the NSF grant PHY-2309270.
The research of LS has been funded by the French National Research Agency (ANR)
under project ANR-24-CE31-1097-01. The work is supported by a PhD Joint Program of the International Research Center for Fundamental Scientific Discovery (IRC Discovery).
﻿

\vspace{-18pt}
\paragraph{Note added:} Shortly after this paper,~\cite{Kobayashi:2025vgl} appeared, which computes the odd-parity magnetic dynamical tidal response, with a slightly different perspective and formalism. Where our results overlap, they are in agreement.

\appendix
﻿
 
\section{Scalar dynamical response running}
\label{app:scalarkl}
﻿
Here we compute the coefficient of the logarithmic running of the conservative dynamical response at  $O(\omega^2)$ for a scalar field at generic $\ell$, and relate it to the linear-in-$\omega$ dissipative scalar response.
To this end, it is convenient to  recall the relation between the Legendre functions $Q_\ell$ and $P_\ell$:
\begin{equation}
        Q_\ell(x)=\frac12\log\left(\frac{x+1}{x-1}\right)P_\ell(x)-W_{\ell-1}(x) ,
        \qquad\quad
W_{\ell-1}(x)\equiv \sum_{n=0}^{\ell-1}c_n(x-1)^n ,
\label{Qell}
\end{equation}
and write $P_\ell$ in polynomial canonical form:
\begin{equation}
        P_\ell(x)=\,{}_2F_1\left[\begin{array}{c}
    \ell+1\,,\,-\ell\\[-3pt]
    1
    \end{array}\Big\rvert \,\frac{1-x}{2}\,\right] 
    =\sum_{n=0}^{\ell}a_n(x-1)^n,
\label{Pell}
\end{equation}
where $c_n$ and $a_n$ are the $\ell$-dependent constants
\be
 a_n = \frac{(\ell+1)_n(\ell+1-n)_n}{2^n(n!)^2}\,, \qquad\qquad c_n = \frac{(\ell+n)!\left(\psi(\ell+1)-\psi(n+1)\right)}{2^n(\ell-n)!(n!)^2}\,, 
\ee
and where $(z)_n$ is the Pochhammer symbol, defined by $(z)_n\equiv \Gamma(z+n)/\Gamma(z)$.
﻿

Let us start by  focusing  on the first integral in~\eqref{eq:psi2GreenInt}. Using~\eqref{Qell}, we find
\begin{equation}
\int^x\D y\frac{(1+y)^3}{y-1}P_\ell(y)Q_\ell(y)   = \int^x \frac{\D y}{2}\frac{(1+y)^3}{y-1}P_\ell(y)^2  \log\left(\frac{y+1}{y-1}\right) 
    -\int^x\D y\frac{(1+y)^3}{y-1}P_\ell(y)W_{\ell-1}(y) .
    \label{eq:psi2GreenInt-2}
\end{equation}
Consider  the first integral in this expression: from the definition of $P_\ell$,~\eqref{Pell}, the first integral in~\eqref{eq:psi2GreenInt-2} boils down to a sum of integrals, each of which has the following form:
\begin{equation}
I_n = \int^x \D y\frac{(1+y)^3}{y-1}(y-1)^n  \log\left(\frac{y+1}{y-1}\right) 
\label{eq:in}
\end{equation}
for some integer $n$. 
Recall that our modest goal here is to simply extract the $\ell$-dependent coefficient of the term  that in~\eqref{eq:psi2GreenInt} scales as $\log(x)/x^{\ell+1}$ at large $x$; all  terms with different scaling in $x$---or with no logarithm---will be ignored in what follows. For instance,   explicitly evaluating  the integral~\eqref{eq:in} for $n=0$ shows that no such $\log(x)/x^{\ell+1}$ is present, so it can be disregarded. Let us thus assume $n>0$.
From a straightforward integration by parts, we can write:
\be
\begin{aligned}
I_n &= \int^x \D y  \left[ \frac{\D}{\D y} \int^y\D z \frac{(1+z)^3}{z-1}(z-1)^n  \right] \log\left(\frac{y+1}{y-1}\right) 
\\
&=  p_n(x) \log\left(\frac{x+1}{x-1}\right) 
+2 \int^x \D y  \frac{p_n(y)}{(y+1)(y-1)} ,
\end{aligned}
\label{inintegrals}
\ee
where we have introduced the polynomials
\begin{equation}
p_n(x) \equiv \left(\frac{(x-1)^3}{n+3}+\frac{6 (x-1)^2}{n+2}+\frac{12 (x-1)}{n+1}+\frac{8}{n}\right) (x-1)^n .
\end{equation}
Using the integral representation of the hypergeometric function, we can write each piece contributing to the last term of~\eqref{inintegrals} as
\begin{equation}
\int^x_1 \D y \frac{(y-1)^m}{y+1} 
=\frac{(x-1)^{m+1}}{2(m+1)}\,{}_2F_1\left[\begin{array}{c}
    1\,,\,m+1\\[-3pt]
    m+2
    \end{array}\Big\rvert \,\frac{1-x}{2}\,\right] \,,
\label{eq:intF21}
\end{equation}	
with $m$ a non-negative integer.\footnote{In~\eqref{eq:intF21}, the lower bound in the integral (which we chose to be $1$ for convenience) is completely immaterial, as this simply amounts to a redefinition of the integration constants in~\eqref{eq:psi2GreenInt}. Note that no $\log(x)/x^{\ell+1}$ scaling is contained in the homogeneous solutions of~\eqref{eq:Sch_IZO2}.} From the properties of the hypergeometric function, each of these terms  is of the form $\text{pol}_1(x)+\text{pol}_2(x)\times\log(x)$, at large $x$, where $\text{pol}_1(x)$ and $\text{pol}_2(x)$ are generic ordinary polynomials with positive powers of $x$. In conclusion, no $\log(x)/x^{\ell+1}$ term results from the $x\rightarrow\infty$ expansion of the $I_n$ integrals~\eqref{inintegrals}, for all $n$. 
The second integral in~\eqref{eq:psi2GreenInt-2} can also  be computed:
\begin{equation}
    -\int^x\D y\frac{(1+y)^3}{y-1}P_\ell(y)W_{\ell-1}(y) 
    = - \sum_{k=0}^{2\ell-1} d_k  \, q_k(x) -\sum_{k=1}^{2\ell-1}  \frac{8}{k} d_k \, (x-1)^k+8 \, c_0 \log(x - 1),
\end{equation}
where $q_k$ denotes the polynomial
\begin{equation}
q_k(y)\equiv\left(\frac{12(y-1)}{1+k}+\frac{6(y-1)^2}{2+k}+\frac{(y-1)^3}{3+k}\right) (y-1)^k\,,
\end{equation}
and the coefficients in the series expansion are
\begin{equation}
d_k = \frac{1}{2^k} \sum_{n=0}^{k} 
\frac{\Gamma(\ell+1+n)\, \Gamma(\ell+1+k-n)}
     {\Gamma(\ell+1-n)\, \Gamma(\ell+1-(k-n)) \, (n!)^2 \, ((k-n)!)^2} 
\bigl(\psi(\ell+1) - \psi(k-n+1)\bigr).
\end{equation}
From this expression, we can now make the following consideration. As noted before for the first integral, we argue that
the same conclusion is trivially  true for the second integral in~\eqref{eq:psi2GreenInt-2}, since both $P_\ell(y)$ and $W_{\ell-1}(y)$ are  polynomials in $(y-1)$. Therefore, there is no $\log(x)/x^{\ell+1}$ type of term resulting from the first integral in \eqref{eq:psi2GreenInt}. 
Let us then focus on the second integral in~\eqref{eq:psi2GreenInt}:
\begin{equation}
\frac{B}{4}\int^x\D y\frac{(1+y)^3}{y-1}P_\ell(y)^2  .
    \label{eq:psi2GreenExplicit-2}
\end{equation}
Using the series representation~\eqref{Pell} of the Legendre polynomials, the only term of the sum~\eqref{Pell} that can generate a logarithm in ~\eqref{eq:psi2GreenExplicit-2} is the one with the $a_0$ coefficient. All the other terms in the series produce just polynomials in $x$ after integration. Explicitly,
\begin{align}
\frac{B}{4}\int^x\D y\frac{(1+y)^3}{y-1}P_\ell(y)^2 &= \frac{B}{4} \left( \sum_{m=0}^{2 \ell} f_m  \, q_m(x)+\sum_{m=1}^{2\ell} \frac{8}{m} f_m (x-1)^m +8 \log(x-1) \right) \nonumber \\
&=
2B \log(x-1) + \text{pol}(x),
    \label{eq:psi2GreenExplicit-3}
\end{align}
where the coefficients $f_m$ from the convolution of the series expansions of $P_{\ell}^2$ take the form
\begin{equation}
f_m=\sum_{n=0}^{m} a_n \, a_{m-n}.
\label{eq:fk}
\end{equation}
From the large-$x$ expansion of the $Q_\ell(x)$ functions,
\begin{equation}
Q_\ell(x) \underset{x\to\infty}{\sim} \frac{2^\ell (\ell!)^2}{(2\ell+1)!}\frac{1}{x^{\ell+1}} ,
\end{equation}
combined with~\eqref{eq:psi2GreenExplicit-3}, we find that the second integral in~\eqref{eq:psi2GreenInt} produces at infinity 
\begin{equation}
\lim_{x\to\infty}\frac{B}{4}\int^x\D y\frac{(1+y)^3}{y-1}P_\ell(y)^2 Q_\ell(x)  \supset
2B   \frac{2^\ell (\ell!)^2}{(2\ell+1)!}\frac{\log(x) }{x^{\ell+1}}  + \cdots\,,
    \label{eq:psi2GreenExplicit-4}
\end{equation}
where we omitted in $\cdots$  all contributions that scale differently from the one that we wrote explicitly.\footnote{The omitted terms are not necessarily subleading---they are not necessarily suppressed by more powers of~$x$.}
﻿
Putting everything together in~\eqref{eq:psi_expansion}, and once again focusing only on the asymptotic falloffs $x^\ell$ and $x^{-\ell-1}$, the expanded intermediate-zone solution, up to order $\epsilon^2$, reads 
\begin{equation}
\lim_{x\to\infty}\phi_{\text{IZ}} \supset B\frac{(2\ell)!}{2^\ell(\ell!)^2} x^\ell \left(1 +O(\epsilon) \right) 
+ B\frac{2^\ell (\ell!)^2}{(2\ell+1)!}\frac{1}{x^{\ell+1}} \left( 2i \epsilon +2\epsilon^2 \log(x) + O(\epsilon^3)  \right) +\dots 
    \label{eq:psi_expansion-largeL}
\end{equation}
where we used that $P_\ell(x) \underset{x\to\infty}{\sim} \frac{2^{-\ell} (2 \ell)!}{(\ell!)^2}x^\ell$.
From the definitions of  $x$ and $\epsilon$ (see~\eqref{defxrs}), the ratio of the two falloffs $r^{-\ell-1}$ and $r^\ell$  (involving only the logarithmic term at order $\epsilon^2$) is
\begin{equation}
    \kappa_\ell=\bigg(i\omega r_s+\omega^2 r_s^2\log\left(\frac{r}{r_s}\right)\bigg)\frac{(\ell!)^4}{(2\ell)!(2\ell+1)!}+ \cdots
    \label{Sch_dyn_resp}
\end{equation}
in agreement with previous results~\cite{Charalambous:2021mea,Perry:2023wmm}. Note that when expressed as $\log r_s$, the relative ratio is precisely $-1$, so that the dynamical Love number running is  (minus) the dissipative response.

\newpage
\section{Far zone}
\label{app:farzone}

Here we study the far zone approximation of the Klein--Gordon equation~\eqref{eqKGPhi} (see also~\cite{Unruh:1976fm}).  Before taking the far-zone limit, it is convenient to  redefine the field in a way that removes the first-derivative term from the equation. We introduce
\begin{equation}
    \xi_\ell(r)\equiv\sqrt{\Delta(r)}\phi_\ell(r),
\label{eq:xiphiFZ}
\end{equation}
where  ${\Delta}(r)\equiv {r(r-r_s)}$.
In terms of $\xi$,~\eqref{eqKGPhi} then becomes
\begin{equation}
    \partial_r^2 \xi_\ell(r)+\left(\frac{\omega^2r^4}{\Delta^2}-\frac{\ell(\ell+1)}{\Delta}- \frac{2\Delta\Delta''-(\Delta')^2}{4\Delta^2}\right) \xi_\ell(r)=0.
    \label{eq:Sch_KG_xi}
\end{equation}

A possible way to obtain a far-zone  solution would be  to  expand the potential in \eqref{eq:Sch_KG_xi} at large $r$,  keeping  terms up to the desired order.\footnote{For instance, up to the order $O(r^{-2})$, one gets
\begin{equation}
    \partial_r^2 \xi_\ell(r)+\left( \omega^2+\frac{2r_s\omega^2}{r}-\frac{\ell(\ell+1)-3\omega^2r_s^2}{r^2} \right)\xi_\ell(r)=0.
    \label{eq:approximated_TE}
\end{equation}
Neglecting the $O(\omega^2/r^2)$ term in the potential, as in \cite{Unruh:1976fm}, yields precisely the Coulomb wave equation.}
Although the solutions to the equation expanded this way reproduce asymptotically the standard Coulomb functions, care must be taken to ensure that the range of validity of the approximation extends sufficiently into the intermediate zone, so that matching with the perturbative solution in Section~\ref{sec:intermediate-zone-scalar} can be carried out.
In particular, we need a solution that remains valid from $\omega r\sim O(1)$ all the way up to $\omega r\to\infty$. To this end, we will expand assuming $r_s\ll r$ but without imposing any hierarchy between $\omega$ and $r$.

Concretely, it is convenient to define the quantities
\begin{equation}
    z\equiv \omega r, \qquad\qquad \epsilon\equiv \omega r_s\,.
    \label{Coulomb_variables}
\end{equation}
We then expand equation~\eqref{eq:Sch_KG_xi} for small $\frac{\epsilon}{z}=\frac{r_s}{r}$, up to order $O(\frac{\epsilon^2}{z^2})$, without making any assumption about the variable $z$, which can range from $O(1)$  to $\infty$.  This yields
\begin{equation}
    \partial_z^2\xi_\ell+\left(1-\frac{\ell(\ell+1)}{z^2}\right)\xi_\ell=-\epsilon \left( \frac{2}{z}-\frac{\ell(\ell+1)}{z^3}\right)\xi_\ell -\epsilon^2\left( \frac{3}{z^2}+\frac{\frac14-\ell(\ell+1)}{z^4}\right)\xi_\ell +O(\epsilon^3)\,.
    \label{eq:Sch_FZ}
\end{equation}
In the following, we solve this equation to linear order in $\epsilon$. We then use the result to compute the scattering amplitude of a scalar field off the black hole, and finally compare it with the solution obtained from the  Mano--Suzuki--Takasugi (MST) formalism~\cite{Mano:1996vt,Sasaki:2003xr}.

\subsection{Far-zone scalar solution}

We look for a perturbative solution in $\epsilon$, i.e., we expand the field as 
\begin{equation}
    \xi_{\ell}(r(z))=\xi_{\ell}^{(0)}(z)+\epsilon\xi_{\ell}^{(1)}(z)+\epsilon^2\xi_{\ell}^{(2)}(z)+\cdots.
\end{equation}
At zeroth order, $\xi_{\ell}^{(0)}$ solves \eqref{eq:Sch_FZ} with $\epsilon=0$.  The two independent solutions can be expressed in terms of the standard Coulomb functions as
\begin{equation}
    \xi_{\ell}^{(0)}(z)=c_1^{(0)}F_\ell(0,z)+c_2^{(0)}G_\ell(0,z),
    \label{FZ_xi0_solution}
\end{equation}
with one of the parameters set to zero. 
Using the well-known Coulomb wave functions' asymptotics,
\begin{equation}
    \begin{aligned}
        F_\ell(\eta,z)&\underset{z\to0}{\sim}C_\ell(\eta)z^{\ell+1}\underset{\eta=0}{=}\frac{1}{(2\ell+1)!!}z^{\ell+1},\\
        G_\ell(\eta,z)&\underset{z\to0}{\sim}\frac{z^{-\ell}}{(2\ell+1)C_\ell(\eta)}\underset{\eta=0}{=}\frac{(2\ell+1)!!}{2\ell+1}z^{-\ell}=(2\ell-1)!!\,z^{-\ell},
    \end{aligned}
\end{equation}
with the constant parameters
\begin{equation}
    C_\ell(\eta)\equiv\frac{2^\ell \E^{-\pi\eta/2}\vert\Gamma(\ell+1+i\eta)\vert}{(2\ell+1)!},
\qquad
(2\ell+1)!! =\frac{(2\ell+1)!}{2^\ell \ell!} ,
\end{equation}
we can compute the $z\to0$ limit of the FZ solution:
\begin{equation}
    \phi_{\ell,\text{FZ}}^{(0)}\underset{z\to0}{\sim} \frac{c_1^{(0)}}{(2\ell+1)!!}\omega z^\ell+c_2^{(0)}\frac{ (2\ell+1)!!}{2\ell+1}\omega z^{-\ell-1},
\end{equation}
where we used the definition \eqref{eq:xiphiFZ} and have expanded $1/\sqrt{\Delta}$ in powers of $\epsilon$ before taking the small-$z$ limit,\footnote{To match with the intermediate-zone solution, we require $\epsilon\ll z$.}
\begin{equation}
\begin{aligned}
   \frac{1}{\sqrt{\Delta}}&=\frac{\omega}{\left(z(z-\epsilon)\right)^{1/2}}\\
   &=\frac{\omega}{z}\left(1+\frac{\epsilon}{2z}+\frac{3\epsilon^2}{8z^2}\right)+O\left(\frac{\epsilon^3}{z^3}\right).
\end{aligned}
\end{equation}
Note that we are doing an expansion in $\epsilon$, and not literally in $\omega$. Hence, the integration constants $c_1^{(0)}$ and $c_2^{(0)}$ can (and in general will) have a nontrivial dependence on $\omega$. In particular, the FZ solution at a specific order in $\epsilon$ may involve different orders in $\omega$.

\noindent
{\it Matching at leading order:}
Comparing with the intermediate-zone solution at order $O(\epsilon^0)$,  which takes the form
\begin{equation}
    \phi_{\ell,\text{IZ}}^{(0)}=B P_\ell\left(2r/r_s-1\right)\underset{r\to\infty}{\sim}B\frac{(2\ell)!}{(\ell!)^2}\left(\frac{r}{r_s}\right)^\ell+\cdots ,
\end{equation}
and matching at leading order in $r/r_s$ yields
\begin{equation}
    c_2^{(0)}=0, \qquad\qquad c_1^{(0)}=\frac{B}{\omega^{\ell+1}r_s^\ell}\frac{(2\ell)!(2\ell+1)!!}{(\ell!)^2}.
    \label{FZ_int_csts0}
\end{equation}

\noindent
{\it Matching at sub-leading order:}
Let us now consider the matching with the  intermediate zone at linear order in $\epsilon$. The $O(\epsilon)$ term on the right-hand side of \eqref{eq:Sch_FZ}  plays the role of a source, which we evaluate on the $O(\epsilon^0)$ solution $\xi_{\ell}^{(0)}$.
 Using standard Green's function methods, we find the following  general solution\footnote{Recall the Wronskian $W[F_\ell(0,z),G_\ell(0,z)]\equiv F_\ell(0,z)\partial_z G_\ell(0,z) - G_\ell(0,z) \partial_z F_\ell(0,z)  =-1$.}
\begin{equation}
    \xi_{\ell}^{(1)}(z)=\left(c_1^{(1)}-I_{F}(z)\right)F_\ell(0,z)+\left(c_2^{(1)}+I_{G}(z)\right)G_\ell(0,z) , 
    \label{xi_1_gal_sol}
\end{equation}
where we defined the indefinite integrals
\begin{equation}
\begin{aligned}
    I_{F}(z)& \equiv \int^z\D t \, G_\ell(0,t)\xi_{\ell}^{(0)}(t)\left(\frac{2}{t}-\frac{\ell(\ell+1)}{t^3}\right) , \\
    I_{G}(z)&\equiv \int^z\D t \, F_\ell(0,t)\xi_{\ell}^{(0)}(t)\left(\frac{2}{t}-\frac{\ell(\ell+1)}{t^3}\right).
\end{aligned}
\end{equation}
These integrals can be computed in terms of (generalized) hypergeometric functions ${_p}F_q$ as 
\begin{equation}
\begin{aligned}
    I_{F}(z)&=\frac{c_1^{(0)}}{2\ell+1}\bigg(\frac{\ell(\ell+1)}{z} \,{}_1F_2\bigg[\begin{array}{c}
   -\tfrac{1}{2}\\
     \tfrac{1}{2}-\ell\,,\,\tfrac{3}{2}+\ell
    \end{array}\bigg\rvert -z^2\,\bigg]+2z \,{}_2F_3\bigg[\begin{array}{c}
    \tfrac{1}{2}\,,\,\tfrac{1}{2}\\
     \tfrac{3}{2}\,,\,\tfrac{1}{2}-\ell \,,\, \tfrac{3}{2}+\ell
    \end{array}\bigg\rvert -z^2\,\bigg] \bigg)\, \\[2pt]
    I_{G}(z)&=\frac{-\sqrt{\pi}c_1^{(0)}z^{2\ell}}{4}\bigg((\ell+1)! \,{}_1\tilde F_2\bigg[\!\begin{array}{c}
   \ell\\[-2pt]
     \tfrac{3}{2}+\ell\,,\,2+2\ell
    \end{array}\bigg\rvert -z^2\,\bigg]\\
    &\hspace{3cm}
    -2(\ell!)^2z^2 {}_2\tilde F_3\bigg[\!\begin{array}{c}
    \ell+1\,,\,\ell+1\\[-1pt]
     \tfrac{3}{2}+\ell\,,\,2+\ell \,,\, 2+2\ell
    \end{array}\bigg\rvert -z^2\,\bigg]\bigg)\ ,
\end{aligned}
\end{equation}
where $\tilde{F}$ denotes the regularized function, defined by
\begin{equation}
    \,{}_p\tilde F_q\left[\begin{array}{c}
   a_1\,,\, \cdots\,,\, a_p\\[-2pt]
     b_1\,,\, \cdots\,,\, b_q
    \end{array}\bigg\rvert\, x\,\right]
    \equiv \Big[\Gamma(b_1)\cdots\Gamma(b_q)\Big]^{-1} \,{}_p F_q\left[\begin{array}{c}
   a_1\,,\, \cdots\,,\, a_p\\[-2pt]
     b_1\,,\, \cdots\,,\, b_q
    \end{array}\bigg\rvert\, x\,\right].
\end{equation}
To match with the intermediate-zone solution, we need the small-$z$ expansion of the solution \eqref{xi_1_gal_sol}. First, let us use the series representation of the generalized hypergeometric functions,
\begin{equation}
    \,{}_p F_q\left[\begin{array}{c}
   a_1\,,\, \cdots\,,\, a_p\\[-2pt]
     b_1\,,\, \cdots\,,\, b_q
    \end{array}\bigg\rvert\, x\,\right]
    =\sum_{n=0}^\infty\frac{(a_1)_n\cdots (a_p)_n}{(b_1)_n\cdots (b_q)_n}\frac{x^n}{n!},
\end{equation}
which we use to write $I_{F}(z)$ and $I_{G}(z)$ as
\begin{equation}
\begin{aligned}
    I_{F}(z)&=\frac{c_1^{(0)}}{2\ell+1}\sum_{k=0}^\infty\left[ \ell(\ell+1)f_{1,k} z^{2k-1}+2f_{2,k} z^{2k+1} \right]\\
    &=\frac{c_1^{(0)}}{2\ell+1}\left(\sum_{k=1}^\infty \underbrace{\left[\ell(\ell+1)f_{1,k}+2f_{2,k-1}\right]}_{\zeta_k^\ell} z^{2k-1}+\frac{\ell(\ell+1)f_{1,0}}{z} \right),
\end{aligned}
\end{equation}
and 
\begin{equation}
\begin{aligned}
    I_{G}(z)=& -\frac{\sqrt{\pi}c_1^{(0)}}{4}z^{2\ell}\sum_{k=0}^\infty\left[(\ell+1)!g_{1,k} z^{2k} -2(\ell!)^2g_{2,k} z^{2k+2}\right]\\
    =& -\frac{\sqrt{\pi}c_1^{(0)}}{4}z^{2\ell}\left(\sum_{k=1}^\infty \underbrace{\left[(\ell+1)!g_{1,k}-2(\ell!)^2 g_{2,k-1}\right]}_{\gamma_k^\ell}z^{2k}+(\ell+1)!g_{1,0} \right),
\end{aligned}
\end{equation}
where we have introduced the combinations
\begin{equation}
    f_{1,k}=\frac{(-1)^k(-\frac12)_k}{(\frac12-\ell)_k(\frac32+\ell)_kk!}, \qquad f_{2,k}=\frac{(-1)^k((\frac12)_k)^2}{(\frac32)_k(\frac12-\ell)_k(\frac32+\ell)_kk!},
\end{equation}
\begin{equation}
    g_{1,k}=\frac{(-1)^k(\ell)_k}{\Gamma(\ell+k+\frac32)\Gamma(2\ell+k+2)k!}, \qquad g_{2,k}=\frac{(-1)^k((\ell+1)_k)^2}{\Gamma(\ell+k+\frac32)\Gamma(2\ell+k+2)\Gamma(\ell+k+2)k!}.
\end{equation}
To expand the Coulomb functions, we can then use their relations to spherical Bessel functions:
\begin{equation}
    F_\ell(0,z)=zj_\ell(z), \qquad\qquad G_\ell(0,z)=-zy_\ell(z),
\end{equation}
where  $j_\ell$ and $y_\ell$ admit the following series representation:
\begin{equation}
    j_\ell(z)=z^\ell \sum_{n=0}^\infty j_n^\ell z^{2n}, \qquad y_\ell(z)=-z^{-\ell-1}\sum_{n=0}^\infty y_n^\ell z^{2n},
\end{equation}
with the parameters
\begin{equation}
\begin{aligned}
    j_n^\ell &= \frac{(-1)^n}{2^nn!(2\ell+2n+1)!!}, \\
    y_n^\ell &= \Theta(\ell-n)\frac{(2\ell-2n-1)!!}{2^nn!}+\Theta(n-(\ell+1))\frac{(-1)^{\ell+n}}{2^nn!(2n-2\ell-1)!!},
\end{aligned}
\end{equation}
where $\Theta(k)=1$ for $\mathbb{N}\ni k \geq 0$, and $\Theta(k)=0$ otherwise.
Finally, the expansion of~\eqref{xi_1_gal_sol} for small $z$ reads
\begin{equation}
\begin{aligned}
    \xi_{\ell}^{(1)}(z)=&\left[c_1^{(1)}z^{\ell+1}-\frac{c_1^{(0)}}{2\ell+1}\left( \sum_{k=1}^\infty \zeta_k^\ell z^{\ell+2k}+\ell(\ell+1)z^\ell \right)\right]\sum_{n=0}^\infty j_n^\ell z^{2n} \\
    &+\left[c_2^{(1)}z^{-\ell}-\frac{c_1^{(0)}\sqrt{\pi}}{4}\left( \sum_{k=1}^\infty\gamma_k^\ell z^{\ell+2k}+(\ell+1)!g_{1,0} z^\ell \right)\right]\sum_{n=0}^\infty y_n^\ell z^{2n}. 
\end{aligned}
\end{equation}
Plugging this into the definition of $\phi^{(1)}_{\ell,\text{FZ}}=\frac{\omega}{z}(\xi^{(1)}_{\ell}+\frac{1}{2z}\xi^{(0)}_{\ell})$, we obtain the following $z\rightarrow0$ behavior for $\phi_{\ell,\text{FZ}}^{(1)}$\footnote{We use $y_0^\ell=(2\ell-1)!!=\frac{(2\ell+1)!!}{2\ell+1}$ to simplify the expressions.}:
\begin{align}
    \phi_{\ell,\text{FZ}}^{(1)}(z)\underset{z\to0}{\sim}& \, \omega \Bigg[ c_2^{(1)} z^{-\ell-1}\left(y_0^\ell+\sum_{n=1}^{\ell}y_n^\ell z^{2n}\right) -c_1^{(0)}z^{\ell-1}\left( \frac{\ell(\ell+1)}{2\ell+1}j_0^\ell+ \frac{\sqrt{\pi}(\ell+1)!}{4}g_{1,0}y_0^\ell-\frac{1}{2(2\ell+1)!!} \right) \nonumber \\
    & \quad +c_1^{(1)}j_0^\ell z^\ell+O(z^{\ell+1}) \Bigg] \label{phi1_FZ_sol}\\
    =& \, \, \omega c_2^{(1)}\frac{(2\ell+1)!!}{2\ell+1}z^{-\ell-1}\left( 1+\sum_{n=1}^{\ell-1}\frac{y_n^\ell}{y_0^\ell}z^{2n}\right)-\omega c_1^{(0)}\frac{\ell}{2(2\ell+1)!!}z^{\ell-1} +\frac{\omega c_1^{(1)}}{(2\ell+1)!!}z^{\ell}+O(z^{\ell+1})\nonumber.
\end{align}
The integration constants $c_1^{(1)}$ and $c_2^{(1)}$ can be determined by comparing $\phi_{\ell,\text{FZ}}^{(1)}$ with the intermediate-zone solution:
\begin{equation}
\begin{aligned}
    \phi_{\ell,\text{IZ}}^{(1)}&=b_1^{(1)}P_\ell\left({2r}/{r_s}-1\right)+b_2^{(1)}Q_\ell({2r}/{r_s}-1) \\[2pt]
    &\!\!\!\!
    \underset{\frac{r}{r_s}\to\infty}{\sim}b_1^{(1)}\frac{(2\ell)!}{(\ell!)^2}\left( \frac{r}{r_s}\right)^\ell+b_2^{(1)}\frac{(\ell!)^2}{2(2\ell+1)!}\left( \frac{r}{r_s}\right)^{-\ell-1}+ O\left(\left( {r}/{r_s}\right)^{\ell-1}\right),
\end{aligned}
\end{equation}
where $b_1^{(1)}$ and  $b_2^{(1)}$ are the intermediate-zone integration constants, which are fixed in terms of the amplitude $B$ via~\eqref{eq:phi1_cst}.
From the comparison of the $r^{-\ell-1}$ and $r^\ell$ falloffs, we find
\begin{equation}
    c_1^{(1)}=\frac{b_1^{(1)}}{\omega^{\ell+1}r_s^\ell}\frac{(2\ell)!(2\ell+1)!!}{(\ell!)^2},
     \qquad\quad c_2^{(1)}=\omega^\ell r_s^{\ell+1}b_2^{(1)}\frac{(\ell!)^2}{2(2\ell)!(2\ell+1)!!}.
    \label{FZ_integration_constants}
\end{equation}
Once these coefficients are fixed, it is straightforward to verify that all the remaining terms in~\eqref{phi1_FZ_sol} that contribute at the given order match automatically.
 First, note that upon substituting~\eqref{FZ_integration_constants} into~\eqref{phi1_FZ_sol}, the subleading terms proportional to $c_2^{(1)}$ are at least of order $O(\omega^{2})$. Since we are only interested in terms up to first order in $\omega$, we can discard them and focus on comparing the term proportional to $c_1^{(0)}$. Substituting the expression for $c_1^{(0)}$ from~\eqref{FZ_int_csts0} into $ \epsilon \phi_{\ell,\text{FZ}}^{(1)}$, one finds
\begin{equation}
    -\epsilon \omega c_1^{(0)}\frac{\ell}{2(2\ell+1)!!}z^{\ell-1}=-B\frac{\ell(2\ell)!}{2(\ell!)^2}\left(\frac{r}{r_s}\right)^{\ell-1}.
\end{equation}
This contribution is  frequency independent. It must therefore match the subleading contribution in $\phi_{\ell,\text{IZ}}^{(0)}$ as $r\to\infty$. To check this, we can use the series representation of the  Legendre polynomials,
\begin{equation}
P_\ell\left(\frac{2r}{r_s}-1\right)=\sum_{n=0}^\ell\sum_{k=0}^n\alpha_{nk}\left(\frac{r}{r_s}\right)^{n-k},
\qquad
    \alpha_{nk}=(-1)^k\binom{n}{k}\frac{(\ell+n)!}{(\ell-n)!(n!)^2},
\end{equation}
to express the zeroth-order intermediate-zone solution as
\begin{equation}
    \phi_{\ell,\text{IZ}}^{(0)}=BP_\ell\left(\frac{2r}{r_s}-1\right)\underset{\frac{r}{r_s}\to\infty}{\sim}B\alpha_{\ell0}\left( \frac{r}{r_s}\right)^\ell+B\left(\alpha_{(\ell-1)0}+\alpha_{\ell 1}\right)\left( \frac{r}{r_s}\right)^{\ell-1}+O\left(\left( r/r_s\right)^{\ell-2}\right), 
\end{equation}
where 
\begin{equation}
    \alpha_{(\ell-1)0}+\alpha_{\ell 1}=-\frac{\ell}{2}\alpha_{\ell0}=-\frac{\ell(2\ell)!}{2(\ell!)^2}.
\end{equation}
As expected, this  reproduces the $O(\omega^0r^{\ell-1}$) part of $\phi_{\ell,\text{FZ}}^{(1)}$.

\subsection{Comparison with the MST solution}

The MST formalism is a systematic approach to compute black hole scattering amplitudes order by order in frequency, constructed by expanding the field solution as an infinite series of hypergeometric functions (see~\cite{Mano:1996vt,Sasaki:2003xr} for details, and also~\cite{Charalambous:2021mea,Ivanov:2024sds,Ivanov:2025ozg,Glazer:2024eyi}). The goal of this section is to show that the far-zone solution  obtained above correctly reproduces the MST result for the scalar scattering amplitude at the considered order of approximation.

For this purpose, it is convenient  to first solve a slightly different, more general problem, which we will later connect to the far zone above and to MST. Unlike before, let us  not make any assumption on $\epsilon$, which can now take any value, and solve~\eqref{eq:Sch_FZ} for large $z$. One could call this the {\it Very Far Zone}.
Truncated  at order $O(z^{-2})$,~\eqref{eq:Sch_FZ} reduces to a standard Coulomb wave equation
\begin{equation}
    \partial_z^2 \xi_{\ell}(z)+\left( 1+\frac{2\epsilon}{z}-\frac{\ell(\ell+1)-3\epsilon^2}{z^2} \right)\xi_{\ell}(z)=0,
    \label{eq:TE_FZ2}
\end{equation}
whose general solutions can be written as 
\begin{equation}
    \xi_{\ell,\epsilon}(z)=d_1 F_L(-\epsilon,z)+d_2 G_L(-\epsilon, z),
    \label{eq:FZ2_solution}
\end{equation}
where $d_1$ and $d_2$ are generic integration constants and $L$ a  parameter defined by
\begin{equation}
\begin{aligned}
    L\equiv &-\frac12+\frac12\sqrt{(2\ell+1)^2-4\epsilon^2\delta \ell}.
\end{aligned}
\end{equation}
Note that the FZ solution $\xi_\ell$ derived in the previous section can be recovered by taking the $\epsilon\to0$ limit of \eqref{eq:FZ2_solution}.

The scalar scattering amplitude $\mathcal{A}$ is given by the ratio of the coefficients of the reflected and ingoing waves~\cite{Ivanov:2022qqt,Saketh:2023bul},
\begin{equation}
    1-i\mathcal{A}=(-1)^{\ell+1}\frac{\phi_\ell^\text{ref}}{\phi_\ell^\text{in}} ,
\end{equation}
where $\phi_\ell^\text{ref}$ and $\phi_\ell^\text{in}$ are defined by
\begin{equation}
     \underset{\omega r\to\infty}{\lim}\phi_{\ell,\text{FZ}}= \underset{\omega r\to\infty}{\lim}\frac{\xi_{\ell,\epsilon}(\omega r)}{r}=\phi_\ell^{\text{ref}}\frac{\E^{+i\omega r_\star}}{r}+\phi_\ell^{\text{in}}\frac{\E^{-i\omega r_\star}}{r},
     \label{phi_FZ_falloff}
\end{equation}
where we introduced the tortoise coordinate $r_\star$, defined as
\begin{equation}
    \omega r_\star =z+\epsilon\log\left(\frac z\epsilon - 1\right) \underset{z\to\infty}{\sim} z+\epsilon\log(z)-\epsilon\log(\epsilon).
\end{equation}
Using the large-$z$ limit of the Coulomb wave functions
\begin{equation}
\begin{aligned}
     F_L(\eta,z)&\underset{z\to\infty}{\sim}\sin\left(z-\eta\log(2z)-\frac{L\pi}{2}+\arg(\Gamma(L+1+i\eta)\right) , \\
    G_L(\eta,z)&\underset{z\to\infty}{\sim}\cos\left(z-\eta\log(2z)-\frac{L\pi}{2}+\arg(\Gamma(L+1+i\eta)\right),
\end{aligned}
\end{equation}
we find the following expressions for the fall-off coefficients,\footnote{We used the following relations for the $\Gamma$-functions with complex argument:
\begin{equation}
   \Gamma(a^*)=\Gamma(a)^*, \quad \quad \arg\Gamma(a)=-\frac{i}{2}\log\frac{\Gamma(a)}{\Gamma(a^*)}.
\end{equation}}
\begin{equation}
\begin{aligned}
    \phi_\ell^{\text{ref}}&=\frac{\E^{-i\frac{\pi L}{2}}}{2}\E^{i\epsilon\log(2\epsilon)} \left(d_2-id_1\right)\sqrt{\frac{\Gamma(L-i\epsilon+1)}{\Gamma(L+i\epsilon+1)}} , \\
    \phi_\ell^{\text{in}}&=\frac{\E^{i\frac{\pi L}{2}}}{2}\E^{-i\epsilon\log(2\epsilon)} \left(d_2+id_1\right)\sqrt{\frac{\Gamma(L+i\epsilon+1)}{\Gamma(L-i\epsilon +1)}},
\end{aligned}
\end{equation}
which finally yields 
\begin{equation}
    (-1)^{L+1}\frac{\phi^{\text{ref}}}{\phi^{\text{in}}}=-\frac{\Gamma(L-i\epsilon+1)}{\Gamma(L+i\epsilon+1)} \E^{2i\epsilon\log(2\epsilon)}\frac{d_2-id_1}{d_2+id_1}.
    \label{FZ_ratio}
\end{equation}

In order to compare~\eqref{FZ_ratio} with the MST result, we take  the expression for the same quantity obtained via the MST approach (see  Appendix E of~\cite{Glazer:2024eyi} for details). In the notation of~\cite{Glazer:2024eyi}, and for small $\epsilon$,
\begin{equation}
\begin{aligned}
    (-1)^{L+1}\frac{R^{(\text{ref.})}}{R^{(\text{in.})}}=&\frac{A_-^\nu}{A_+^\nu} \E^{i\epsilon\left(2\log\epsilon-(1-\kappa)\right)}
    \frac{K_{\nu}-i\E^{i\pi\nu} K_{-\nu-1}}{K_{\nu}+i \E^{-i\pi\nu} \frac{\sin(\nu + i\epsilon)}{\sin(\nu-i\epsilon)} K_{-\nu-1}} \\
    =&\frac{A_-^\nu}{A_+^\nu} \E^{i\epsilon\left(2\log\epsilon-(1-\kappa)\right)}\bigg[1-i\E^{i\pi L}\left(1+\E^{-2i\pi L}\frac{\sin(\pi(L+i\epsilon)}{\sin(\pi(L-i\epsilon)}\right)\frac{K_{-\nu-1}}{K_\nu} \bigg]+O(\epsilon^{2L+3}),
    \label{MST_ratio}
\end{aligned}
\end{equation}
which we adapted here to the case of a Schwarzschild black hole. In particular, $\nu=L+\omega^2 \nu_2$
and $L$ is the analytic continuation of $\ell$ to non-integer values. Note also that $\frac{K_{-\nu-1}}{K_\nu}=O(\epsilon^{2L+1})$. 
The expression \eqref{MST_ratio} has been shown to reflect a  near-far factorization of the scattering amplitude~\cite{Ivanov:2022qqt,Saketh:2023bul}. In particular,  the term in square brackets arises from the physics of the near zone and its matching to the far region,  while the prefactor captures  the far-zone contribution. 
This factorization is not yet  apparent from~\eqref{FZ_ratio}. To make it manifest, it is useful to re-express the ratio $\frac{d_2-id_1}{d_2+id_1}$ by extracting  the $\epsilon$-dependence from $d_1$ and $d_2$. To do so, we first match the asymptotic solution $\xi_{\ell,\epsilon}$ in \eqref{eq:FZ2_solution} to the  FZ solution $\xi_\ell$ derived in the previous section for  small $\epsilon$. The $d_{1,2}$ constants then become linear combinations of the FZ integration constants $c_{1,2}^{(n)}$, where $n$ is the order in $\epsilon$ we are considering. Given the form of the constants at linear order in $\epsilon$ (see~\eqref{FZ_int_csts0} and \eqref{FZ_integration_constants}), it is convenient to parametrize $d_1$ and $d_2$ as
\begin{equation}
    d_1=\epsilon^\ell d_1^+ +\frac{d_1^-}{\epsilon^{\ell+1}}, \qquad\qquad d_2=\epsilon^\ell d_2^+ +\frac{d_2^-}{\epsilon^{\ell+1}},
\end{equation}
where $d_{1,2}^\pm$ have a polynomial dependence on $\epsilon$.
Plugging these into~\eqref{FZ_ratio}, one finds
\begin{equation}
    (-1)^{L+1}\frac{\phi^{\text{ref}}}{\phi^{\text{in}}}=-\frac{\Gamma(L-i\epsilon+1)}{\Gamma(L+i\epsilon+1)} e^{2i\epsilon\log(2\epsilon)}\frac{d_2^--id_1^-}{d_2^-+i d_1^-}\left[1+2i\frac{d_2^+d_1^- -d_2^- d_1^+}{(d_2^-)^2+(d_1^-)^2}\epsilon^{2\ell+1}+O(\epsilon^{4\ell+2})\right],
    \label{CS_ratio}
\end{equation}
which we expressed so as to recover the form of the near-far factorization of~\eqref{MST_ratio}. We can now perform the comparison by separately expanding  the near-zone and  far-zone terms at linear order in $\epsilon$.
First, we fix the  constants $d_{1,2}^{(1)}$ at this order by matching the asymptotic solution with the one from the previous section. Imposing
\begin{equation}
    \underset{z\to\infty}{\lim}\xi_{\ell}(z)-\underset{\epsilon\to0}{\lim}\left(\underset{z\to\infty}{\lim}\xi_{\ell,\epsilon}(z)\right)=O(\epsilon^2) , 
    \label{FZ_matching}
\end{equation} 
 where 
\begin{equation}
\begin{aligned}
    \underset{z\to\infty}{\lim}\xi_{\ell}(z)=& \,  \epsilon\cos\left(\frac{\pi\ell}{2}-z\right) \left( c_2^{(1)}+c_1^{(0)}\left(\log(2z)-\psi(\ell+1)-\frac12\right) \right) \\
    &-\left(c_1^{(0)}+ c_1^{(1)} \epsilon+c_1^{(0)}\frac{\pi}{2}\epsilon\right) \sin \left(\frac{\pi \ell}{2}-z\right)+O(\epsilon^2),
\end{aligned}
\end{equation}
and
\begin{equation}
\begin{aligned}
    \underset{\epsilon\to0}{\lim}\left(\underset{z\to\infty}{\lim}\xi_{\ell,\epsilon}(z)\right)=& \, \cos\left(\frac{\pi  L}{2}-z\right) (-d_1 \epsilon  \psi(L+1)+d_1 \epsilon  \log (2 z)+d_2)\\
    &-\sin \left(\frac{\pi L}{2}-z\right) (d_1+d_2 \epsilon  \psi(L+1)-d_2 \epsilon \log(2z))+O(\epsilon^2),
\end{aligned}
\end{equation}
and using $L=\ell+O(\epsilon^2)$, we obtain
\begin{equation}
\begin{aligned}
    d_1^- &= \epsilon^{\ell+1} \left( c_1^{(0)}+\left(c_1^{(1)}+c_1^{(0)}\frac{\pi}{2}\right)\epsilon +O(\epsilon^2)\right), & d_1^+&=O(\epsilon^2), \\
    d_2^- &= -\epsilon^{\ell+1}\frac{ c_1^{(0)}\epsilon+O(\epsilon^2)}{2}, & d_2^+&=\frac{ c_2^{(1)}\epsilon+O(\epsilon^2)}{\epsilon^{\ell}},
    \label{BC_FZ2}
\end{aligned}
\end{equation} 
where the  constants $c$'s are given in~\eqref{FZ_int_csts0} and \eqref{FZ_integration_constants}.

Let us now compare  the two ``far-zone'' contributions in~\eqref{MST_ratio} and \eqref{CS_ratio}. In the MST result, we can take  the integer-$L$ limit, which is smooth, and simply replace $L \rightarrow \ell\in\mathbb{N}$. Expanding the definitions, we find that
\begin{equation}
    -\frac{\Gamma(L-i\epsilon+1)}{\Gamma(L+i\epsilon+1)} \E^{2i\epsilon\log(2\epsilon)}\frac{d_2^--id_1^-}{d_2^-+i d_1^-}-\frac{A_-^\nu}{A_+^\nu} \E^{i\epsilon\left(2\log\epsilon-(1-\kappa)\right)} = O(\epsilon^2).
\end{equation}
The ``near-zone'' term---corresponding to the square bracket in~\eqref{MST_ratio}---appears instead to be singular in the integer-$\ell$ limit. However, by expressing everything in terms of trigonometric functions, one can easily verify that the divergences cancel at first order in $\epsilon$, yielding
\begin{equation}
    1-i\frac{K_{-\nu-1}}{K_\nu}\left(i\E^{i\pi L}+\E^{-i\pi L}\frac{\sin(\pi(L+i\epsilon)}{\sin(\pi(L-i\epsilon)}\right)\underset{\underset{\epsilon\to0}{L\to\ell\in\mathbb{N}}}{\sim}1-\epsilon^{2\ell+1}\left( \frac{2^{2\ell+1}(\ell!)^6}{((2\ell)!)^2((2\ell+1)!)^2}\epsilon+O(\epsilon^2)\right),
\end{equation}
where the term in parenthesis matches the prefactor of the  $\epsilon^{2\ell+1}$ term in the square brackets of~\eqref{CS_ratio}, up to subleading orders in $\epsilon$, which we ignored here.

\newpage
\section{Gravitational $\ell=3,4$ solutions}
\label{app:l=3-4_solution}

Here we collect the final expressions for the $\ell=3$ and $\ell=4$ solutions of the Regge–Wheeler and Zerilli fields up to second order in frequency. We first present the results from a full general relativistic calculation, in which the Regge–Wheeler and Zerilli equations are solved order by order in frequency. We then provide the corresponding expressions for the Born-series solutions obtained from the point-particle EFT for the black hole.

\subsection{General relativistic solutions}
\label{app:grsols}

We first write the results of a general relativity computation of the tidal field.

\noindent
{\bf Odd sector:}
Solving the Regge--Wheeler equation~\eqref{eq:ddimRWeq} perturbatively up to second-order in $\omega$,  as described in Section~\ref{sec:GRsols}, we find for $\ell=3$
\begin{equation}
\begin{aligned}
\Psi^{\ell=3}_{\text{RW,IZ}}(r)
&\underset{\frac{r}{r_s}\to\infty}{=}
B\bigg(
6\frac{r^4}{r_s^4}
-5\frac{r^3}{r_s^3}
\bigg)
+iB\omega r_s\bigg(
\frac{137}{10}\frac{r^4}{r_s^4}
-\frac{137}{12}\frac{r^3}{r_s^3}
+\frac{1}{42}\frac{r_s^3}{r^3}
\bigg)\\
&~\hspace{1cm}
+B\omega^2 r_s^2 \bigg[
-\frac{1}{3}\frac{r^6}{r_s^6}
-\frac{r^5}{r_s^5}
-\frac{r^4}{r_s^4}\!\left(\pi^2+\frac{100\,243}{4\,200}-\frac{13}{7}\log\frac{r_s}{r}\right)
\\
&~\hspace{2.75cm}+\frac{5}{6}\frac{r^3}{r_s^3}\!\left(\pi^2+\frac{38\,981}{1\,400}-\frac{13}{7}\log\frac{r_s}{r}\right)-\frac{109}{210}\frac{r^2}{r_s^2}
-\frac{13}{280}\frac{r}{r_s}
\\&~\hspace{2.75cm}+\frac{19}{280}
+\frac{337}{2\,800}\frac{r_s}{r}
+\frac{1}{6}\frac{r_s^2}{r^2}
+\frac{1}{42}\frac{r_s^3}{r^3}\!\left(\frac{223}{42}-\log\frac{r_s}{r}\right)
\bigg]+O\!\left(\frac{r_s^4}{r^4}\right),
\end{aligned}
\end{equation}
and for $\ell = 4$
\begin{equation}
\begin{aligned}
\Psi^{\ell=4}_{\text{RW,IZ}}(r)
&\underset{\frac{r}{r_s}\to\infty}{=}
B\bigg(
28\frac{r^5}{r_s^5}
-42\frac{r^4}{r_s^4}
+15\frac{r^3}{r_s^3}
\bigg)
+iB\omega r_s\bigg(
\frac{413}{5}\frac{r^5}{r_s^5}
-\frac{1\,239}{10}\frac{r^4}{r_s^4}
+\frac{177}{4}\frac{r^3}{r_s^3}
+\frac{1}{252}\frac{r_s^4}{r^4}
\bigg)\\
&~\hspace{1cm}+B\omega^2 r_s^2 \bigg[
-\frac{14}{11}\frac{r^7}{r_s^7}
-\frac{147}{55}\frac{r^6}{r_s^6}
-\frac{r^5}{r_s^5}\!\left(\frac{14\pi^2}{3}+\frac{3\,696\,439}{23\,100}-\frac{3\,142}{495}\log\frac{r_s}{r}\right)\\
&~\hspace{2.75cm}
+\frac{r^4}{r_s^4}\!\left(7\pi^2+\frac{35\,319\,721}{138\,600}-\frac{1\,571}{165}\log\frac{r_s}{r}\right)
\\
&~\hspace{2.75cm}
-\frac{r^3}{r_s^3}\!\left(\frac{5\pi^2}{2}+\frac{762\,181}{7\,920}-\frac{1\,571}{462}\log\frac{r_s}{r}\right)+\frac{3\,373}{4\,158}\frac{r^2}{r_s^2}
+\frac{4\,759}{27\,720}\frac{r}{r_s}
\\
&~\hspace{2.75cm}
+\frac{11\,689}{138\,600}
+\frac{57\,889}{831\,600}\frac{r_s}{r}
+\frac{1}{14}\frac{r_s^2}{r^2}
+\frac{9}{112}\frac{r_s^3}{r^3}
\\
&~\hspace{2.75cm}
+\frac{1}{252}\frac{r_s^4}{r^4}\!\left(\frac{9\,767}{504}-\log\frac{r_s}{r}\right)
\bigg]+O\!\left(\frac{r_s^5}{r^5}\right).
\end{aligned}
\end{equation}

\noindent
{\bf Even sector:}
Similarly we tabulate the solutions to the Zerilli equation in the even sector. The solution for $\ell=3$ is

\begin{align}
&\Psi_\text{Z,IZ}^{\ell=3}(r) \underset{\frac{r}{r_s}\to\infty}{=}
\frac{C}{50}\bigg(300\frac{r^4}{r_s^4}-190\frac{r^3}{r_s^3}-93\frac{r^2}{r_s^2}+\frac{279}{10}\frac{r}{r_s}+\frac{663}{100}-\frac{1\,989}{1\,000}\frac{r_s}{r}+\frac{5\,967}{10\,000}\frac{r_s^2}{r^2}-\frac{17\,901}{100\,000}\frac{r_s^3}{r^3}
\bigg) \nonumber \\
& ~\hspace{.1cm}
+iC\omega r_s\bigg(14\frac{r^4}{r_s^4}-\frac{133}{15}\frac{r^3}{r_s^3}-\frac{217}{50}\frac{r^2}{r_s^2}+\frac{651}{500}\frac{r}{r_s}+\frac{1\,547}{5\,000}-\frac{4\,641}{50\,000}\frac{r_s}{r}
+\frac{13\,923}{500\,000}\frac{r_s^2}{r^2}+\frac{1\,622\,851}{105\,000\,000}\frac{r_s^3}{r^3}
\bigg) \nonumber \\
& ~\hspace{.1cm}
+C\omega^2 r_s^2 \Bigg[-\frac{1}{3}\frac{r^6}{r_s^6}-\frac{11}{10}\frac{r^5}{r_s^5}-\frac{r^4}{r_s^4}\!\left(\pi^2+\frac{51\,917}{2\,100}-\frac{13}{7}\log\frac{r_s}{r}\right)
+\frac{r^3}{r_s^3}\!\left(\frac{19\pi ^2}{30}+\frac{399\,451}{21\,000}-\frac{247}{210} \log\frac{r_s}{r}\right) \nonumber \\
&~\hspace{1.85cm}+\frac{r^2}{r_s^2}\!\left(\frac{31\pi ^2}{100}+\frac{1\,634\,147}{210\,000}-\frac{403}{700} \log\frac{r_s}{r}\right)
-\frac{r}{10^3 r_s}\!\left(93\pi ^2+\frac{2\,039\,147}{700}-\frac{1\,209}{7}\log \frac{r_s}{r}\right) \nonumber \\
&~\hspace{1.85cm}
-\frac{221\pi^2}{10\,000}-\frac{10\,422\,677}{21\,000\,000}+\frac{2\,873}{70\,000}\log\frac{r_s}{r}
+\frac{r_s}{10^5 r}\!\left(663\pi^2+\frac{21\,972\,677}{700}-\frac{8\,619}{7}\log \!\left(\frac{r_s}{r}\right)\right) \nonumber \\
&~\hspace{1.85cm}+\frac{r_s^2}{10^6 r^2}\!\left(-1\,989\pi^2+\frac{220\,495\,907}{2\,100}+\frac{25\,857}{7}\log \frac{r_s}{r}\right) \nonumber  \\
&~\hspace{1.85cm}+\frac{r_s^3}{10^7 r^3}\!\left(5\,967\pi^2+\frac{58\,238\,757\,859}{44\,100}-\frac{5\,232\,713}{21}\log \frac{r_s}{r}\right)
\Bigg]+O\!\left(\frac{r_s^4}{r^4}\right),
\end{align}
and the solution for $\ell=4$ is
%
\begin{align}
&\Psi^{\ell=4}_{\text{Z,IZ}}(r) \underset{\frac{r}{r_s}\to\infty}{=}C \bigg(
28\frac{r^5}{r_s^5}-\frac{119}{3}\frac{r^4}{r_s^4}+\frac{179}{18}\frac{r^3}{r_s^3}+\frac{361}{108}\frac{r^2}{r_s^2}-\frac{361}{648}\frac{r}{r_s}
\nonumber \\
&~\hspace{1.55cm} 
-\frac{287}{3\,888}+\frac{287}{23\,328}\frac{r_s}{r}-\frac{287}{139\,968}\frac{r_s^2}{r^2}+\frac{287}{839\,808}\frac{r_s^3}{r^3}-\frac{287}{5\,038\,848}\frac{r_s^4}{r^4}
\bigg)\nonumber \\
& ~\hspace{0.1cm}+Ci\omega r_s \Bigg[
\frac{1\,246}{15}\frac{r^{5}}{r_s^{5}}-\frac{10\,591}{90}\frac{r^{4}}{r_s^{4}}+\frac{15\,931}{540}\frac{r^{3}}{r_s^{3}}+\frac{32\,129}{3\,240}\frac{r^{2}}{r_s^{2}}-\frac{32\,129}{19\,440}\frac{r}{r_s}
\nonumber\\
&~\hspace{1.2cm}\!
-\frac{25\,543}{116\,640} +\frac{25\,543}{699\,840}\frac{r_s}{r}-\frac{25\,543}{4\,199\,040}\frac{r_s^{2}}{r^{2}}+\frac{25\,543}{25\,194\,240}\frac{r_s^{3}}{r^{3}}+\frac{4\,020\,239}{1\,058\,158\,080}\frac{r_s^{4}}{r^{4}}
\Bigg]\nonumber\\
&~\hspace{0.1cm}
+C\omega^2 r_s^2 \Bigg[-\frac{14}{11}\frac{r^7}{r_s^7}-\frac{931}{330}\frac{r^6}{r_s^6}-\frac{r^5}{r_s^5}\!\left(\frac{14 \pi ^2}{3}+\frac{2\,798\,434}{17\,325}-\frac{3\,142}{495} \log\frac{r_s}{r}\right) \nonumber \\
&~\hspace{.8cm}
+\frac{r^4}{r_s^4}\!\left(\frac{119 \pi ^2}{18}+\frac{202\,591\,567}{831\,600}-\frac{26\,707}{2\,970} \log\frac{r_s}{r}\right)-\frac{r^3}{r_s^3}\!\left(\frac{179 \pi ^2}{108}+\frac{332\,182\,607}{4\,989\,600}-\frac{281\,209}{124\,740} \log\frac{r_s}{r})\right) \nonumber \\
&~\hspace{.8cm}-\frac{r^2}{r_s^2}\!\left(\frac{361 \pi ^2}{648}+\frac{603\,088\,753}{29\,937\,600}-\frac{567\,131}{748\,440} \log\frac{r_s}{r}\right)+\frac{r}{r_s}\!\left(\frac{361 \pi ^2}{3\,888}+\frac{703\,291\,153}{179\,625\,600}-\frac{567\,131}{4\,490\,640} \log\frac{r_s}{r}\right) \nonumber \\
&~\hspace{.8cm}+\frac{81\,592\,409}{153\,964\,800}
+\frac{287\pi^2}{23\,328}-\frac{64\,411}{3\,849\,120} \log\frac{r_s}{r}-\frac{r_s}{r}\!\left(\frac{287 \pi ^2}{139\,968}+\frac{12\,308\,249}{923\,788\,800}-\frac{64\,411}{23\,094\,720} \log\frac{r_s}{r}\right) \nonumber \\
&~\hspace{.8cm}+\frac{r_s^2}{r^2}\!\left(\frac{287 \pi ^2}{839\,808}+\frac{3\,234\,737\,903}{38\,799\,129\,600}-\frac{64\,411}{138\,568\,320}\log \frac{r_s}{r}\right) \nonumber \\
&~\hspace{.8cm}+\frac{r_s^3}{r^3}\!\left(-\frac{287 \pi ^2}{5\,038\,848}+\frac{17\,904\,629\,137}{232\,794\,777\,600}+\frac{64\,411}{831\,409\,920} \log\frac{r_s}{r}\right) \nonumber \\
&~\hspace{.8cm}+\frac{r_s^4}{r^4}\!\left(\frac{287 \pi ^2}{30\,233\,088}+\frac{738\,910\,509\,641}{9\,777\,380\,659\,200}-\frac{139\,019\,197}{34\,919\,216\,640} \log\frac{r_s}{r}\right)
\Bigg]+\cdots\,,
\end{align}
%
where the subleading terms that we have not displayed are $O(r_s^5/r^5)$.

\subsection{EFT solutions}

Here we present the particular solutions obtained from computing the Born series for $\ell=3,4$.

\noindent
{\bf Odd sector:}
We begin with the odd sector. The $\ell=3$ solution is
\begin{equation}
\begin{aligned}
    \Psi_{\text{RW}}^{\ell=3}(r)=& \, \, r^4 B_\text{reg} \left(1+\frac{361 \bar{G}^2 \omega^2}{441}-\frac{13 \bar{G}^2 \omega ^2}{42 \varepsilon}-\frac{13 \bar{G}^2 \omega^2}{14}\log(\mu r)\right)\\
   &+\frac{B_\text{reg}}{r^3} \left(\frac{287\,552 \bar{G}^9 \omega ^2}{19\,845} +\frac{8 \bar{G}^9 \omega ^2}{63 \varepsilon}+\frac{136 \bar{G}^9 \omega^2}{63}\log(\mu r) \right)\\
   &+\frac{B_\text{irr}}{r^3} \left(\frac17+\frac{1\,108\bar{G}^2 \omega^2}{3\,087} +\frac{13 \bar{G}^2 \omega^2}{294 \varepsilon }+\frac{65 \bar{G}^2 \omega^2}{294}\log(\mu r)\right),
\end{aligned}
\end{equation}
while the $\ell=4$ solution is
\begin{equation}
\begin{aligned}
    \Psi_{\text{RW}}^{\ell=4}(r)=& \, \, r^5 B_\text{reg} \left(1+\frac{4\,472\,159 \bar{G}^2 \omega^2}{4\,802\,490}-\frac{1\,571 \bar{G}^2 \omega ^2}{6\,930 \varepsilon}-\frac{1\,571 \bar{G}^2 \omega^2}{2\,310}\log(\mu r)\right)\\
   &+\frac{B_\text{reg}}{r^4} \left(\frac{2\,291\,068 \bar{G}^{11} \omega ^2}{363\,825} +\frac{32 \bar{G}^{11} \omega ^2}{2\,205 \varepsilon}+\frac{32 \bar{G}^{11} \omega^2}{105}\log(\mu r) \right)\\
   &+\frac{B_\text{irr}}{r^4} \left(\frac19+\frac{62\,860\,681\bar{G}^2 \omega^2}{216\,112\,050} +\frac{1\,571 \bar{G}^2 \omega^2}{62\,370 \varepsilon }+\frac{1\,571 \bar{G}^2 \omega^2}{12\,474}\log(\mu r)\right).
\end{aligned}
\end{equation}
To subtract the infinities, we introduce the renormalized coefficients $\Bar{B}_\text{reg}$ and $\Bar{B}_\text{irr}$ as
\begin{equation}
    B_\text{reg}=\Bar{B}_\text{reg}(1+\omega^2\delta_{11})+\Bar{B}_\text{irr}\omega^2\delta_{12}, \qquad\quad   B_\text{irr}=\Bar{B}_\text{irr}(1+\omega^2\delta_{22})+\Bar{B}_\text{reg}\omega^2\delta_{21},
\label{eq:rencoeffsRW-2}
\end{equation}
where the $\delta$ parameters are (arranged in a matrix as in~\eqref{eq:deltamatrixdef})
\begin{equation}
    \delta^{\ell=3}=
    \begin{pmatrix}
       \frac{13 \bar{G}^2}{42\varepsilon}  & 0 \\
       -\frac{8 \bar{G}^9}{9\varepsilon} &-\frac{13 \bar{G}^2}{42\varepsilon}
    \end{pmatrix}, \qquad 
    \delta^{\ell=4}=
    \begin{pmatrix}
       \frac{1571 \bar{G}^2}{6\,930\varepsilon}  & 0 \\
       -\frac{32 \bar{G}^11}{245\varepsilon} &-\frac{1\,571 \bar{G}^2}{6\,930\varepsilon}
    \end{pmatrix}.
\end{equation}
Plugging the renormalized coefficients~\eqref{eq:rencoeffsRW-2} back into the expression for $\Psi_\text{RW}$ before taking the $\varepsilon \to 0$ limit yields the following renormalized solutions. For $\ell=3$ we have
\begin{equation}
\begin{aligned}
    \Psi_\text{RW}^{\text{R},\ell=3} & =\Bar{B}_\text{reg}\left[r^4+r^4\bar{G}^2\omega^2\left(\frac{361}{441}-\frac{26}{21}\log(\mu r)\right)+\frac{\bar{G}^9\omega^2}{r^3}\left(\frac{278\,512}{19\,845}+\frac{128}{63}\log(\mu r)\right)\right]\\
    &~~~+\Bar{B}_\text{irr}\left[\frac{1}{7r^3}+\frac{\bar{G}^2\omega^2}{r^3}\left(\frac{1\,069}{3\,087}+\frac{26}{147}\log(\mu r)\right)\right],
\end{aligned}
\end{equation}
while the renormalized solution for $\ell=4$ is
\begin{align}
\nonumber
    \Psi_\text{RW}^{\text{R},\ell=4} &=\Bar{B}_\text{reg}\left[r^5+r^5\bar{G}^2\omega^2\left(\frac{4\,472\,159}{4\,802\,490}-\frac{3\,142}{3\,465}\log(\mu r)\right)+\frac{\bar{G}^{11}\omega^2}{r^4}\left(\frac{47\,616\,488}{7\,640\,325}+\frac{128}{441}\log(\mu r)\right)\right]\\
    &~~~+\Bar{B}_\text{irr}\left[\frac{1}{9r^4}+\frac{\bar{G}^2\omega^2}{r^4}\left(\frac{20\,550\,337}{72\,037\,350}+\frac{3\,142}{31\,185}\log(\mu r)\right)\right].
\end{align}
Comparing with the solutions $\Psi_\text{RW,IZ}^{\ell}$  from Section~\ref{app:grsols}, we can fix the renormalized  constants  $\Bar{B}_\text{reg}$ and $\Bar{B}_\text{irr}$ as
\begin{equation}
\begin{aligned}
    \Bar{B}_\text{reg}^{\ell=3} =&\, B\left[ 6+i\omega r_s\frac{137}{10}-\omega^2r_s^2\left(\frac{737\,801}{29\,400}+\pi^2-\frac{13}{7}\log(\mu r_s) \right) \right],\\
    \Bar{B}_\text{irr}^{\ell=3} =& \, B\left[\frac{i\omega r_s^8}{6}-\omega^2r_s^9\left(\frac{4\,027}{15\,120}+\frac16\log (\mu r_s)\right)\right],\\
    \Bar{B}_\text{reg}^{\ell=4} =&\, B\left[ 28+i\omega r_s\frac{413}{5}-\omega^2r_s^2\left(\frac{1\,142\,563\,973}{6\,860\,700}+\frac{14\pi^2}{3}-\frac{3\,142}{495}\log(\mu r_s) \right) \right],\\
    \Bar{B}_\text{irr}^{\ell=4} =& \, B\left[\frac{i\omega r_s^{10}}{28}-\omega^2r_s^{11} \left( \frac{580\,211}{7\,761\,600}+\frac{1}{28}\log (\mu r_s)\right)\right]. 
\end{aligned}
\end{equation}
From the ratio of these coefficients, 
we then obtain the response functions~\eqref{eq:magl12}.

\noindent
{\bf Even sector:}
In the even sector, the Born series computations are similar, but the expressions are considerably more lengthy. To simplify things, we exploit the absence of decaying fall-off in the static limit of the full black hole solution. In the following expressions, we will therefore take $B_\text{irr}\equiv \omega \tilde{B}_\text{irr}$, where $\tilde{B}_\text{irr}$ starts at order $O(\omega^0)$. Then the solutions are
\begin{align}
    \Psi_{\text{Z}}^{\ell=3}(r)=& \, \, r^4 B_\text{reg} \left(1+\frac{8\,563 \bar{G}^2 \omega^2}{11\,025}-\frac{13 \bar{G}^2 \omega ^2}{42 \varepsilon}-\frac{13 \bar{G}^2 \omega^2}{14}\log(\mu r)\right)\\ \nonumber
   &
   \hspace{-1.7cm}
   +\frac{B_\text{reg}}{r^3} \left(-\frac{5\,967 \bar{G}^7}{78\,125}+\frac{68\,705\,719\,849 \bar{G}^9 \omega ^2}{5\,167\,968\,750} +\frac{1\,482\,713 \bar{G}^9 \omega ^2}{9\,843\,750 \varepsilon}+\frac{25\,206\,121 \bar{G}^9 \omega^2}{9\,843\,750}\log(\mu r) \right)+\frac{\omega\tilde{B}_\text{irr}}{7r^3}\,,
\end{align}
and for $\ell=4$
\begin{equation}
\begin{aligned}
    \Psi_{\text{Z}}^{\ell=4}(r)=& \, \, r^5 B_\text{reg} \left(1+\frac{8\,999\,989\bar{G}^2 \omega^2}{9\,604\,980}-\frac{1\,571 \bar{G}^2 \omega ^2}{6\,930\varepsilon}-\frac{1\,571 \bar{G}^2 \omega^2}{2\,310}\log(\mu r)\right)\\
   &+\frac{B_\text{reg}}{r^4} \bigg(-\frac{41\bar{G}^9}{39\,366}+\frac{40\,726\,113\,886\,309 \bar{G}^{11} \omega ^2}{6\,616\,918\,746\,900} +\frac{28\,264\,541 \bar{G}^{11} \omega ^2}{1\,909\,644\,660 \varepsilon}\\
   &\hspace{1.5cm}+\frac{28\,164\,541\bar{G}^{11} \omega^2}{90\,935\,460}\log(\mu r) \bigg)+\frac{\omega\tilde{B}_\text{irr}}{9r^4}.
\end{aligned}
\end{equation}
In this notation, the renormalized coefficients read
\begin{equation}
    B_\text{reg}=\Bar{B}_\text{reg}(1+\omega^2\delta_{11})+\Bar{B}_\text{irr}\omega^2\delta_{12}, \qquad\quad \omega \tilde{B}_\text{irr}=\Bar{B}_\text{irr}(1+\omega^2\delta_{22})+\Bar{B}_\text{reg}\omega^2\delta_{21},
\label{eq:rencoeffsRW-3}
\end{equation}
with
\begin{equation}
    \delta^{\ell=3}=
    \begin{pmatrix}
       \frac{13 \bar{G}^2}{42\varepsilon}  & 0 \\
       -\frac{8 \bar{G}^9}{9\varepsilon} &0
    \end{pmatrix},
\qquad\qquad
    \delta^{\ell=4}=
    \begin{pmatrix}
       \frac{1\,571 \bar{G}^2}{6\,930\varepsilon}  & 0 \\
       -\frac{32 \bar{G}^{11}}{245\varepsilon} &0
    \end{pmatrix}.
\end{equation}
Then, the renormalized EFT solutions are
\begin{equation}
\begin{aligned}
    \Psi_\text{Z}^{\text{R},\ell=3}=&\,\Bar{B}_\text{reg}\bigg[r^4+r^4\bar{G}^2\omega^2\left(\frac{8\,563}{11\,025}-\frac{26}{21}\log(\mu r)\right)\bigg] \\
    &+\Bar{B}_\text{reg}\bigg[-\frac{5\,967 \bar{G}^7}{78\,125 r^3}+\frac{\bar{G}^9\omega^2}{r^3}\left(\frac{33\,532\,060\,907}{2\,583\,984\,375}+\frac{10\,465\,426}{4\,921\,875}\log(\mu r)\right)\bigg]+\frac{\Bar{B}_\text{irr}}{7r^3},
\end{aligned}
\end{equation}
\begin{equation}
\begin{aligned}
    \Psi_\text{Z}^{\text{R},\ell=4}=&\,\Bar{B}_\text{reg}\bigg[r^5+r^5\bar{G}^2\omega^2\left(\frac{8\,999\,989}{9\,604\,980}-\frac{3\,142}{3\,465}\log(\mu r)\right)\bigg] \\
    &+\Bar{B}_\text{reg}\bigg[-\frac{41 \bar{G}^9}{39\,366 r^4}+\frac{\bar{G}^{11}\omega^2}{r^4}\left(\frac{575\,411\,524\,549}{94\,527\,410\,670}+\frac{139\,019\,197}{477\,411\,165}\log(\mu r)\right)\bigg]
    +\frac{\Bar{B}_\text{irr}}{9r^4}.
\end{aligned}
\end{equation}
Matching with $\Psi_\text{Z,IZ}^{\ell}$, we find
\begin{equation}
\begin{aligned}
    \Bar{B}_\text{reg}^{\ell=3} &=\, B\left[ 6+14i\omega r_s-\omega^2r_s^2\left(\pi^2+\frac{76\,109}{2\,940}-\frac{13}{7}\log\mu r_s \right) \right],\\
    \Bar{B}_\text{irr}^{\ell=3} &= \, B\left[\frac{i\omega r_s^8}{6}-\omega^2r_s^9 \left( \frac{139}{560}+\frac16\log (\mu r_s)\right)\right],\\
    \Bar{B}_\text{reg}^{\ell=4} &=\, B\left[ 28+i\omega r_s\frac{1\,246}{15}-\omega^2r_s^2\left(\frac{14\pi^2}{3}+\frac{1\,153\,179\,809}{6\,860\,700}-\frac{3\,142}{495}\log\mu r_s \right) \right],\\
    \Bar{B}_\text{irr}^{\ell=4} &= \, B\left[\frac{i\omega r_s^{10}}{28}-\omega^2r_s^{11} \left( \frac{23\,263}{323\,400}+\frac{1}{28}\log (\mu r_s)\right)\right].
\end{aligned}
\end{equation}
The ratios of these fall-offs lead to the response functions~\eqref{eq:elecl12}.

\newpage
\section{Tidal field in de Donder gauge}
\label{app:tidalfield}
﻿
In Section~\ref{sec:GR-EFT}, we made use of specific properties of the tidal field in de Donder gauge to simplify some expressions. Here, we present the details of those properties.

In particular, we will be interested in the last term in~\eqref{eq:BBBbar-2}. Since it is already of $O(\omega)$,
it will be sufficient for our purposes to work out the tidal field solution $\bar{h}^{ij}$ in the static limit. 
We  begin with the expression of the tidal field in Regge--Wheeler (RW) gauge~\cite{Regge:1957td}, which we will transform to  the de Donder (dD) gauge. 
In Minkowski space, this transformation can be written as
\begin{equation}
h_{\mu\nu}^\text{dD} = h_{\mu\nu}^\text{RW} + \partial_\mu \xi_\nu + \partial_\nu \xi_\mu ,
\label{eq:gaugetrsDDRW}
\end{equation}
where $\xi^\mu$ defines the gauge parameter relating $h_{\mu\nu}^\text{dD}$ and $h_{\mu\nu}^\text{RW}$ (we omit for simplicity the  bar over $h_{\mu\nu}$ to denote the background tidal field from now on). By definition, $h_{\mu\nu}^\text{dD}$ satisfies
\begin{equation}
\partial^\mu h_{\mu\nu}^\text{dD} = \frac{1}{2} \partial_\nu h^\text{dD},
\end{equation}
and the equation of motion
\begin{equation}
\square h_{\mu\nu}^\text{dD} =0 .
\end{equation}
Denoting by $h_{\mu\nu}^{\text{odd}}$ ($h_{\mu\nu}^{\text{even}}$) the odd (even) components of $h_{\mu\nu}$, we have $h^{\text{odd}}=0$,\footnote{This holds in any gauge---recall that the trace $h\equiv h^\mu_\mu$ is a scalar and vanishes, by definition, identically in the odd sector.}  and,  from the trace of~\eqref{eq:gaugetrsDDRW},  $\partial^\mu \xi^{\text{odd}}_\mu=0$, with  $\xi^{\text{odd}}_\mu$ the odd part of the gauge transformation. Since $\xi^{\text{odd}}_0=0$ identically, we then have
\begin{equation}
\partial^i \xi^{\text{odd}}_i=0 .
\label{eq:staticxi}
\end{equation}
This implies that~$\xi^{\text{odd}}_i \equiv \epsilon_{ijk}\partial^j V^k $, for some vector $V^k$.
Let us now recall that, after solving the constraint equations in the Regge--Wheeler gauge, one finds $h^{{\text{odd}},\text{RW}}_{ij}=0$.\footnote{The only non-vanishing components of $h^{{\text{odd}},\text{RW}}_{\mu\nu}$ are $h^{{\text{odd}},\text{RW}}_{0i}$ in the static limit. See e.g.,~\cite{Iteanu:2024dvx}.}
Then the previous conditions imply
\begin{equation}
\vec{\nabla}^2\xi^{\text{odd}}_i=0 .
\label{eq:nabla2xi}
\end{equation}
Expanding $\xi^{\text{odd}}_i$ in powers of $\vec{x}$, from~\eqref{eq:staticxi} and~\eqref{eq:nabla2xi} it follows that
\begin{equation}
\xi^{\text{odd}}_i = \frac{1}{\ell+1} c_{i ( j_1 \cdots j_{\ell+1} )_T} x^{j_1 }\cdots x^{j_{\ell+1}},
\end{equation}
where $c_{i ( j_1 \cdots j_{\ell+1} )_T}$ is a generic constant tensor, symmetric and traceless in its last $\ell+1$ indices,  and satisfying ${c^i}_{( i j_1 \cdots j_{\ell})_T}=0$.

Since $h^{{\text{odd}},\text{RW}}_{ij}=0$ in Regge--Wheeler gauge in the static regime, then
\begin{equation}
\begin{aligned}
h_{ij}^{{\text{odd}},\text{dD}} & =  \left(  c_{i ( j j_1 \cdots j_{\ell} )_T } + c_{j ( i j_1 \cdots j_{\ell} )_T} \right) x^{j_1 }\cdots x^{j_{\ell}}
\\
& \equiv \bar c_{ ( i j j_1 \cdots j_{\ell} )_T} x^{j_1 }\cdots x^{j_{\ell}},
\end{aligned}
\end{equation}
where $c_{ ( i j j_1 \cdots j_{\ell} )_T}$ is---by construction---symmetric and traceless in all its indices.
Plugging into \eqref{eq:BBBbar-2}, we can clearly see that 
\begin{equation}
\partial^{[k} \bar{h}_{\text{dD}}^{j] i}
= \ell \left[ \bar c^{ (  j i k j_2 \cdots j_{\ell})_T}   - \bar c^{ (  k i j j_2 \cdots j_{\ell} )_T}    \right] x_{j_2 }\cdots x_{j_{\ell}} =0,
\label{eq:oddcond1}
\end{equation}
which is the property we used in the main text.

\newpage
\linespread{.95}
\addcontentsline{toc}{section}{References}
\bibliographystyle{utphys}
{\small
\bibliography{eftbib}

\providecommand{\href}[2]{#2}\begingroup\raggedright\begin{thebibliography}{100}

\bibitem{Goldberger:2004jt}
W.~D. Goldberger and I.~Z. Rothstein, ``{An Effective field theory of gravity
  for extended objects},''
  \href{http://dx.doi.org/10.1103/PhysRevD.73.104029}{{\em Phys. Rev. D}
  {\bfseries 73} (2006) 104029},
  \href{http://arxiv.org/abs/hep-th/0409156}{{\ttfamily arXiv:hep-th/0409156}}.

\bibitem{Goldberger:2005cd}
W.~D. Goldberger and I.~Z. Rothstein, ``{Dissipative effects in the worldline
  approach to black hole dynamics},''
  \href{http://dx.doi.org/10.1103/PhysRevD.73.104030}{{\em Phys. Rev. D}
  {\bfseries 73} (2006) 104030},
  \href{http://arxiv.org/abs/hep-th/0511133}{{\ttfamily arXiv:hep-th/0511133}}.

\bibitem{10.1093/mnras/69.6.476}
A.~E.~H. Love, ``The yielding of the earth to disturbing forces,''
  \href{http://dx.doi.org/10.1093/mnras/69.6.476}{{\em Monthly Notices of the
  Royal Astronomical Society} {\bfseries 69} no.~6, (04, 1909) 476--480}.

\bibitem{Poisson_Will_2014}
E.~Poisson and C.~M. Will,
  \href{http://dx.doi.org/10.1017/CBO9781139507486}{{\em Gravity: Newtonian,
  Post-Newtonian, Relativistic}}.
\newblock Cambridge University Press, 2014.

\bibitem{Fang:2005qq}
H.~Fang and G.~Lovelace, ``{Tidal coupling of a Schwarzschild black hole and
  circularly orbiting moon},''
  \href{http://dx.doi.org/10.1103/PhysRevD.72.124016}{{\em Phys. Rev.}
  {\bfseries D72} (2005) 124016},
\href{http://arxiv.org/abs/gr-qc/0505156}{{\ttfamily arXiv:gr-qc/0505156
  [gr-qc]}}.

\bibitem{Damour:2009vw}
T.~Damour and A.~Nagar, ``{Relativistic tidal properties of neutron stars},''
  \href{http://dx.doi.org/10.1103/PhysRevD.80.084035}{{\em Phys. Rev. D}
  {\bfseries 80} (2009) 084035},
  \href{http://arxiv.org/abs/0906.0096}{{\ttfamily arXiv:0906.0096 [gr-qc]}}.

\bibitem{Binnington:2009bb}
T.~Binnington and E.~Poisson, ``{Relativistic theory of tidal Love numbers},''
  \href{http://dx.doi.org/10.1103/PhysRevD.80.084018}{{\em Phys. Rev.}
  {\bfseries D80} (2009) 084018},
\href{http://arxiv.org/abs/0906.1366}{{\ttfamily arXiv:0906.1366 [gr-qc]}}.

\bibitem{Kol:2011vg}
B.~Kol and M.~Smolkin, ``{Black hole stereotyping: Induced gravito-static
  polarization},'' \href{http://dx.doi.org/10.1007/JHEP02(2012)010}{{\em JHEP}
  {\bfseries 02} (2012) 010},
\href{http://arxiv.org/abs/1110.3764}{{\ttfamily arXiv:1110.3764 [hep-th]}}.

\bibitem{Hui:2020xxx}
L.~Hui, A.~Joyce, R.~Penco, L.~Santoni, and A.~R. Solomon, ``{Static response
  and Love numbers of Schwarzschild black holes},''
  \href{http://dx.doi.org/10.1088/1475-7516/2021/04/052}{{\em JCAP} {\bfseries
  04} (2021) 052}, \href{http://arxiv.org/abs/2010.00593}{{\ttfamily
  arXiv:2010.00593 [hep-th]}}.

\bibitem{Hui:2021vcv}
L.~Hui, A.~Joyce, R.~Penco, L.~Santoni, and A.~R. Solomon, ``{Ladder symmetries
  of black holes. Implications for love numbers and no-hair theorems},''
  \href{http://dx.doi.org/10.1088/1475-7516/2022/01/032}{{\em JCAP} {\bfseries
  01} no.~01, (2022) 032}, \href{http://arxiv.org/abs/2105.01069}{{\ttfamily
  arXiv:2105.01069 [hep-th]}}.

\bibitem{Hui:2022vbh}
L.~Hui, A.~Joyce, R.~Penco, L.~Santoni, and A.~R. Solomon, ``{Near-zone
  symmetries of Kerr black holes},''
  \href{http://dx.doi.org/10.1007/JHEP09(2022)049}{{\em JHEP} {\bfseries 09}
  (2022) 049}, \href{http://arxiv.org/abs/2203.08832}{{\ttfamily
  arXiv:2203.08832 [hep-th]}}.

\bibitem{Rai:2024lho}
M.~Rai and L.~Santoni, ``{Ladder symmetries and Love numbers of
  Reissner-Nordstr\"om black holes},''
  \href{http://dx.doi.org/10.1007/JHEP07(2024)098}{{\em JHEP} {\bfseries 07}
  (2024) 098}, \href{http://arxiv.org/abs/2404.06544}{{\ttfamily
  arXiv:2404.06544 [gr-qc]}}.

\bibitem{LeTiec:2020spy}
A.~Le~Tiec and M.~Casals, ``{Spinning Black Holes Fall in Love},''
  \href{http://dx.doi.org/10.1103/PhysRevLett.126.131102}{{\em Phys. Rev.
  Lett.} {\bfseries 126} no.~13, (2021) 131102},
  \href{http://arxiv.org/abs/2007.00214}{{\ttfamily arXiv:2007.00214 [gr-qc]}}.

\bibitem{LeTiec:2020bos}
A.~Le~Tiec, M.~Casals, and E.~Franzin, ``{Tidal Love Numbers of Kerr Black
  Holes},'' \href{http://dx.doi.org/10.1103/PhysRevD.103.084021}{{\em Phys.
  Rev. D} {\bfseries 103} no.~8, (2021) 084021},
  \href{http://arxiv.org/abs/2010.15795}{{\ttfamily arXiv:2010.15795 [gr-qc]}}.

\bibitem{Charalambous:2021mea}
P.~Charalambous, S.~Dubovsky, and M.~M. Ivanov, ``{On the Vanishing of Love
  Numbers for Kerr Black Holes},''
  \href{http://dx.doi.org/10.1007/JHEP05(2021)038}{{\em JHEP} {\bfseries 05}
  (2021) 038}, \href{http://arxiv.org/abs/2102.08917}{{\ttfamily
  arXiv:2102.08917 [hep-th]}}.

\bibitem{Gurlebeck:2015xpa}
N.~G\"urlebeck, ``{No-hair theorem for Black Holes in Astrophysical
  Environments},'' \href{http://dx.doi.org/10.1103/PhysRevLett.114.151102}{{\em
  Phys. Rev. Lett.} {\bfseries 114} no.~15, (2015) 151102},
  \href{http://arxiv.org/abs/1503.03240}{{\ttfamily arXiv:1503.03240 [gr-qc]}}.

\bibitem{Rodriguez:2023xjd}
M.~J. Rodriguez, L.~Santoni, A.~R. Solomon, and L.~F. Temoche, ``{Love numbers
  for rotating black holes in higher dimensions},''
  \href{http://dx.doi.org/10.1103/PhysRevD.108.084011}{{\em Phys. Rev. D}
  {\bfseries 108} no.~8, (2023) 084011},
  \href{http://arxiv.org/abs/2304.03743}{{\ttfamily arXiv:2304.03743
  [hep-th]}}.

\bibitem{Charalambous:2023jgq}
P.~Charalambous and M.~M. Ivanov, ``{Scalar Love numbers and Love symmetries of
  5-dimensional Myers-Perry black holes},''
  \href{http://dx.doi.org/10.1007/JHEP07(2023)222}{{\em JHEP} {\bfseries 07}
  (2023) 222}, \href{http://arxiv.org/abs/2303.16036}{{\ttfamily
  arXiv:2303.16036 [hep-th]}}.

\bibitem{Charalambous:2024tdj}
P.~Charalambous, ``{Love numbers and Love symmetries for p-form and
  gravitational perturbations of higher-dimensional spherically symmetric black
  holes},'' \href{http://dx.doi.org/10.1007/JHEP04(2024)122}{{\em JHEP}
  {\bfseries 04} (2024) 122}, \href{http://arxiv.org/abs/2402.07574}{{\ttfamily
  arXiv:2402.07574 [hep-th]}}.

\bibitem{Charalambous:2025ekl}
P.~Charalambous, S.~Dubovsky, and M.~M. Ivanov, ``{Love numbers of black
  p-branes: fine tuning, Love symmetries, and their geometrization},''
  \href{http://dx.doi.org/10.1007/JHEP06(2025)180}{{\em JHEP} {\bfseries 06}
  (2025) 180}, \href{http://arxiv.org/abs/2502.02694}{{\ttfamily
  arXiv:2502.02694 [hep-th]}}.

\bibitem{Glazer:2024eyi}
D.~Glazer, A.~Joyce, M.~J. Rodriguez, L.~Santoni, A.~R. Solomon, and L.~F.
  Temoche, ``{Higher-dimensional spinning black holes and effective field
  theory},'' \href{http://dx.doi.org/10.1007/JHEP03(2026)036}{{\em JHEP}
  {\bfseries 03} (2026) 036}, \href{http://arxiv.org/abs/2412.21090}{{\ttfamily
  arXiv:2412.21090 [hep-th]}}.

\bibitem{Vines:2017hyw}
J.~Vines, ``{Scattering of two spinning black holes in post-Minkowskian
  gravity, to all orders in spin, and effective-one-body mappings},''
  \href{http://dx.doi.org/10.1088/1361-6382/aaa3a8}{{\em Class. Quant. Grav.}
  {\bfseries 35} no.~8, (2018) 084002},
  \href{http://arxiv.org/abs/1709.06016}{{\ttfamily arXiv:1709.06016 [gr-qc]}}.

\bibitem{Chung:2018kqs}
M.-Z. Chung, Y.-T. Huang, J.-W. Kim, and S.~Lee, ``{The simplest massive
  S-matrix: from minimal coupling to Black Holes},''
  \href{http://dx.doi.org/10.1007/JHEP04(2019)156}{{\em JHEP} {\bfseries 04}
  (2019) 156}, \href{http://arxiv.org/abs/1812.08752}{{\ttfamily
  arXiv:1812.08752 [hep-th]}}.

\bibitem{Arkani-Hamed:2019ymq}
N.~Arkani-Hamed, Y.-t. Huang, and D.~O'Connell, ``{Kerr black holes as
  elementary particles},''
  \href{http://dx.doi.org/10.1007/JHEP01(2020)046}{{\em JHEP} {\bfseries 01}
  (2020) 046}, \href{http://arxiv.org/abs/1906.10100}{{\ttfamily
  arXiv:1906.10100 [hep-th]}}.

\bibitem{BenAchour:2022uqo}
J.~Ben~Achour, E.~R. Livine, S.~Mukohyama, and J.-P. Uzan, ``{Hidden symmetry
  of the static response of black holes: applications to Love numbers},''
  \href{http://dx.doi.org/10.1007/JHEP07(2022)112}{{\em JHEP} {\bfseries 07}
  (2022) 112}, \href{http://arxiv.org/abs/2202.12828}{{\ttfamily
  arXiv:2202.12828 [gr-qc]}}.

\bibitem{Charalambous:2021kcz}
P.~Charalambous, S.~Dubovsky, and M.~M. Ivanov, ``{Hidden Symmetry of Vanishing
  Love Numbers},'' \href{http://dx.doi.org/10.1103/PhysRevLett.127.101101}{{\em
  Phys. Rev. Lett.} {\bfseries 127} no.~10, (2021) 101101},
  \href{http://arxiv.org/abs/2103.01234}{{\ttfamily arXiv:2103.01234
  [hep-th]}}.

\bibitem{Charalambous:2022rre}
P.~Charalambous, S.~Dubovsky, and M.~M. Ivanov, ``{Love symmetry},''
  \href{http://dx.doi.org/10.1007/JHEP10(2022)175}{{\em JHEP} {\bfseries 10}
  (2022) 175}, \href{http://arxiv.org/abs/2209.02091}{{\ttfamily
  arXiv:2209.02091 [hep-th]}}.

\bibitem{Berens:2022ebl}
R.~Berens, L.~Hui, and Z.~Sun, ``{Ladder symmetries of black holes and de
  Sitter space: love numbers and quasinormal modes},''
  \href{http://dx.doi.org/10.1088/1475-7516/2023/06/056}{{\em JCAP} {\bfseries
  06} (2023) 056}, \href{http://arxiv.org/abs/2212.09367}{{\ttfamily
  arXiv:2212.09367 [hep-th]}}.

\bibitem{Combaluzier-Szteinsznaider:2024sgb}
O.~Combaluzier-Szteinsznaider, L.~Hui, L.~Santoni, A.~R. Solomon, and S.~S.~C.
  Wong, ``{Symmetries of vanishing nonlinear Love numbers of Schwarzschild
  black holes},'' \href{http://dx.doi.org/10.1007/JHEP03(2025)124}{{\em JHEP}
  {\bfseries 03} (2025) 124}, \href{http://arxiv.org/abs/2410.10952}{{\ttfamily
  arXiv:2410.10952 [gr-qc]}}.

\bibitem{Lupsasca:2025pnt}
A.~Lupsasca, ``{Why there is no Love in black holes},''
  \href{http://arxiv.org/abs/2506.05298}{{\ttfamily arXiv:2506.05298 [gr-qc]}}.

\bibitem{Parra-Martinez:2025bcu}
J.~Parra-Martinez and A.~Podo, ``{Naturalness of vanishing black-hole tides},''
  \href{http://arxiv.org/abs/2510.20694}{{\ttfamily arXiv:2510.20694
  [hep-th]}}.

\bibitem{Berens:2025okm}
R.~Berens, L.~Hui, D.~McLoughlin, R.~Penco, and J.~Staunton, ``{Geometric
  Symmetries for the Vanishing of the Black Hole Tidal Love Numbers},''
  \href{http://arxiv.org/abs/2510.18952}{{\ttfamily arXiv:2510.18952
  [hep-th]}}.

\bibitem{Berens:2025jfs}
R.~Berens, L.~Hui, D.~McLoughlin, A.~R. Solomon, and J.~Staunton, ``{Ladder
  Symmetries of Higher Dimensional Black Holes},''
  \href{http://arxiv.org/abs/2510.26748}{{\ttfamily arXiv:2510.26748
  [hep-th]}}.

\bibitem{Poisson:2020vap}
E.~Poisson, ``{Compact body in a tidal environment: New types of relativistic
  Love numbers, and a post-Newtonian operational definition for tidally induced
  multipole moments},''
  \href{http://dx.doi.org/10.1103/PhysRevD.103.064023}{{\em Phys. Rev. D}
  {\bfseries 103} no.~6, (2021) 064023},
  \href{http://arxiv.org/abs/2012.10184}{{\ttfamily arXiv:2012.10184 [gr-qc]}}.

\bibitem{Poisson:2021yau}
E.~Poisson, ``{Tidally induced multipole moments of a nonrotating black hole
  vanish to all post-Newtonian orders},''
  \href{http://dx.doi.org/10.1103/PhysRevD.104.104062}{{\em Phys. Rev. D}
  {\bfseries 104} no.~10, (2021) 104062},
  \href{http://arxiv.org/abs/2108.07328}{{\ttfamily arXiv:2108.07328 [gr-qc]}}.

\bibitem{Riva:2023rcm}
M.~M. Riva, L.~Santoni, N.~Savi\'c, and F.~Vernizzi, ``{Vanishing of nonlinear
  tidal Love numbers of Schwarzschild black holes},''
  \href{http://dx.doi.org/10.1016/j.physletb.2024.138710}{{\em Phys. Lett. B}
  {\bfseries 854} (2024) 138710},
  \href{http://arxiv.org/abs/2312.05065}{{\ttfamily arXiv:2312.05065 [gr-qc]}}.

\bibitem{Iteanu:2024dvx}
S.~Iteanu, M.~M. Riva, L.~Santoni, N.~Savi{\'c}, and F.~Vernizzi, ``{Vanishing
  of quadratic Love numbers of Schwarzschild black holes},''
  \href{http://dx.doi.org/10.1007/JHEP02(2025)174}{{\em JHEP} {\bfseries 02}
  (2025) 174}, \href{http://arxiv.org/abs/2410.03542}{{\ttfamily
  arXiv:2410.03542 [gr-qc]}}.

\bibitem{DeLuca:2023mio}
V.~De~Luca, J.~Khoury, and S.~S.~C. Wong, ``{Nonlinearities in the tidal Love
  numbers of black holes},''
  \href{http://dx.doi.org/10.1103/PhysRevD.108.024048}{{\em Phys. Rev. D}
  {\bfseries 108} no.~2, (2023) 024048},
  \href{http://arxiv.org/abs/2305.14444}{{\ttfamily arXiv:2305.14444 [gr-qc]}}.

\bibitem{Gounis:2024hcm}
L.~R. Gounis, A.~Kehagias, and A.~Riotto, ``{The Vanishing of the Non-linear
  Static Love Number of Kerr Black Holes and the Role of Symmetries},''
  \href{http://arxiv.org/abs/2412.08249}{{\ttfamily arXiv:2412.08249 [gr-qc]}}.

\bibitem{Kehagias:2024rtz}
A.~Kehagias and A.~Riotto, ``{Black Holes in a Gravitational Field: The
  Non-linear Static Love Number of Schwarzschild Black Holes Vanishes},''
  \href{http://arxiv.org/abs/2410.11014}{{\ttfamily arXiv:2410.11014 [gr-qc]}}.

\bibitem{Hanson:1974qy}
A.~J. Hanson and T.~Regge, ``{The Relativistic Spherical Top},''
  \href{http://dx.doi.org/10.1016/0003-4916(74)90046-3}{{\em Annals Phys.}
  {\bfseries 87} (1974) 498}.

\bibitem{Bailey:1975fe}
I.~Bailey and W.~Israel, ``{Lagrangian Dynamics of Spinning Particles and
  Polarized Media in General Relativity},''
  \href{http://dx.doi.org/10.1007/BF01609434}{{\em Commun. Math. Phys.}
  {\bfseries 42} (1975) 65--82}.

\bibitem{Porto:2005ac}
R.~A. Porto, ``{Post-Newtonian corrections to the motion of spinning bodies in
  NRGR},'' \href{http://dx.doi.org/10.1103/PhysRevD.73.104031}{{\em Phys. Rev.
  D} {\bfseries 73} (2006) 104031},
  \href{http://arxiv.org/abs/gr-qc/0511061}{{\ttfamily arXiv:gr-qc/0511061}}.

\bibitem{Steinhoff:2011sya}
J.~Steinhoff, ``{Canonical formulation of spin in general relativity},''
  \href{http://dx.doi.org/10.1002/andp.201000178}{{\em Annalen Phys.}
  {\bfseries 523} (2011) 296--353},
  \href{http://arxiv.org/abs/1106.4203}{{\ttfamily arXiv:1106.4203 [gr-qc]}}.

\bibitem{Delacretaz:2014oxa}
L.~V. Delacr\'etaz, S.~Endlich, A.~Monin, R.~Penco, and F.~Riva,
  ``{(Re-)Inventing the Relativistic Wheel: Gravity, Cosets, and Spinning
  Objects},'' \href{http://dx.doi.org/10.1007/JHEP11(2014)008}{{\em JHEP}
  {\bfseries 11} (2014) 008}, \href{http://arxiv.org/abs/1405.7384}{{\ttfamily
  arXiv:1405.7384 [hep-th]}}.

\bibitem{Goldberger:2020fot}
W.~D. Goldberger, J.~Li, and I.~Z. Rothstein, ``{Non-conservative effects on
  spinning black holes from world-line effective field theory},''
  \href{http://dx.doi.org/10.1007/JHEP06(2021)053}{{\em JHEP} {\bfseries 06}
  (2021) 053}, \href{http://arxiv.org/abs/2012.14869}{{\ttfamily
  arXiv:2012.14869 [hep-th]}}.

\bibitem{Porto:2016pyg}
R.~A. Porto, ``{The effective field theorist\textquoteright{}s approach to
  gravitational dynamics},''
  \href{http://dx.doi.org/10.1016/j.physrep.2016.04.003}{{\em Phys. Rept.}
  {\bfseries 633} (2016) 1--104},
  \href{http://arxiv.org/abs/1601.04914}{{\ttfamily arXiv:1601.04914
  [hep-th]}}.

\bibitem{Levi:2018nxp}
M.~Levi, ``{Effective Field Theories of Post-Newtonian Gravity: A comprehensive
  review},'' \href{http://dx.doi.org/10.1088/1361-6633/ab12bc}{{\em Rept. Prog.
  Phys.} {\bfseries 83} no.~7, (2020) 075901},
  \href{http://arxiv.org/abs/1807.01699}{{\ttfamily arXiv:1807.01699
  [hep-th]}}.

\bibitem{Chakrabarti:2013lua}
S.~Chakrabarti, T.~Delsate, and J.~Steinhoff, ``{New perspectives on neutron
  star and black hole spectroscopy and dynamic tides},''
  \href{http://arxiv.org/abs/1304.2228}{{\ttfamily arXiv:1304.2228 [gr-qc]}}.

\bibitem{Hinderer:2016eia}
T.~Hinderer {\em et~al.}, ``{Effects of neutron-star dynamic tides on
  gravitational waveforms within the effective-one-body approach},''
  \href{http://dx.doi.org/10.1103/PhysRevLett.116.181101}{{\em Phys. Rev.
  Lett.} {\bfseries 116} no.~18, (2016) 181101},
  \href{http://arxiv.org/abs/1602.00599}{{\ttfamily arXiv:1602.00599 [gr-qc]}}.

\bibitem{Steinhoff:2016rfi}
J.~Steinhoff, T.~Hinderer, A.~Buonanno, and A.~Taracchini, ``{Dynamical Tides
  in General Relativity: Effective Action and Effective-One-Body
  Hamiltonian},'' \href{http://dx.doi.org/10.1103/PhysRevD.94.104028}{{\em
  Phys. Rev. D} {\bfseries 94} no.~10, (2016) 104028},
  \href{http://arxiv.org/abs/1608.01907}{{\ttfamily arXiv:1608.01907 [gr-qc]}}.

\bibitem{Pratten:2021pro}
G.~Pratten, P.~Schmidt, and N.~Williams, ``{Impact of Dynamical Tides on the
  Reconstruction of the Neutron Star Equation of State},''
  \href{http://dx.doi.org/10.1103/PhysRevLett.129.081102}{{\em Phys. Rev.
  Lett.} {\bfseries 129} no.~8, (2022) 081102},
  \href{http://arxiv.org/abs/2109.07566}{{\ttfamily arXiv:2109.07566
  [astro-ph.HE]}}.

\bibitem{Pitre:2023xsr}
T.~Pitre and E.~Poisson, ``{General relativistic dynamical tides in binary
  inspirals without modes},''
  \href{http://dx.doi.org/10.1103/PhysRevD.109.064004}{{\em Phys. Rev. D}
  {\bfseries 109} no.~6, (2024) 064004},
  \href{http://arxiv.org/abs/2311.04075}{{\ttfamily arXiv:2311.04075 [gr-qc]}}.

\bibitem{Chakraborty:2023zed}
S.~Chakraborty, E.~Maggio, M.~Silvestrini, and P.~Pani, ``{Dynamical tidal Love
  numbers of Kerr-like compact objects},''
  \href{http://dx.doi.org/10.1103/PhysRevD.110.084042}{{\em Phys. Rev. D}
  {\bfseries 110} no.~8, (2024) 084042},
  \href{http://arxiv.org/abs/2310.06023}{{\ttfamily arXiv:2310.06023 [gr-qc]}}.

\bibitem{Perry:2023wmm}
M.~Perry and M.~J. Rodriguez, ``{Dynamical Love Numbers for Kerr Black
  Holes},'' \href{http://arxiv.org/abs/2310.03660}{{\ttfamily arXiv:2310.03660
  [gr-qc]}}.

\bibitem{Saketh:2023bul}
M.~V.~S. Saketh, Z.~Zhou, and M.~M. Ivanov, ``{Dynamical tidal response of Kerr
  black holes from scattering amplitudes},''
  \href{http://dx.doi.org/10.1103/PhysRevD.109.064058}{{\em Phys. Rev. D}
  {\bfseries 109} no.~6, (2024) 064058},
  \href{http://arxiv.org/abs/2307.10391}{{\ttfamily arXiv:2307.10391
  [hep-th]}}.

\bibitem{Jakobsen:2023pvx}
G.~U. Jakobsen, G.~Mogull, J.~Plefka, and B.~Sauer, ``{Tidal effects and
  renormalization at fourth post-Minkowskian order},''
  \href{http://dx.doi.org/10.1103/PhysRevD.109.L041504}{{\em Phys. Rev. D}
  {\bfseries 109} no.~4, (2024) L041504},
  \href{http://arxiv.org/abs/2312.00719}{{\ttfamily arXiv:2312.00719
  [hep-th]}}.

\bibitem{Mandal:2023hqa}
M.~K. Mandal, P.~Mastrolia, H.~O. Silva, R.~Patil, and J.~Steinhoff,
  ``{Renormalizing Love: tidal effects at the third post-Newtonian order},''
  \href{http://dx.doi.org/10.1007/JHEP02(2024)188}{{\em JHEP} {\bfseries 02}
  (2024) 188}, \href{http://arxiv.org/abs/2308.01865}{{\ttfamily
  arXiv:2308.01865 [hep-th]}}.

\bibitem{Ivanov:2024sds}
M.~M. Ivanov, Y.-Z. Li, J.~Parra-Martinez, and Z.~Zhou, ``{Gravitational Raman
  Scattering in Effective Field Theory: A Scalar Tidal Matching at O(G3)},''
  \href{http://dx.doi.org/10.1103/PhysRevLett.132.131401}{{\em Phys. Rev.
  Lett.} {\bfseries 132} no.~13, (2024) 131401},
  \href{http://arxiv.org/abs/2401.08752}{{\ttfamily arXiv:2401.08752
  [hep-th]}}. [Erratum: Phys.Rev.Lett. 134, 159901 (2025)].

\bibitem{Katagiri:2024wbg}
T.~Katagiri, K.~Yagi, and V.~Cardoso, ``{Relativistic dynamical tides:
  Subtleties and calibration},''
  \href{http://dx.doi.org/10.1103/PhysRevD.111.084080}{{\em Phys. Rev. D}
  {\bfseries 111} no.~8, (2025) 084080},
  \href{http://arxiv.org/abs/2409.18034}{{\ttfamily arXiv:2409.18034 [gr-qc]}}.

\bibitem{HegadeKR:2024agt}
A.~Hegade K.~R., J.~L. Ripley, and N.~Yunes, ``{Dynamical tidal response of
  nonrotating relativistic stars},''
  \href{http://dx.doi.org/10.1103/PhysRevD.109.104064}{{\em Phys. Rev. D}
  {\bfseries 109} no.~10, (2024) 104064},
  \href{http://arxiv.org/abs/2403.03254}{{\ttfamily arXiv:2403.03254 [gr-qc]}}.

\bibitem{Yu:2024uxt}
H.~Yu, P.~Arras, and N.~N. Weinberg, ``{Dynamical tides during the inspiral of
  rapidly spinning neutron stars: Solutions beyond mode resonance},''
  \href{http://dx.doi.org/10.1103/PhysRevD.110.024039}{{\em Phys. Rev. D}
  {\bfseries 110} no.~2, (2024) 024039},
  \href{http://arxiv.org/abs/2404.00147}{{\ttfamily arXiv:2404.00147 [gr-qc]}}.

\bibitem{DeLuca:2024ufn}
V.~De~Luca, A.~Garoffolo, J.~Khoury, and M.~Trodden, ``{Tidal Love numbers and
  Green{\textquoteright}s functions in black hole spacetimes},''
  \href{http://dx.doi.org/10.1103/PhysRevD.110.064081}{{\em Phys. Rev. D}
  {\bfseries 110} no.~6, (2024) 064081},
  \href{http://arxiv.org/abs/2407.07156}{{\ttfamily arXiv:2407.07156 [gr-qc]}}.

\bibitem{Katagiri:2024fpn}
T.~Katagiri, V.~Cardoso, T.~Ikeda, and K.~Yagi, ``{Tidal response beyond vacuum
  general relativity with a canonical definition},''
  \href{http://dx.doi.org/10.1103/PhysRevD.111.084081}{{\em Phys. Rev. D}
  {\bfseries 111} no.~8, (2025) 084081},
  \href{http://arxiv.org/abs/2410.02531}{{\ttfamily arXiv:2410.02531 [gr-qc]}}.

\bibitem{Bhatt:2024rpx}
R.~P. Bhatt, S.~Chakraborty, and S.~Bose, ``{Response of a Kerr black hole to a
  generic tidal perturbation},''
  \href{http://arxiv.org/abs/2412.15117}{{\ttfamily arXiv:2412.15117 [gr-qc]}}.

\bibitem{Saketh:2024juq}
M.~V.~S. Saketh, Z.~Zhou, S.~Ghosh, J.~Steinhoff, and D.~Chatterjee,
  ``{Investigating tidal heating in neutron stars via gravitational Raman
  scattering},'' \href{http://dx.doi.org/10.1103/PhysRevD.110.103001}{{\em
  Phys. Rev. D} {\bfseries 110} (2024) 103001},
  \href{http://arxiv.org/abs/2407.08327}{{\ttfamily arXiv:2407.08327 [gr-qc]}}.

\bibitem{Chakraborty:2025wvs}
S.~Chakraborty, V.~De~Luca, L.~Gualtieri, and P.~Pani, ``{Dynamical Love
  numbers of black holes: theory and gravitational waveforms},''
  \href{http://arxiv.org/abs/2507.22994}{{\ttfamily arXiv:2507.22994 [gr-qc]}}.

\bibitem{Pitre:2025qdf}
T.~Pitre and E.~Poisson, ``{Impact of nonlinearities on relativistic dynamical
  tides in compact binary inspirals},''
  \href{http://arxiv.org/abs/2506.08722}{{\ttfamily arXiv:2506.08722 [gr-qc]}}.

\bibitem{Kobayashi:2025swn}
H.~Kobayashi, S.~Mukohyama, N.~Oshita, K.~Takahashi, and V.~Yingcharoenrat,
  ``{Parametrized tidal dissipation numbers of nonrotating black holes},''
  \href{http://dx.doi.org/10.1103/yclz-lbbs}{{\em Phys. Rev. D} {\bfseries 112}
  no.~8, (2025) 084061}, \href{http://arxiv.org/abs/2505.19725}{{\ttfamily
  arXiv:2505.19725 [gr-qc]}}.

\bibitem{Chia:2020yla}
H.~S. Chia, ``{Tidal deformation and dissipation of rotating black holes},''
  \href{http://dx.doi.org/10.1103/PhysRevD.104.024013}{{\em Phys. Rev. D}
  {\bfseries 104} no.~2, (2021) 024013},
  \href{http://arxiv.org/abs/2010.07300}{{\ttfamily arXiv:2010.07300 [gr-qc]}}.

\bibitem{Chia:2024bwc}
H.~S. Chia, Z.~Zhou, and M.~M. Ivanov, ``{Bring the Heat: Tidal Heating
  Constraints for Black Holes and Exotic Compact Objects from the
  LIGO-Virgo-KAGRA Data},'' \href{http://arxiv.org/abs/2404.14641}{{\ttfamily
  arXiv:2404.14641 [gr-qc]}}.

\bibitem{Mano:1996vt}
S.~Mano, H.~Suzuki, and E.~Takasugi, ``{Analytic solutions of the Teukolsky
  equation and their low frequency expansions},''
  \href{http://dx.doi.org/10.1143/PTP.95.1079}{{\em Prog. Theor. Phys.}
  {\bfseries 95} (1996) 1079--1096},
  \href{http://arxiv.org/abs/gr-qc/9603020}{{\ttfamily arXiv:gr-qc/9603020}}.

\bibitem{Sasaki:2003xr}
M.~Sasaki and H.~Tagoshi, ``{Analytic black hole perturbation approach to
  gravitational radiation},'' \href{http://dx.doi.org/10.12942/lrr-2003-6}{{\em
  Living Rev. Rel.} {\bfseries 6} (2003) 6},
  \href{http://arxiv.org/abs/gr-qc/0306120}{{\ttfamily arXiv:gr-qc/0306120}}.

\bibitem{Caron-Huot:2025tlq}
S.~Caron-Huot, M.~Correia, G.~Isabella, and M.~Solon, ``{Gravitational Wave
  Scattering via the Born Series: Scalar Tidal Matching to $\mathcal{O}(G^7)$
  and Beyond},'' \href{http://arxiv.org/abs/2503.13593}{{\ttfamily
  arXiv:2503.13593 [hep-th]}}.

\bibitem{Correia:2024jgr}
M.~Correia and G.~Isabella, ``{The Born regime of gravitational amplitudes},''
  \href{http://dx.doi.org/10.1007/JHEP03(2025)144}{{\em JHEP} {\bfseries 03}
  (2025) 144}, \href{http://arxiv.org/abs/2406.13737}{{\ttfamily
  arXiv:2406.13737 [hep-th]}}.

\bibitem{10.2307/78902}
S.~Chandrasekhar and S.~Detweiler, ``The quasi-normal modes of the
  schwarzschild black hole,'' {\em Proceedings of the Royal Society of London.
  Series A, Mathematical and Physical Sciences} {\bfseries 344} no.~1639,
  (1975) 441--452. \url{http://www.jstor.org/stable/78902}.

\bibitem{1975RSPSA.343..289C}
S.~Chandrasekhar, ``{On the Equations Governing the Perturbations of the
  Schwarzschild Black Hole},''
  \href{http://dx.doi.org/10.1098/rspa.1975.0066}{{\em Proceedings of the Royal
  Society of London Series A} {\bfseries 343} (1975) 289--298}.

\bibitem{Chandrasekhar:1985kt}
S.~Chandrasekhar, ``{The mathematical theory of black holes},'' in {\em
  {Oxford, UK: Clarendon (1992) 646 p., Oxford, UK: Clarendon (1985) 646 P.}}
\newblock
1985.
\newblock

\bibitem{MR1392976}
A.~Ronveaux, ed., {\em Heun's differential equations}.
\newblock Oxford Science Publications. The Clarendon Press, Oxford University
  Press, New York, 1995.
\newblock With contributions by F. M. Arscott, S. Yu. Slavyanov, D. Schmidt, G.
  Wolf, P. Maroni and A. Duval.

\bibitem{slavjanov2000special}
S.~Slavjanov and W.~Lay, {\em Special Functions: A Unified Theory Based on
  Singularities}.
\newblock Oxford Science Publications. Oxford University Press, 2000.

\bibitem{Bateman:100233}
H.~Bateman and A.~Erdélyi, {\em {Higher transcendental functions}}.
\newblock Calif. Inst. Technol. Bateman Manuscr. Project. McGraw-Hill, New
  York, NY, 1955.
\newblock \url{https://cds.cern.ch/record/100233}.

\bibitem{Aminov:2020yma}
G.~Aminov, A.~Grassi, and Y.~Hatsuda, ``{Black Hole Quasinormal Modes and
  Seiberg\textendash{}Witten Theory},''
  \href{http://dx.doi.org/10.1007/s00023-021-01137-x}{{\em Annales Henri
  Poincare} {\bfseries 23} no.~6, (2022) 1951--1977},
  \href{http://arxiv.org/abs/2006.06111}{{\ttfamily arXiv:2006.06111
  [hep-th]}}.

\bibitem{Bonelli:2021uvf}
G.~Bonelli, C.~Iossa, D.~P. Lichtig, and A.~Tanzini, ``{Exact solution of Kerr
  black hole perturbations via CFT2 and instanton counting: Greybody factor,
  quasinormal modes, and Love numbers},''
  \href{http://dx.doi.org/10.1103/PhysRevD.105.044047}{{\em Phys. Rev. D}
  {\bfseries 105} no.~4, (2022) 044047},
  \href{http://arxiv.org/abs/2105.04483}{{\ttfamily arXiv:2105.04483
  [hep-th]}}.

\bibitem{Bonelli:2022ten}
G.~Bonelli, C.~Iossa, D.~Panea~Lichtig, and A.~Tanzini, ``{Irregular Liouville
  Correlators and Connection Formulae for Heun Functions},''
  \href{http://dx.doi.org/10.1007/s00220-022-04497-5}{{\em Commun. Math. Phys.}
  {\bfseries 397} no.~2, (2023) 635--727},
  \href{http://arxiv.org/abs/2201.04491}{{\ttfamily arXiv:2201.04491
  [hep-th]}}.

\bibitem{Lisovyy:2022flm}
O.~Lisovyy and A.~Naidiuk, ``{Perturbative connection formulas for Heun
  equations},'' \href{http://dx.doi.org/10.1088/1751-8121/ac9ba7}{{\em J. Phys.
  A} {\bfseries 55} no.~43, (2022) 434005},
  \href{http://arxiv.org/abs/2208.01604}{{\ttfamily arXiv:2208.01604
  [math-ph]}}.

\bibitem{Aminov:2023jve}
G.~Aminov, P.~Arnaudo, G.~Bonelli, A.~Grassi, and A.~Tanzini, ``{Black hole
  perturbation theory and multiple polylogarithms},''
  \href{http://dx.doi.org/10.1007/JHEP11(2023)059}{{\em JHEP} {\bfseries 11}
  (2023) 059}, \href{http://arxiv.org/abs/2307.10141}{{\ttfamily
  arXiv:2307.10141 [hep-th]}}.

\bibitem{RWhite}
R.~B. White, {\em Asymptotic Analysis of Differential Equations}.
\newblock World Scientific, 2010.

\bibitem{Unruh:1976fm}
W.~G. Unruh, ``{Absorption Cross-Section of Small Black Holes},''
  \href{http://dx.doi.org/10.1103/PhysRevD.14.3251}{{\em Phys. Rev. D}
  {\bfseries 14} (1976) 3251--3259}.

\bibitem{Bezerra:2013iha}
V.~B. Bezerra, H.~S. Vieira, and A.~A. Costa, ``{The Klein-Gordon equation in
  the spacetime of a charged and rotating black hole},''
  \href{http://dx.doi.org/10.1088/0264-9381/31/4/045003}{{\em Class. Quant.
  Grav.} {\bfseries 31} no.~4, (2014) 045003},
  \href{http://arxiv.org/abs/1312.4823}{{\ttfamily arXiv:1312.4823 [gr-qc]}}.

\bibitem{Hui:2019aqm}
L.~Hui, D.~Kabat, X.~Li, L.~Santoni, and S.~S.~C. Wong, ``{Black Hole Hair from
  Scalar Dark Matter},''
  \href{http://dx.doi.org/10.1088/1475-7516/2019/06/038}{{\em JCAP} {\bfseries
  06} (2019) 038}, \href{http://arxiv.org/abs/1904.12803}{{\ttfamily
  arXiv:1904.12803 [gr-qc]}}.

\bibitem{Bucciotti:2023bvw}
B.~Bucciotti and E.~Trincherini, ``{interplay between black holes and
  ultralight dark matter: analytic solutions},''
  \href{http://dx.doi.org/10.1007/JHEP11(2023)193}{{\em JHEP} {\bfseries 11}
  (2023) 193}, \href{http://arxiv.org/abs/2309.02482}{{\ttfamily
  arXiv:2309.02482 [hep-th]}}.

\bibitem{Hui:2022sri}
L.~Hui, Y.~T.~A. Law, L.~Santoni, G.~Sun, G.~M. Tomaselli, and E.~Trincherini,
  ``{Black hole superradiance with dark matter accretion},''
  \href{http://dx.doi.org/10.1103/PhysRevD.107.104018}{{\em Phys. Rev. D}
  {\bfseries 107} no.~10, (2023) 104018},
  \href{http://arxiv.org/abs/2208.06408}{{\ttfamily arXiv:2208.06408 [gr-qc]}}.

\bibitem{NIST:DLMF}
``{\it NIST Digital Library of Mathematical Functions}.''
  \url{https://dlmf.nist.gov/}, release 1.2.4 of 2025-03-15.
\newblock \url{https://dlmf.nist.gov/}. F.~W.~J. Olver, A.~B. {Olde Daalhuis},
  D.~W. Lozier, B.~I. Schneider, R.~F. Boisvert, C.~W. Clark, B.~R. Miller,
  B.~V. Saunders, H.~S. Cohl, and M.~A. McClain, eds.

\bibitem{Magnus:1966}
R.~P.~S. Wilhelm~Magnus, Fritz~Oberhettinger,
  \href{http://dx.doi.org/10.1007/978-3-662-11761-3}{{\em {Formulas and
  Theorems for the Special Functions of Mathematical Physics}}}.
\newblock Grundlehren der mathematischen Wissenschaften. Springer Berlin,
  Heidelberg, Berlin, 1966.

\bibitem{Ivanov:2025ozg}
M.~M. Ivanov, Y.-Z. Li, J.~Parra-Martinez, and Z.~Zhou, ``{Resummation of
  Universal Tails in Gravitational Waveforms},''
  \href{http://arxiv.org/abs/2504.07862}{{\ttfamily arXiv:2504.07862
  [hep-th]}}.

\bibitem{Schwinger:1960qe}
J.~S. Schwinger, ``{Brownian motion of a quantum oscillator},''
  \href{http://dx.doi.org/10.1063/1.1703727}{{\em J. Math. Phys.} {\bfseries 2}
  (1961) 407--432}.

\bibitem{Keldysh:1964ud}
L.~V. Keldysh, ``{Diagram technique for nonequilibrium processes},'' {\em Zh.
  Eksp. Teor. Fiz.} {\bfseries 47} (1964) 1515--1527.

\bibitem{kamenev2011field}
A.~Kamenev, {\em Field Theory of Non-Equilibrium Systems}.
\newblock Cambridge University Press, 2011.

\bibitem{Akyuz:2023lsm}
C.~O. Akyuz, G.~Goon, and R.~Penco, ``{The Schwinger-Keldysh coset
  construction},'' \href{http://dx.doi.org/10.1007/JHEP06(2024)004}{{\em JHEP}
  {\bfseries 06} (2024) 004}, \href{http://arxiv.org/abs/2306.17232}{{\ttfamily
  arXiv:2306.17232 [hep-th]}}.

\bibitem{Haehl:2015foa}
F.~M. Haehl, R.~Loganayagam, and M.~Rangamani, ``{The Fluid Manifesto: Emergent
  symmetries, hydrodynamics, and black holes},''
  \href{http://dx.doi.org/10.1007/JHEP01(2016)184}{{\em JHEP} {\bfseries 01}
  (2016) 184}, \href{http://arxiv.org/abs/1510.02494}{{\ttfamily
  arXiv:1510.02494 [hep-th]}}.

\bibitem{Crossley:2015evo}
M.~Crossley, P.~Glorioso, and H.~Liu, ``{Effective field theory of dissipative
  fluids},'' \href{http://dx.doi.org/10.1007/JHEP09(2017)095}{{\em JHEP}
  {\bfseries 09} (2017) 095}, \href{http://arxiv.org/abs/1511.03646}{{\ttfamily
  arXiv:1511.03646 [hep-th]}}.

\bibitem{Liu:2018kfw}
H.~Liu and P.~Glorioso, ``{Lectures on non-equilibrium effective field theories
  and fluctuating hydrodynamics},''
  \href{http://dx.doi.org/10.22323/1.305.0008}{{\em PoS} {\bfseries TASI2017}
  (2018) 008}, \href{http://arxiv.org/abs/1805.09331}{{\ttfamily
  arXiv:1805.09331 [hep-th]}}.

\bibitem{Haehl:2024pqu}
F.~M. Haehl and M.~Rangamani, ``{Records from the S-Matrix Marathon:
  Schwinger-Keldysh Formalism},''
\newblock 10, 2024.
\newblock \href{http://arxiv.org/abs/2410.10602}{{\ttfamily arXiv:2410.10602
  [hep-th]}}.

\bibitem{Glorioso:2016gsa}
P.~Glorioso and H.~Liu, ``{The second law of thermodynamics from symmetry and
  unitarity},'' \href{http://arxiv.org/abs/1612.07705}{{\ttfamily
  arXiv:1612.07705 [hep-th]}}.

\bibitem{Weinberg:2005vy}
S.~Weinberg, ``{Quantum contributions to cosmological correlations},''
  \href{http://dx.doi.org/10.1103/PhysRevD.72.043514}{{\em Phys. Rev. D}
  {\bfseries 72} (2005) 043514},
  \href{http://arxiv.org/abs/hep-th/0506236}{{\ttfamily arXiv:hep-th/0506236}}.

\bibitem{Ivanov:2022hlo}
M.~M. Ivanov and Z.~Zhou, ``{Revisiting the matching of black hole tidal
  responses: A systematic study of relativistic and logarithmic corrections},''
  \href{http://dx.doi.org/10.1103/PhysRevD.107.084030}{{\em Phys. Rev. D}
  {\bfseries 107} no.~8, (2023) 084030},
  \href{http://arxiv.org/abs/2208.08459}{{\ttfamily arXiv:2208.08459
  [hep-th]}}.

\bibitem{Feynman:1963fq}
R.~P. Feynman and F.~L. Vernon, Jr., ``{The Theory of a general quantum system
  interacting with a linear dissipative system},''
  \href{http://dx.doi.org/10.1016/0003-4916(63)90068-X}{{\em Annals Phys.}
  {\bfseries 24} (1963) 118--173}.

\bibitem{Caldeira:1982iu}
A.~O. Caldeira and A.~J. Leggett, ``{Path integral approach to quantum Brownian
  motion},'' \href{http://dx.doi.org/10.1016/0378-4371(83)90013-4}{{\em Physica
  A} {\bfseries 121} (1983) 587--616}.

\bibitem{Calzetta:1986cq}
E.~Calzetta and B.~L. Hu, ``{Nonequilibrium Quantum Fields: Closed Time Path
  Effective Action, Wigner Function and Boltzmann Equation},''
  \href{http://dx.doi.org/10.1103/PhysRevD.37.2878}{{\em Phys. Rev. D}
  {\bfseries 37} (1988) 2878}.

\bibitem{Kamenev:2009jj}
A.~Kamenev and A.~Levchenko, ``{Keldysh technique and nonlinear sigma-model:
  Basic principles and applications},''
  \href{http://dx.doi.org/10.1080/00018730902850504}{{\em Adv. Phys.}
  {\bfseries 58} (2009) 197}, \href{http://arxiv.org/abs/0901.3586}{{\ttfamily
  arXiv:0901.3586 [cond-mat.other]}}.

\bibitem{Salcedo:2025ezu}
S.~A. Salcedo, T.~Colas, L.~Dufner, and E.~Pajer, ``{An Open System Approach to
  Gravity},'' \href{http://arxiv.org/abs/2507.03103}{{\ttfamily
  arXiv:2507.03103 [hep-th]}}.

\bibitem{Saketh:2022xjb}
M.~V.~S. Saketh, J.~Steinhoff, J.~Vines, and A.~Buonanno, ``{Modeling horizon
  absorption in spinning binary black holes using effective worldline
  theory},'' \href{http://dx.doi.org/10.1103/PhysRevD.107.084006}{{\em Phys.
  Rev. D} {\bfseries 107} no.~8, (2023) 084006},
  \href{http://arxiv.org/abs/2212.13095}{{\ttfamily arXiv:2212.13095 [gr-qc]}}.

\bibitem{Cheung:2018wkq}
C.~Cheung, I.~Z. Rothstein, and M.~P. Solon, ``{From Scattering Amplitudes to
  Classical Potentials in the Post-Minkowskian Expansion},''
  \href{http://dx.doi.org/10.1103/PhysRevLett.121.251101}{{\em Phys. Rev.
  Lett.} {\bfseries 121} no.~25, (2018) 251101},
  \href{http://arxiv.org/abs/1808.02489}{{\ttfamily arXiv:1808.02489
  [hep-th]}}.

\bibitem{Bern:2019crd}
Z.~Bern, C.~Cheung, R.~Roiban, C.-H. Shen, M.~P. Solon, and M.~Zeng, ``{Black
  Hole Binary Dynamics from the Double Copy and Effective Theory},''
  \href{http://dx.doi.org/10.1007/JHEP10(2019)206}{{\em JHEP} {\bfseries 10}
  (2019) 206}, \href{http://arxiv.org/abs/1908.01493}{{\ttfamily
  arXiv:1908.01493 [hep-th]}}.

\bibitem{Regge:1957td}
T.~Regge and J.~A. Wheeler, ``{Stability of a Schwarzschild singularity},''
\href{http://dx.doi.org/10.1103/PhysRev.108.1063}{{\em Phys. Rev.} {\bfseries
  108} (1957) 1063--1069}.

\bibitem{Zerilli:1970wzz}
F.~J. Zerilli, ``{Gravitational field of a particle falling in a schwarzschild
  geometry analyzed in tensor harmonics},''
  \href{http://dx.doi.org/10.1103/PhysRevD.2.2141}{{\em Phys. Rev. D}
  {\bfseries 2} (1970) 2141--2160}.

\bibitem{Creci:2021rkz}
G.~Creci, T.~Hinderer, and J.~Steinhoff, ``{Tidal response from scattering and
  the role of analytic continuation},''
  \href{http://dx.doi.org/10.1103/PhysRevD.104.124061}{{\em Phys. Rev. D}
  {\bfseries 104} no.~12, (2021) 124061},
  \href{http://arxiv.org/abs/2108.03385}{{\ttfamily arXiv:2108.03385 [gr-qc]}}.
  [Erratum: Phys.Rev.D 105, 109902 (2022)].

\bibitem{Bhatt:2024yyz}
R.~P. Bhatt, S.~Chakraborty, and S.~Bose, ``{Rotating black holes experience
  dynamical tides},''
  \href{http://dx.doi.org/10.1103/PhysRevD.111.L041504}{{\em Phys. Rev. D}
  {\bfseries 111} no.~4, (2025) L041504},
  \href{http://arxiv.org/abs/2406.09543}{{\ttfamily arXiv:2406.09543 [gr-qc]}}.

\bibitem{Kodama:2000fa}
H.~Kodama, A.~Ishibashi, and O.~Seto, ``{Brane world cosmology: Gauge invariant
  formalism for perturbation},''
  \href{http://dx.doi.org/10.1103/PhysRevD.62.064022}{{\em Phys. Rev.}
  {\bfseries D62} (2000) 064022},
\href{http://arxiv.org/abs/hep-th/0004160}{{\ttfamily arXiv:hep-th/0004160
  [hep-th]}}.

\bibitem{Kodama:2003jz}
H.~Kodama and A.~Ishibashi, ``{A Master equation for gravitational
  perturbations of maximally symmetric black holes in higher dimensions},''
  \href{http://dx.doi.org/10.1143/PTP.110.701}{{\em Prog. Theor. Phys.}
  {\bfseries 110} (2003) 701--722},
\href{http://arxiv.org/abs/hep-th/0305147}{{\ttfamily arXiv:hep-th/0305147
  [hep-th]}}.

\bibitem{Ishibashi:2003ap}
A.~Ishibashi and H.~Kodama, ``{Stability of higher dimensional Schwarzschild
  black holes},'' \href{http://dx.doi.org/10.1143/PTP.110.901}{{\em Prog.
  Theor. Phys.} {\bfseries 110} (2003) 901--919},
\href{http://arxiv.org/abs/hep-th/0305185}{{\ttfamily arXiv:hep-th/0305185
  [hep-th]}}.

\bibitem{Rosen:2020crj}
R.~A. Rosen and L.~Santoni, ``{Black hole perturbations of massive and
  partially massless spin-2 fields in (anti) de Sitter spacetime},''
  \href{http://dx.doi.org/10.1007/JHEP03(2021)139}{{\em JHEP} {\bfseries 03}
  (2021) 139}, \href{http://arxiv.org/abs/2010.00595}{{\ttfamily
  arXiv:2010.00595 [hep-th]}}.

\bibitem{1999physics...8003D}
G.~{Darboux}, ``{On a proposition relative to linear equations},''
  \href{http://dx.doi.org/10.48550/arXiv.physics/9908003}{{\em arXiv e-prints}
  (Aug., 1999) physics/9908003},
  \href{http://arxiv.org/abs/physics/9908003}{{\ttfamily arXiv:physics/9908003
  [physics.hist-ph]}}.

\bibitem{Glampedakis:2017rar}
K.~Glampedakis, A.~D. Johnson, and D.~Kennefick, ``{Darboux transformation in
  black hole perturbation theory},''
  \href{http://dx.doi.org/10.1103/PhysRevD.96.024036}{{\em Phys. Rev. D}
  {\bfseries 96} no.~2, (2017) 024036},
  \href{http://arxiv.org/abs/1702.06459}{{\ttfamily arXiv:1702.06459 [gr-qc]}}.

\bibitem{Hadad:2024lsf}
T.~Hadad, B.~Kol, and M.~Smolkin, ``{Gravito-magnetic polarization of
  Schwarzschild black hole},''
  \href{http://dx.doi.org/10.1007/JHEP06(2024)169}{{\em JHEP} {\bfseries 06}
  (2024) 169}, \href{http://arxiv.org/abs/2402.16172}{{\ttfamily
  arXiv:2402.16172 [hep-th]}}.

\bibitem{Solomon:2023ltn}
A.~R. Solomon, ``{Off-Shell Duality Invariance of Schwarzschild Perturbation
  Theory},'' \href{http://dx.doi.org/10.3390/particles6040061}{{\em Particles}
  {\bfseries 6} no.~4, (2023) 943--974},
  \href{http://arxiv.org/abs/2310.04502}{{\ttfamily arXiv:2310.04502 [gr-qc]}}.

\bibitem{Cooper:1994eh}
F.~Cooper, A.~Khare, and U.~Sukhatme, ``{Supersymmetry and quantum
  mechanics},'' \href{http://dx.doi.org/10.1016/0370-1573(94)00080-M}{{\em
  Phys. Rept.} {\bfseries 251} (1995) 267--385},
\href{http://arxiv.org/abs/hep-th/9405029}{{\ttfamily arXiv:hep-th/9405029
  [hep-th]}}.

\bibitem{Kobayashi:2025vgl}
H.~Kobayashi, S.~Mukohyama, N.~Oshita, K.~Takahashi, and V.~Yingcharoenrat,
  ``{Dynamical Tidal Response of Non-rotating Black Holes: Connecting the MST
  Formalism and Worldline EFT},''
  \href{http://arxiv.org/abs/2511.12580}{{\ttfamily arXiv:2511.12580 [gr-qc]}}.

\bibitem{Ivanov:2022qqt}
M.~M. Ivanov and Z.~Zhou, ``{Vanishing of Black Hole Tidal Love Numbers from
  Scattering Amplitudes},''
  \href{http://dx.doi.org/10.1103/PhysRevLett.130.091403}{{\em Phys. Rev.
  Lett.} {\bfseries 130} no.~9, (2023) 091403},
  \href{http://arxiv.org/abs/2209.14324}{{\ttfamily arXiv:2209.14324
  [hep-th]}}.

\end{thebibliography}\endgroup
}
﻿
﻿
﻿
\end{document}